\def\hb{H$\beta$\/}
\def\oiii{[\ion{O}{III}]$\lambda\lambda$4959,5007\/}
\def\oiiionly{[\ion{O}{III}]\/}
\def\oiiiseven{[\ion{O}{III}]$\lambda$5007\/}
\def\civ{\ion{C}{IV}$\lambda$1549\/}
\def\civonly{\ion{C}{IV}\/}
\def\siiv{\ion{Si}{IV}$\lambda$1397\/}

\def\aliii{\ion{Al}{III}$\lambda$1860\/}

\def\feii{\ion{Fe}{II}\/}
\def\kms{km\,s$^{-1}$\/}

\def\mbh{$M_{\rm BH}$\/}
\def\lledd{$L/L_{\rm Edd}$\/}
\def\ledd{$L_{\rm Edd}$\/}
\def\civonly{\ion{C}{IV}\/}

\def\c14{$c$(1/4)\/}
\documentclass{aa}  
\usepackage{ulem}
\usepackage{lipsum}
\usepackage{ragged2e}
\usepackage{natbib}
\usepackage[colorlinks=true,citecolor=blue, urlcolor=blue,linkcolor=blue]{hyperref}%

\usepackage{graphicx}
\usepackage{txfonts}
\usepackage{xcolor} 
\usepackage{rotating}
\usepackage{pdflscape}
\usepackage{multirow}
\usepackage{booktabs, makecell, longtable}

\begin{document} 
\renewcommand{\tabcolsep}{2pt}
\defcitealias{Deconto-Machado_2023}{Paper I}

   \title{Exploring the links between quasar winds and radio emission along the Main Sequence at high redshift\thanks{Based on observations collected at ESO under   programs 086.B-0774(A),   085.B-0162(A), and  083.B-0273(A).}}

    \titlerunning{Quasar winds and radio emission along the MS}
    \authorrunning{A. Deconto-Machado,  A. del Olmo, P. Marziani}
   \author{A. Deconto-Machado
          \inst{1}
          \and
          A. del Olmo\inst{1}
         \and
          P. Marziani\inst{2}
           }

   \institute{Instituto de Astrofísica de Andalucía, IAA-CSIC, E-18008, Granada, Spain\\
              \email{adeconto@iaa.es, chony@iaa.es}
         \and
             INAF, Osservatorio Astronomico di Padova, IT 35122, Padova, Italy\\
             \email{paola.marziani@inaf.it}
             }

   \date{Received ; accepted }

 
  \abstract
   {Despite the increasing prevalence of radio-loud (RL) sources at cosmic noon, our understanding of the underlying physics that governs the accretion disc outflows in these particular sources {and its dissimilarity with radio-quiet (RQ) quasars} remains somewhat limited.}
   {Disentangling the real impact of the radio-loudness and accretion on the outflow parameters remains a challenge to this day. 
   We present  10 new spectra of high-redshift and high-luminosity quasars and combine it with previous data at both high and low redshift with the aim to evaluate the role of  the   feedback from RL and RQ AGN. The final high-redshift {(1.5 $\lesssim$ z $\lesssim$ 3.9) high-luminosity (47.1 $\leq \mathrm{log(L)} \leq$ 48.5)} sample consists of a combination of {60 quasars from our ISAAC and the Hamburg-ESO surveys}. The low-redshift {(z $\leq$ 0.8)} sample has 84 quasars with optical and Faint Object Spectrograph (FOS) reanalyzed data.}
   {We perform a multicomponent analysis of optical and UV emission line profiles along the quasar main sequence, and provide a  relation to estimate the outflow main parameters (mass rate, thrust and kinetic power) {in both the BLR and NLR through the analysis of the \oiiiseven{} and \civ\  emission lines.} 
   }
   {{Spectrophotometric properties and line profile measurements are presented for \hb{}+\oiii{}, \siiv{}+\ion{O}{IV}]$\lambda$1402, \civ{}+\ion{He}{II}$\lambda$1640, and the 1900\AA\ blend.} High-ionization lines as such \civ{} and \oiiiseven{} usually present a significant asymmetry towards the blue especially in radio-quiet sources that is strong evidence of outflow motions. {In the ISAAC sample, 72\% of the quasars where \oiiionly{} is clearly detected {{present  significant outflows, with centroid velocity at half intensity blueshifted to values 
   greater than $\sim$250\,km\,s$^{-1}$. }} Radio-loud quasars tend to present slightly more modest blueshifted components in both the UV and optical ranges. The behavior of \oiiiseven{} mirrors the one of \civ{}, with blueshift amplitudes between the two lines showing a high degree of correlation that  appears unaffected by the presence of radio emission.}}
   {In contrast to low redshift,   both RL and RQ AGN outflow parameters at high luminosity appear in the range needed to provide feedback effects on their host galaxies.  
   Both high- and low-$z$ RL quasars exhibit smaller outflows compared to RQ, suggesting a potential role of radioloudness in mitigating outflow effects. Nevertheless, the radio-loudness effect on the AGN feedback is much less significant than the one of accretion {that emerges as the main driver of the nuclear outflows}.}
 

   \maketitle
%
\section{Introduction}
\par It is now an established result that quasar spectra do not scatter randomly around an average and that a systematic scheme is needed to organize their spectral diversity. One of the most successful tools to analyze such objects  makes use of independent observational properties obtained from the optical and UV emission lines, as well as from soft X-rays (the 4D Eigenvector 1 correlation space; \citealt{boroson_1992,sulenticetal00c}). As part of the 4DE1, it is possible to identify  a sequence of quasars in the plane defined by two optical parameters: the full width at half maximum of the \hb{} emission line (FWHM(H$\beta$)) and the ratio between the intensities of the blend of \ion{Fe}{II} emission lines at 4570$\AA$ and \hb{} ($R_{\rm \ion{Fe}{II}}$). This relation  has become to be known as the Main Sequence of quasars \citep[MS; ][]{Sulentic_2000, sulenticetal00c,marziani_2001,shen_2014, Marziani2018}. Several multi-frequency correlations associated with the MS are well established at low redshift and low luminosity \citep[e.g. see the summary tables in][]{sulentic_2011,fraix-burnetetal17}.

\par  At low-redshift, it appears that the jetted\footnote{Here we define ``jetted'' sources the ones that present a radio loudness parameter $R_{\textrm{K}}\gtrsim 1.8$ \citep[
also called here as ``radio-loud'',][see also \S \ref{sec:radiodata}]{Ganci_2019}.} (radio-loud, RL) sources show a preference for the Pop. B domain in the MS, while the radio-quiet (RQ) are distributed equally between Pop. A and Pop. B \citep{Zamfir_2008}. This suggests potential differences in spectral and physical properties between RL and RQ sources. The phenomenology involves broader lines for RL with respect to RQ (as most RL are in Pop. B), and an optical spectrum showing lower \feii\ and a higher  ionization degree with respect to full samples of RQ quasars in both the BLR and the NLR \citep{marzianietal03b,zamfir_2010,Kovacevic_2015,Coziol_2017,Ganci_2019}. Consequently, the low-$z$ radio-quiet sources seem to follow different distributions of Eddington ratio and black hole masses than the radio-loud ones \citep{woourry02,Marziani_2003,sikoraetal07, fraix-burnetetal17}. In general,  at low-$z$, RQ sources are the ones that usually present smaller masses and larger Eddington ratio.

Outflows appear to be ubiquitous in both the BLR and NLR \citep[e.g.,][and references therein]{Coatman_2019, vietrietal20,Marziani_2022ApJ}. However,  at low $z$\ mild-ionization gas outflows observed in jetted sources appear to be weaker than in the RQ ones \citep{marziani_1996,Bachev_2004,Punsly_2010,richards_2011}. This may not be at odds with the higher ionization degree revealed in RLs \citep[e.g., ][]{Buttiglione_2010,Mengistue_2023}, although a full physical explanation is still missing, mainly due to the still-limited  scope of detailed studies of accretion disk  outflows in RL quasars. This difference between RQ and RL outflows is observed at both high- and low-redshift. A key distinction is  however that at high redshift, the outflows appear significantly stronger compared to those at low redshift  \citep{richards_2011,sulentic_2017,Deconto-Machado_2023}. At low-redshift ranges, many authors found ionized gas outflows with kiloparsec scales, however their impact is only in the central region of the galaxies \citep[e.g.][and references therein]{Kim_2023}.



\par More and more RL sources are being discovered at the cosmic noon \citep[e.g.,][ and references therein]{Patil_2022,breiding_2023}. Their rest-frame optical properties remain poorly studied since they require high-S/N moderate dispersion IR spectroscopy. 
For this reason, we obtained new IR spectroscopic observations for other ten high-redshift ($z \sim 1.5 - 2.5$) and high-luminosity ($M_\mathrm{B} \lesssim -27$) quasars. The targets are intended to cover the rest-frame \hb\ range, and include both RQ and jetted (or RL) objects, completing the  sample reported in \citet[][hereafter \citetalias{Deconto-Machado_2023}]{Deconto-Machado_2023}. {Our main aim is to clarify the origin of several phenomenological differences between RL and RQ and in particular the relation between accretion status,  accretion disk  outflows, and the presence of powerful relativistic ejections, exploiting the MS and its correlation with the  \civ\ and \oiii\ emission. 
} 


In addition to the rest-frame optical data, we collected archival rest-frame UV data of the targets in the present paper and in \citetalias{Deconto-Machado_2023}, and we defined comparison samples for which both rest-frame UV and optical data are available (Section \ref{samples}).  The data analysis includes discussion about UV and optical regions for the two different quasar populations and the spectral types  that have been defined along the quasar MS (Section \ref{section:obs_data_analysis}). The measurements carried out in both the optical (\feii, \hb{}, +\oiii{}) and UV (\siiv{}, \civ{}, and the 1900$\AA$\ blend) regions are reported in Section \ref{section:results} whereas  their interpretation  in terms of the outflow dynamical parameters is presented in Section    \ref{section:discussion}. We show that the RQ and RL population are both associated with high Eddington ratio at high luminosity, and that BLR and NLR scale  outflows are consistently powerful, albeit being somewhat weaker in RL sources. This difference  between RQ and RL outflows might be due to an apparently minor  effect related to the propagation of  the relativistic jet.  In Section \ref{section:conclusions}, we summarize the key findings and discuss some implications of  our results.  
   
\begin{table}[t!]
\caption{Source identification.}
    \centering
    \resizebox{\linewidth}{!}{
    \begin{tabular}{lcccccc}
     \hline
    \hline
    \noalign{\smallskip}
    Source & RA & DEC & $z$ & $\delta$$z$ & $m_{\textrm{H}}$ &  $M_{\textrm{i}}$ \\
    (1) & (2) & (3) & (4) & (5) & (6) & (7)\\ 
    \noalign{\smallskip}
    \hline
    \noalign{\smallskip}
PKS0226-038 & 02 28 53.26 & -03 37 05.30 & 2.0692
& 0.0002 & 15.97 & -28.55\\
PKS0237-23 & 02 40 08.17 & -23 09 15.72 & 2.2298 & 0.0008 & 14.52 & -29.79\\
BZQJ0544-2241 & 05 44 08.63 & -22 40 37.10 & 1.5547 & 0.0003 & 14.90 & -27.99\\
PKS0858-279 & 09 00 40.03  & -28 08 20.35 & 2.1725 & 0.0007 & 13.65 & -31.26\\
CTSJ01.03 & 09 39 51.10 & -18 32 15.00 & 2.3754
 & 0.0004 & 14.98 & -30.07\\
WB J0948+0855 & 09 48 53.60 & +08 55 14.45 & 1.9842
& 0.0016 & 15.17 & -29.01\\
CTSJ03.14 & 10 18 21.75 & -21 40 07.80 & 2.4493
& 0.0003 & 15.57 & -28.51\\
PKS1448-232 & 14 51 02.50 & -23 29 31.08 & 2.2264 & 0.0002 & 14.97 & -29.13\\
$[\textrm{HB89}]$1559+088 & 16 02 22.73 & +08 45 38.42 & 2.2837 & 0.0007 & 15.01 & -29.02\\
FBQS J2149-0811 & 21 49 48.18 & -08 10 16.60 & 2.1295 & 0.0002 & 15.74 & -28.48\\
\noalign{\smallskip}
\hline
    \end{tabular}    }
\tablefoot{{ Column (1) Source identification according to the different catalogues. Columns (2) and (3) Right ascension (hh mm ss) and Declination (dd mm ss) respectively, at J2000 coordinates. (4) Redshift estimated as explained in Section \ref{isaac2}. (5) Redshift uncertainty. (6) $H$-band apparent magnitude ($m_{\rm H}$) from the 2-MASS catalogue. (7) $i$-band absolute magnitude $M_{\rm i}$.} }
    \label{tab:identification}
\end{table}

\section{Samples}
\label{samples}
\subsection{The new ISAAC2 sample}
\label{isaac2}
\par New near-IR spectroscopic observations for ten quasars (hereafter ISAAC2 sample) were obtained to cover the high-redshift ($1.55 \leq z \leq 2.45$) and high-luminosity ($47.18 \leq \log L_{\textrm{bol}} \leq 48.14$ [erg s$^{-1}$]) ranges. The redshift range of this sample allows for the detection and observation of the \hb{}+\oiii{} region through the transparent window in the near-infrared with the ISAAC spectrograph at VLT. 
These spectra together with the ones from \citetalias{Deconto-Machado_2023} complete {what we will call from now on} the ISAAC sample (32 sources in total), which is described in more details in \S \ref{complete_ISAAC}. 

\par  {Table \ref{tab:identification} lists the main properties of the ISAAC2 sample.  
The redshift was estimated using a similar approach to the one of \citetalias{Deconto-Machado_2023}. For five sources (PKS0226-038, PKS0858-279, CTSJ01.03, CTSJ03.14, and PKS1448-232), the redshift estimation relied on the observed wavelength of the narrow component (NC) of \hb{}, which is also consistent with the central wavelength of the \oiiionly{} NC.
For PKS0237-23, BZQJ0544-2241, $[\textrm{HB89}]$1559+088, and FBQS J2149-0811, the redshift was determined using the broad component (BC) of \hb{}, whose peak wavelength  agrees with the peak of a very faint \hb{} NC in all cases except PKS0237-23 and $[\textrm{HB89}]$1559+088 where the NC is only barely detected. } In the case of WB J0948+0855, the presence of strong telluric bands, combined with the fact that the \hb{} region is at the edge of the spectrum, significantly affect the estimation of the \hb{} peak centroid, leading to large uncertainties. Therefore, we used \ion{Mg}{II}$\lambda \lambda$2796,2803 doublet from the SDSS UV spectrum to determine the redshift for this source (see Appendix \ref{app:SDSSJ0948}).
{The luminosity distance was estimated from the redshift using the approximation reported in \citet{sulentic_2006}, valid for $\Omega_{\textrm{M}} =0.3$, $\Omega_{\Lambda}=0.7$, and H$_{\textrm{0}}=70$ km s$^{-1}$ Mpc$^{-1}$. }
\par 


\begin{figure}
    \centering
    \includegraphics[width=\columnwidth]{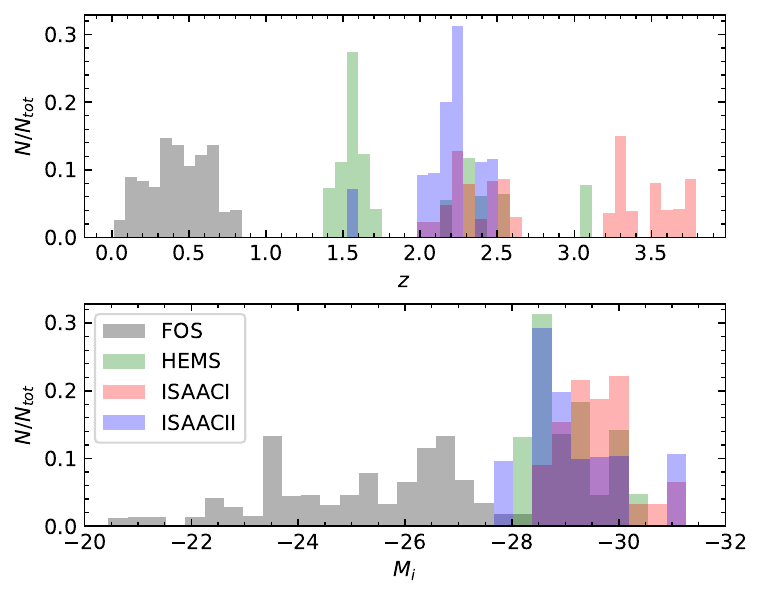}
    \caption{Distributions of $z$ and $M_{\textrm{i}}$ for FOS, HEMS, ISAACI from \citetalias{Deconto-Machado_2023}, and the new ISAAC2 data. Each distribution was normalised separately.}
    \label{fig:hist_z_Mi}
\end{figure}

\subsection{The complete ISAAC sample}
\label{complete_ISAAC}
\par Our complete ISAAC sample comprises 32 quasars, consisting of 22 sources from \citetalias{Deconto-Machado_2023}, along with the ten new objects presented in this work. These quasars are characterized by high redshift values, ranging from $z=1.55$ to $z=3.80$, and high luminosity ($47.18 \lesssim \log L_{\textrm{bol}} \lesssim 48.36$ [erg s$^{-1}$]), including both radio-loud (10) and radio-quiet (22) sources. The ISAAC sample was selected from the catalogue of \citet[VCV]{veron_2010}, among the type 1 quasars observable from the Paranal Observatory, with declination $\delta\le 25^o$, and with a redshift between 1 and 4 according to VCV (from where the first estimation of the redshift was obtained), that would allow for good coverage of the \hb{} and \oiiionly{} region through the NIR windows with the ISAAC spectrograph at VLT.  Fig. \ref{fig:hist_z_Mi} shows the distribution in $z$ and $M_{\textrm{i}}$ of the complete ISAAC sample, along with the other samples considered in this work (described in \S \ref{sect:comparison_samples}). The ISAAC2 sample exhibits a very similar $M_{\textrm{i}}$ range with the ISAAC1, despite ISAAC1 having a more extended distribution in the $z$ context, with some cases reaching redshifts as high as $\sim$ 3.8. 
 
\begin{table}[t!]
\caption{Log of optical observations with VLT/ISAAC.}
    \centering
    \resizebox{0.85\linewidth}{!}{
    \begin{tabular}
    {lrccrc}
     \hline
    \hline
    \noalign{\smallskip}
    Source & Date obs. & Band & DIT & N$_{\textrm{exp}}$ & Airmass\\
    &  (start) & & (s) & & start-end \\
    (1) & (2) & (3) & (4) & (5) & (6)\\ 
    \noalign{\smallskip}
    \hline
    \noalign{\smallskip}
PKS0226-038 & 2011-01-27 & sH & 160 & 16 & 1.50-1.87 \\
PKS0237-23 & 2010-11-17 & sH & 140 & 16 & 1.48-1.78 \\
BZQJ0544-2241 & 2010-11-24 & J & 145 & 24 & 1.01-1.04\\
PKS0858-279 & 2010-10-13 & sH & 140 & 8 & 1.79-1.46\\
CTSJ01.03 & 2010-10-27 & sH & 180 & 12 & 1.65-1.41 \\
WB J0948+0855 & 2011-01-26 & sH & 160 & 12 & 1.27-1.21\\
CTSJ03.14 & 2010-11-23 & sH & 160 & 12 & 1.34-1.18\\
PKS1448-232 & 2011-01-31 & sH & 160 & 16 & 2.01-1.58 \\
$[\textrm{HB89}]$1559+088 & 2011-02-01 & sH & 170 & 16 & 2.19-1.70\\
FBQS J2149-0811 & 2010-10-26 & sH & 160 & 16 & 1.09-1.23\\
\noalign{\smallskip}
\hline
    \end{tabular}
    }
    \label{tab:log_obs}
    \tablefoot{ {(1) Source identification. (2) Date of observation. (3) Grating used. (4) Individual Detector Integration Time (DIT). (5) Number of exposures with single exposure time equal to DIT. (6) Range of air mass of the observations.}}
\end{table}

\subsection{Other samples}
\label{sect:comparison_samples}
\par 
{This work includes two additional samples, at high and low redshift (including RL and RQ sources),  previously published, with UV and optical spectra available that have been analysed using the same approach for the decomposition of the emission line profiles. They consist of:} 



\paragraph{\textit{High-redshift Hamburg-ESO sample (hereafter, HEMS):}} The HEMS sample consist of the 28 high-luminosity and high-redshift objects with measures reported by \citet{sulentic_2017}, where a detailed analysis (with the same methodology used for our ISAAC sample) of the \hb{} and \civ{} broad emission lines is presented. This sample was selected for having additional UV spectra from the original 52 sources with ISAAC \hb{} observations of the Hamburg-ESO (HE) sample discussed in \citet{Sulentic_2004, sulentic_2006} and \citet{Marziani_2009}.  These sources are extremely luminous ($47.5 \lesssim \log L_{\textrm{bol}} \lesssim 48$\ [erg s$^{-1}$]) and are located in a redshift range of $1.4 \lesssim z \lesssim 3.1$.  {The radio properties of the HEMS sample are listed in Table \ref{tab:radio_fos_hems} in Appendix \ref{app:radio-hems-fos}, together with the UV and optical full profile measured parameters for this sample and used in this paper. }
From the 28 quasars, 24 are classified as RQ while four are RL sources. We combine these data with our ISAAC1+ISAAC2 sources to build a high-luminosity sample to be analysed in the present work. As can be seen in Fig. \ref{fig:hist_z_Mi}, the HEMS sample   shares similar $i$-band absolute magnitudes with the ISAAC sample.

\paragraph{\textit{Low-redshift FOS data:}} {The low-luminosity  sample was selected from \citet{Sulentic_2007} who analyzed the \civ{} emission line parameters of 130 low-redshift sources observed with the Faint Object Spectrograph (FOS) on board HST.} 84 out of these 130 quasars have optical spectra available in the recent literature and we have reanalyzed them following the same approach we have performed for the ISAAC data. We use this sample as comparison sample at low $L$ for both optical and UV spectral ranges.  {Table \ref{tab:radio_fos_hems} also lists the radio properties and the UV and optical full profile parameters (velocity centroids at 1/2 and 1/4 intensity of \civonly{} and \oiiionly{}) of the FOS sample.} This sample includes both RQ and RL (50 and 34 sources, respectively) and has a typical bolometric luminosity of $\sim 45.6$ $[\textrm{erg s}^{-1}]$ and a redshift $z \leq 0.8$, as shown in Fig. \ref{fig:hist_z_Mi}. 
\\

\par {In the analysis carried out along this paper (sections \S\ref{section:results},\S\ref{section:discussion}),  our high-$z$ high-luminosity sample (including RQ and RL sources) will consist of the combination of the full ISAAC (ISAAC1+ISAAC2) and HEMS samples, since both of them share similar redshift and luminosity ranges. This high-$z$ sample consists of 60 sources in total (32 from ISAAC and 28 from HEMS), from which we have \oiiionly{} detected for  {58 objects. In the case of SDSSJ005700.18+143737.7, \oiiionly{} is located very close to the edge of the spectrum, and for WB J0948+0855, the \oiiionly{} region is completely affected by absorption lines.}
 { From these 58 source,} 34 are identified as blue outliers in \oiiionly{} (i.e. velocity shifts at half intensity $c(1/2)$ higher than $\rm{-250}$\,\kms, see \S \ref{sec:synopsis}), with 20 sources from ISAAC (representing $\sim$ 62\% of the sample) and 14 from HEMS ($\sim$ 50\% of the sample). Regarding \civonly{} at high-$z$, we have data available for 48 (20 from ISAAC and 28 from HEMS), wherein blueshifts are identified in all sources but HE2355-4621.} 

The low-$z$ low-luminosity sample consists of the 84 re-analyzed FOS sources, for which we have both \oiiionly{} and \civonly{} measured data.  {Among them, 11 sources in \oiiionly{} and 21 in \civonly{} present a significant blueshifted component. For both emission lines, a source is classified as a blue-outlier and consequently  considered to have an outflow if its full profile exhibits a velocity centroid at half peak intensity $c(1/2) < -250$\,\kms. Also for the FOS low-z sample, when the \oiiionly{} full profile shows slightly smaller $c(1/2)$ blueshift, the object is identified with an outflow if its blueshifted component (after the spectral fitting, see section \ref{sect:data_analysis}) has a significant intensity ($\gtrsim$ 40\%) relative to the full  profile.  The list of sources identified as outliers is reported in Table \ref{tab:parametros_outflows_oiii_civ}.} 

\par  {{Throughout Sections \ref{section:results} and \ref{section:discussion}, the sample size depends on the analysis done. For the discussions about the general behavior of the profiles or e.g. of the effect of radioloudness on the line profiles both in \civonly{} and \oiiionly{}, the complete samples are considered (i.e. the 60 sources at high-$z$ and 84 ones at low-$z$). However, when addressing outflows and their dynamical parameters, only the objects identified as exhibiting blueshifted components are taking into account.}}


\section{Observations and data analysis}
\label{section:obs_data_analysis}


\subsection{NIR Observations and Data Reduction}
\label{nir}
\par  The new spectra were taken in service mode in 2010 and 2011 under the ESO programme 086.B-0774(A), with the infrared spectrometer ISAAC, mounted at the Nasmyth A focus of VLT-U3 (Melipal) at the ESO Paranal Observatory. 
 {A summary of the observations is listed in Table \ref{tab:log_obs}.}



{{Reductions were performed with an usual methodology in NIR observations using standard IRAF routines, and follow the procedures described in \citetalias{Deconto-Machado_2023}. Pairs of spectra were taken nodding the telescope between two positions, A and B, displaced on the CCD frame and following an ABBA cycle as it allows a better subtraction of the sky.} 
{The wavelength calibration was achieved from xenon/argon arc spectra with rms residuals of 0.4\AA\, in J and 0.6\AA\, in sH. This wavelength calibration was corrected for  {small 0-order offsets} by measuring the centroids of several OH sky lines against the arc calibration. The  {2D wavelength-calibrated spectra (A and B)}  of each pair AB or BA were rebinned to a common linear wavelength scale.} 
{{Afterwards, the sky background was subtracted using the double-subtraction technique. For each rebinned wavelength 2D spectra, we computed a frame (A-B or B-A), so that the reduced image (AB or BA) consist of two spectra of the source, one positive and one negative, positioned as separated by the nod throw and in which the sky has been removed. We then extracted the 1D wavelength-calibrated spectra using the task {\tt apsum}, creating one spectrum for each nodding position. The one-dimensional -B (or -A) spectrum was subtracted again to obtain a final 1D A-(-B) = A+B spectrum for each observing AB and BA sequence of the ABBA cycle. Finally, all double-subtracted spectra were stacked together.
}} 
\par 
The absolute flux calibration was performed through observations, with the same setup that quasar spectra, of standard telluric stars.  The standard star spectral energy distribution (SED) were retrieved from Library of Stellar Spectra for spectrophotometric calibration available at ESO\footnote{\href{https://www.eso.org/sci/observing/tools/standards/IR_spectral_library.html}{https://www.eso.org/sci/observing/tools/standards/\-IR\_spectral\_library.html}}. 
The absolute flux scale for the standard star was provided by the Two Micron All Sky Survey \citep[2MASS,][]{Skrutskie_2006} magnitudes.  Each standard star SED spectrum was then divided by its corresponding  spectrum in order to correct for the atmospheric absorption features (IRAF routine {\tt telluric}). Spectra were also corrected for galactic extinction.



As a final step, we evaluate the absolute flux calibration uncertainty on the quasar spectra by performing a comparison between the $J/H$-band magnitudes estimated by convolving the $J/H$ 2MASS filter with the observed spectrum and the $J/H$ magnitudes in the NASA/IPAC Extragalactic Database (NED). The average difference between 2MASS and our flux estimations is 0.026 mag, with the largest value of 0.091 $\pm$ 0.018 mag found for CTSJ01.03.

\begin{table}[t!]
    \caption{UV spectra information.}
    \centering
    \begin{tabular}{lc}
    \hline
    \hline
    \noalign{\smallskip}
    Source & UV Spect.\\
    (1) & (2)\\
    \noalign{\smallskip}
    \hline
    \noalign{\smallskip}
     PKS0226-038 & BOSS\\
PKS0237-23 &  \citet{Wilkes_1983}\\
BZQJ0544-2241 &  \citet{Perlman_1998}\\
PKS0858-279 &  \citet{Stickel_1993}\\
CTSJ01.03 &  \citet{Tytler_2004}\\
WB J0948+0855 & BOSS\\
CTSJ03.14 & 6dFGS\\
PKS1448-232 & \ldots\\
$[\textrm{HB89}]$1559+088 & SDSS\\
FBQS J2149-0811 & \ldots \\
\noalign{\smallskip}
\hline
    \end{tabular}
    \label{tab:uv_spectra}
    \tablefoot{ {(1) Source identification. (2) Database of reference from which the UV spectra were obtained.}}
\end{table}

\subsection{Optical data}

\par Eight out of the ten new sources from ISAAC2 sample have available rest-frame UV spectra that cover at least one of the three regions of our interest (i.e. \ion{Si}{IV}$\lambda$1397+\ion{O}{IV}]$\lambda$1402, \civ{}+\ion{He}{II}$\lambda$1402, and the 1900\AA\ blend regions). In Table \ref{tab:uv_spectra} we report the database or reference from which each UV spectrum was obtained. For four quasars (PKS0237-23, BZQJ0544-2241, PKS0858-279, and CTSJ01.03) it was necessary to digitize the spectra from the respective references reported in Table \ref{tab:uv_spectra}. { This process has been performed using the \texttt{WebPlotDigitizer} facility \footnote{\href{https://automeris.io/}{https://automeris.io/}}. }The only sources for which we could not find useful UV spectra are PKS1448-232 and FBQS J2149-0811. {For these two quasars, there were UV spectra available, but not with the enough S/N required to perform the fittings.} 

\subsection{Data analysis}
\label{sect:data_analysis}
\par {The spectral analysis has been carried out in two complementary ways, following the methodology described in \citetalias{Deconto-Machado_2023}. A multicomponent fit was performed, after the spectra were set at rest-frame, using the \texttt{specfit} \citep{kriss_1994} routine from IRAF package. This routine allows for simultaneous minimum-$\chi^2$ fit of the continuum approximated by a power-law, a scalable \ion{Fe}{II} (or \ion{Fe}{III}) pseudo-continuum and the spectral line components yielding FWHM, peak wavelength, and intensity of all line components {(see Fig. 2 of \citetalias{Deconto-Machado_2023} for an illustration of the decomposition analysis of the broad lines)}. 
We also include absorption   lines in case some emission profiles (usually the ones of \ion{C}{IV})  are affected by them. In addition, a study of the full broad emission line profiles has been performed, once the continuum power law, \ion{Fe}{II}, and narrow components have been subtracted.}

\begin{table*}[h!]
    \caption{Measurements on the \hb{} full broad line profile and derived properties of the \hb{} region.} 
    \label{tab:hb_full}

    \centering
    \resizebox{0.98\linewidth}{!}{
    \begin{tabular}{lcccrrrrrrrrrcc}
    \hline
    \hline
    \noalign{\smallskip}
    Source & $F_{\textrm{5100\AA}}$ & $F_{\textrm{tot}}$ & W &\multicolumn{1}{c}{FWHM} &  \multicolumn{1}{c}{A.I.}    & \multicolumn{1}{c}{c(1/4)} &  \multicolumn{1}{c}{c(1/2)} & \multicolumn{1}{c}{c(3/4)} & \multicolumn{1}{c}{c(9/10)} & $R_{\textrm{\ion{Fe}{II}}}$ & ST & $\log L_{\textrm{bol}}$ & $\log M_{\textrm{BH}}$ & \multicolumn{1}{c}{$\log L/L_{\textrm{Edd}}$} \\
    & & & \multicolumn{1}{c}{[\AA]} & \multicolumn{1}{c}{$[\textrm{km\,s}^{-1}]$}  &  & \multicolumn{1}{c}{$[\textrm{km\,s}^{-1}]$} & \multicolumn{1}{c}{$[\textrm{km\,s}^{-1}]$} & \multicolumn{1}{c}{$[\textrm{km\,s}^{-1}]$} &  \multicolumn{1}{c}{$[\textrm{km\,s}^{-1}]$} & & & $[\textrm{erg\,s$^{-1}$}]$ & \multicolumn{1}{c}{$[M_{\odot}]$} & \\
    \multicolumn{1}{c}{(1)} & (2) &  (3) & (4) & \multicolumn{1}{c}{(5)} & \multicolumn{1}{c}{(6)} & \multicolumn{1}{c}{(7)} & \multicolumn{1}{c}{(8)} & \multicolumn{1}{c}{(9)} & \multicolumn{1}{c}{(10)} & (11) & (12) & \multicolumn{1}{c}{(13)} & \multicolumn{1}{c}{(14)} & \multicolumn{1}{c}{(15)} \\
    \noalign{\smallskip}
    \hline
    \noalign{\smallskip}
    \multicolumn{15}{c}{Population A}\\
    \noalign{\smallskip}
    \hline
    \noalign{\smallskip}
BZQJ0544-2241	&	4.53	&	2.52	&	51	&	5191	$\pm$	458	&	-0.05	$\pm$	0.09	&	-266	$\pm$	402	&	-174	$\pm$	93	&	-94	$\pm$	141	&	-57	$\pm$	170	&	0.69 &	A2	&	47.52	&  9.66	&	-0.31 \\
WB J0948+0855$^{(a)}$	&	2.88	&	1.27	&	43	&	5580	$\pm$	494	&	0.00	$\pm$	0.11	&	-1	$\pm$	456	&	-1	$\pm$	92	&	0	$\pm$	149	&	0	$\pm$	184	&	0.89 &	A2	&	47.45	& 9.74 & -0.46		\\
CTSJ01.03	&	1.96	&	1.73	&	81	&	4551	$\pm$	403	&	0.00	$\pm$	0.20	&	320	$\pm$	372	&	321	$\pm$	75	&	321	$\pm$	122	&	321	$\pm$	150	&	0.21 &	A1	&	47.37 &	9.52	& -0.33 	\\
FBQS J2149-0811	&	1.73	&	1.80	&	113	&	4992	$\pm$	442	&	0.00	$\pm$	0.11	&	0	$\pm$	408	&	0	$\pm$	82	&	-1	$\pm$	133	&	-1	$\pm$	165	&	0.23	&	A1 &	47.26 & 9.55	& -0.46		\\
 
\noalign{\smallskip}
\hline
\noalign{\smallskip}
\noalign{\smallskip}
\multicolumn{15}{c}{Population B}\\
\noalign{\smallskip}
\hline    
\noalign{\smallskip}
PKS0226-038	&	1.48	&	0.99	&	66	&	5280 $\pm$ 420			&	0.46 $\pm$ 0.05			&	2874 $\pm$ 646			&	109 $\pm$ 53		&	-45 $\pm$ 140			&	-86 $\pm$ 190			&	0.62	&	B2 &	47.18 & 9.44		& -0.44	\\
PKS0237-23	&	5.19	&	4.48	&	89	&	7986	$\pm$	821	&	0.31	$\pm$	0.11	&	2546	$\pm$	389	&	1374	$\pm$	65	&	730	$\pm$	211	&	574	$\pm$	262	&	0.42	&	B1 &	47.76	& 9.85 & -0.27		\\
PKS0858-279$^{(a)}$	&	12.9	&	7.84	&	56	&	5385	$\pm$	402	&	0.15	$\pm$	0.11	&	790	$\pm$	388	&	299	$\pm$	61	&	191	$\pm$	141	&	159	$\pm$	191	&	0.47	&	B1 &	48.14 & 9.91	& \ 0.05 \\
CTSJ03.14$^{(a)}$	&	3.44	&	3.61	&	102	&	4056	$\pm$	305	&	0.07	$\pm$	0.09	&	271	$\pm$	274	&	100	$\pm$	52	&	62	$\pm$	106	&	50	$\pm$	143	&	0.38	&	B1 &	47.62 & 9.36	& \ 0.08 \\
PKS1448-232	&	3.43	&	4.53	&	129	&	4397	$\pm$	364	&	0.54	$\pm$	0.09	&	3664	$\pm$	745	&	263	$\pm$	47	&	150	$\pm$	114	&	122	$\pm$	153	&	0.07	&	B1 &	47.58 & 9.43	& -0.02		\\
$[\textrm{HB89}]$1559+088	&	3.31	&	3.66	&	110	&	5802	$\pm$	514	&	0.01	$\pm$	0.18	&	71	$\pm$	750	&	-314	$\pm$	115	&	-39	$\pm$	157	&	13	$\pm$	184	&	0.29	&	B1 &	47.57 & 9.56	& -0.16		\\
\noalign{\smallskip}
\hline
    \end{tabular}
    } 
\\
{\small \raggedright {Notes}.  { {(1) Source identification. (2) Flux of the continuum on the 5100\AA\ wavelength, corrected from galactic extinction, in units of $10^{-15}$ erg s$^{-1}$ cm$^{-2}$ \AA$^{-1}$. (3) Full \hb{} line flux (i.e., the flux for all broad line components, BC and VBC, and BLUE whenever appropriate), in units of $10^{-13}$ erg s$^{-1}$ cm$^{-2}$. (4) \hb{} equivalent width $W$. (5) FWHM. (6) Asymmetry index (A.I.). (7), (8), (9), (10) Centroid velocity shifts at $\frac{1}{4}$, $\frac{1}{2}$, $\frac{3}{4}$, and $\frac{9}{10}$ fractional intensities. (11) Ratio $R_{\rm \ion{Fe}{II}}$ (see  \S\ref{sec:optical_plane}). (12) Spectral types according to the quasar main sequence. (13) Bolometric luminosity. (14) Black hole mass. (15) Eddington ratio.}  $^{(a)}$ \hb{} profile may be affected by the correction of   the telluric absorptions present in the spectrum (see Appendix \ref{appendix_1}).} \par}
    
    \end{table*}

\begin{table*}
    \caption{Results from \texttt{specfit} analysis on the broad and narrow profiles of \hb{}.}
    \label{tab:hb_specfit}

    \centering
    \resizebox{0.99\linewidth}{!}{
    \begin{tabular}{lcccccccccccccccccccc}
    \hline
    \hline
    \noalign{\smallskip}
    & & \multicolumn{9}{c}{Full broad profile (BLUE+BC+VBC)} &  & & \multicolumn{7}{c}{Full narrow profile (SBC+NC)}\\
    \noalign{\smallskip}
    \cline{2-12} \cline{14-21}
    \noalign{\smallskip}
    & \multicolumn{3}{c}{BLUE} & & \multicolumn{3}{c}{BC} & & \multicolumn{3}{c}{VBC} & & &\multicolumn{3}{c}{SBC} & & \multicolumn{3}{c}{NC}\\
    \noalign{\smallskip}
    \cline{2-4} \cline{6-8} \cline{10-12} \cline{15-17} \cline{19-21} 
    \noalign{\smallskip}
    Source & $F/F_{\textrm{tot}}$ &  FWHM  & Peak  &  &$F/F_{\textrm{tot}}$ &  FWHM  & Peak  &  & $F/F_{\textrm{tot}}$ &  FWHM  & Peak & & $F_{\textrm{tot}}$ & $F/F_{\textrm{tot}}$ & FWHM & Peak & & $F/F_{\textrm{tot}}$ & FWHM & Peak \\
    & & $[\textrm{km s}^{-1}]$  &  $[\textrm{km s}^{-1}]$&  & & $[\textrm{km s}^{-1}]$ & $[\textrm{km s}^{-1}]$ & & & $[\textrm{km s}^{-1}]$ &  $[\textrm{km s}^{-1}]$ & & & & $[\textrm{km s}^{-1}]$ & $[\textrm{km s}^{-1}]$ & & & $[\textrm{km s}^{-1}]$ & $[\textrm{km s}^{-1}]$\\
    (1) & (2) &  (3) & (4) & & (5) &  (6) &  (7) &  & (8) & (9) & (10) & & (11) & (12) & (13) & (14) & & (15) & (16) & (17)\\
    \noalign{\smallskip}
    \hline
    \noalign{\smallskip}
    \multicolumn{21}{c}{Population A}\\
    \noalign{\smallskip}
    \hline
    \noalign{\smallskip}
  BZQJ0544-2241 & 0.04 & 5580 & -2461 & & 0.96 & 4877 & 0 & &  … &  … &  … & & 0.35 & 0.09 & 1618 & -591 &  & 0.91 & 1187 & 0 \\
WB J0948+0855 & … &  … &  … & & 1.00 & 5579 & 0 & &  … &  … &  … &  & …  & … & … & … & & … & … & …\\
CTSJ01.03$^{(a)}$ & … &  … &  … & & 1.00 & 4546 & 321 & &  … &  … &  … & & 1.00 & 0.18 & 1437 & -1080 & & 0.82 & 1388 & 0 \\
FBQS J2149-0811$^{(b)}$ & … &  … &  … & & 1.00 & 4992 & 0 & &  … &  … &  … & & 0.24 & 0.33 & 1488 & -1263 & & 0.67 & 1258 & 0 \\
\noalign{\smallskip}
\hline
\noalign{\smallskip}
\multicolumn{21}{c}{Population B}\\
\noalign{\smallskip}
\hline    
\noalign{\smallskip}
PKS0226-038 & … &  … &  … & & 0.55 & 4620 & -258 & & 0.45 & 10495 & 6340 & & 0.41 & 0.37 & 1169 & -1051 & & 0.63 & 901 & 0 \\
PKS0237-23 & … &  … &  … & & 0.41 & 5331 & 0 & & 0.59 & 10304 & 3881 & & 0.11 & … & … & … & & 1.00 & 654:: & 230:: \\
PKS0858-279 & … &  … &  … & & 0.56 & 4571 & 6 & & 0.44 & 10342 & 2640 & & 6.85 & 0.55 & 1273 & -1182 & & 0.45 & 1168 & 0 \\
CTSJ03.14 & … &  … &  … & & 0.47 & 3268 & 0 & & 0.53 & 7975 & 644 & & 0.43 & … & … & … & & 1.00 & 693 & 0 \\
PKS1448-232 & … &  … &  … & & 0.38 & 3615 & 16 & & 0.62 & 14381 & 5824 & & 1.31 & 0.61 & 1045 & -724 & & 0.39 & 1007 & 0 \\
$[\textrm{HB89}]$1559+088 & 0.08 & 4702 & -3110 & & 0.33 & 3807 & 0 & & 0.59 & 14817 & 1341 & & 0.44 & 1.00 & 1581 & -759 & & … & … & … \\
\noalign{\smallskip}
\hline
    \end{tabular}
    } \\
{\small \raggedright {Notes}.  { {(1) Source identification. (2), (3), (4) Relative flux $F/F_{\rm tot}$, FWHM, and velocity shift of the peak for the BLUE component. (5), (6), (7) Same for the BC. (8), (9), (10) Same for the VBC. (11) Total flux of the \hb{} narrow profile, in units of $10^{-14}$ erg s$^{-1}$ cm$^{-2}$. (12), (13), (14) Relative flux $F/F_{\rm tot}$, FWHM, and velocity shift of the peak for the SBC component. (15), (16), (17) Same for the NC. } $^{(a)}$  Very difficult to separate SBC and NC (see Fig. \ref{fig:CTSJ0103}). $^{(b)}$ SBC and NC very weak and uncertain (see Fig. \ref{fig:2149}).}\par}
    
    \end{table*}

\subsubsection{\hb{} spectral region }




\noindent{{In the \hb{} region fittings,  apart from the power-law and the scalable \ion{Fe}{II} template, we include for the \hb{}  {BLR broad emission line}: (1) a \textit{broad component} (BC), which is symmetric and typically set at the rest-frame wavelength (Lorentzian-like shape for Pop. A and Gaussian for Pop. B); a {\it very broad component} (VBC), representing the innermost part of the BLR, characterised by a Gaussian shape and only present in  Pop. B;  (3) a \textit{blueshifted component} (BLUE), if needed, represented by a Gaussian profile (with the option of being skewed). In addition, we include a narrow component superimposed to the broad \hb{} that may include up to two components: the proper \textit{narrow component} (NC), usually with FWHM $\leq 1200$ km s$^{-1}$ and fitted as an unshifted Gaussian; and a \textit{semi-broad component} (SBC), represented by a blue-shifted (symmetric or skewed) Gaussian profile with FWHM and shift similar to the \oiiionly{} semi-broad component, if present  {(see below for \oiiionly{} components)}. Final fitting for each spectrum, based on the best-fit model with the minimum $\chi^2$, was obtained after performing a set of  models both as Pop. A and Pop. B}.}

\begin{figure*}
    \centering
     \includegraphics[width=0.69\linewidth]{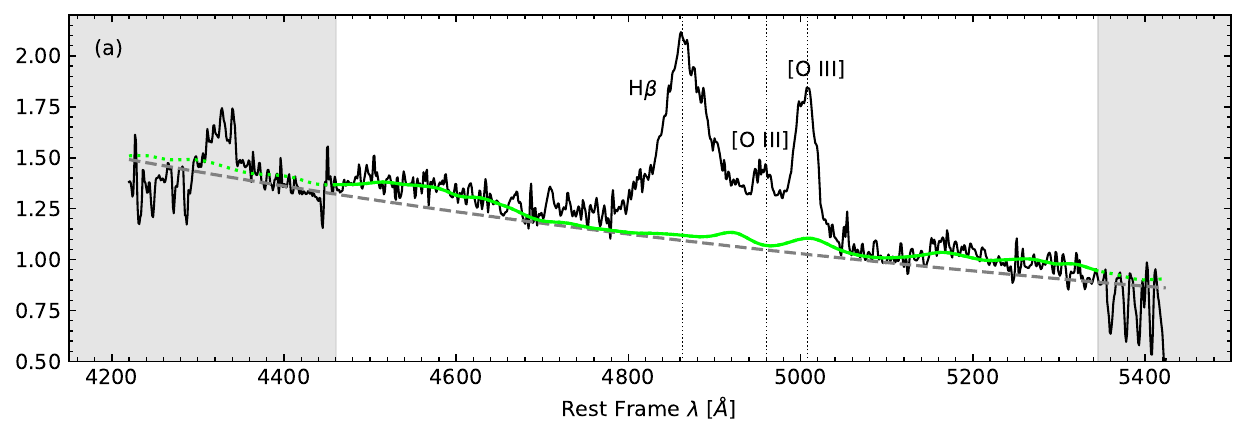}
    \includegraphics[width=0.305\linewidth]{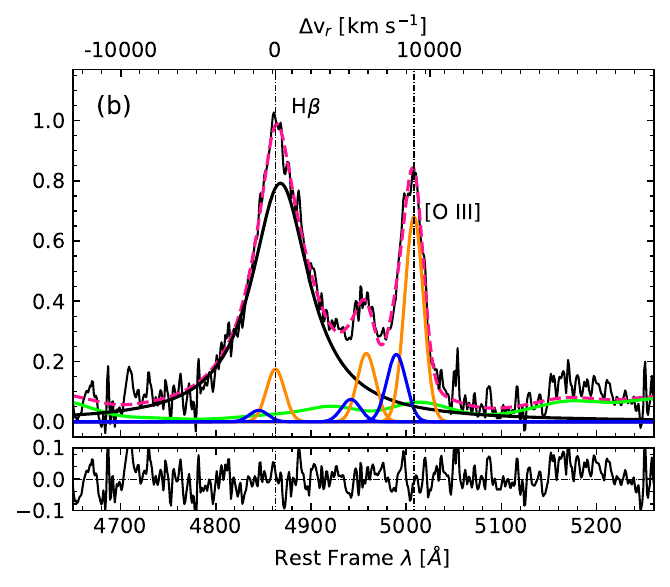}
    \caption{ {(a) Rest-frame optical spectrum of CTSJ01.03. The white area shows the spectral region used in the fitting. The gray dashed line indicates the continuum level obtained with the \texttt{specfit} multicomponent fitting.  \ion{Fe}{II} contributions are represented by the green line.   (b) example of the \hb{}+\oiiionly{} emission line decomposition (upper panel) and its respective residuals (bottom panel). Black, orange, and blue lines indicate the broad, the narrow, and the blueshifted components of the profile, respectively. The final fit is shown by the pink dashed line.}}
    \label{fig:example_decomposition}
\end{figure*}

\par  {Fig. \ref{fig:example_decomposition} shows the multicomponent decomposition of the \hb{}+\oiiionly{} spectral region for one of our sources. The fittings of the complete ISAAC2 sample are shown in Appendix \ref{appendix_1}, along with their respective VLT-ISAAC optical spectra. } 
 {Measurements on the \hb{} full broad profiles of the ISAAC2 sample are listed in Table \ref{tab:hb_full}, where properties derived from the \hb{} region have also been included.} 
$L_{\textrm{bol}}$ has been calculated from the continuum flux at 5100\AA\, and by applying the bolometric correction factor expected for the respective luminosity range as established by \citet{Netzer_2019}. The black hole mass $M_{\textrm{BH}}$ for each quasar was estimated by using the \hb{} scaling law of \citet[Eq. (5)]{Vestergaard_2006}. Eddington ratio (\lledd) has been computed by using the bolometric luminosity and  with \ledd =$1.5 \cdot 10^{38}$\mbh.
 {The corresponding individual components obtained from the \texttt{specfit} decomposition of the broad \hb\ profile are reported in Table \ref{tab:hb_specfit}.}

{The \oiii{} emission line profiles were modelled assuming that each line consist of a narrower centered Gaussian component (NC)  and one (or more) skewed blueshifted Gaussian semi-broad component (SBC) representing bipolar outflow emissions, where the recessing side of the outflow remains obscured {\citep[see e.g.,][and references therein]{zamanovetal02,Kim_2023}}.}  {Similarly to \hb{}, the results obtained for the full profile and the components of \oiiiseven{} are reported in Table \ref{tab:oiii_specfit}.} 

\subsubsection{UV spectral region}

\par The UV spectral analysis is performed for three different regions: one centred in the \ion{Si}{IV}$\lambda$1397+\ion{O}{IV}]$\lambda$1402 lines, one in which the \civ{} dominates the emission together with a less strong \ion{He}{II}$\lambda$1640 line, and another region considering the emission lines from the 1900\AA\ blend ({consisting mainly of} \ion{Al}{III}$\lambda$1860 doublet, \ion{Si}{III}]$\lambda$1892, and \ion{C}{III}]$\lambda$1909). In the three spectral regions, the continuum was modeled locally by a power-law and the \ion{Fe}{III} multiplets (e.g. the strong ones observed on the red side of the 1900\AA\ blend) were modeled using the \citet{Vestergaard_2001} empirical template. For some Pop. A sources, an additional \ion{Fe}{III}$\lambda$1914 {emission line} is needed to fully represent the red side of \ion{C}{III}]$\lambda$1909 {\citep[see e.g.][]{martinezaldama_2018}}.  The broad UV lines (\ion{Si}{IV}$\lambda$1397, \civ{}, \ion{He}{II}$\lambda$1640, and \ion{C}{III}]$\lambda$1909) are fitted by the same three model components employed for \hb\ varying their relative contribution to each line flux, as detailed in \citetalias{Deconto-Machado_2023}.



\par  {Table \ref{tab:civ_full} provides measurements for the \civ{} full broad profile} 
 {and information on the \civ{} and \ion{He}{II}$\lambda1640$ individual components is given in Table \ref{tab:civ_specfit}.} 
The FWHM and peak shift of the \civ{} and \ion{He}{II}$\lambda1640$\ components were suitable for representing the profiles in all cases save CTSJ03.14, for which the \ion{He}{II}$\lambda1640$ BLUE is much narrower than the \civ{} one.

\par  {Table \ref{tab:siiv_full} shows the measurements on the full profile of \ion{Si}{IV}$\lambda$1397 and \ion{O}{IV}]$\lambda$1402.} 
Similar to the approach taken in \citetalias{Deconto-Machado_2023}, we also set the \siiv{} BC at rest-frame and select as initial guess a FWHM similar to the one determined for \civ{} BC. 

\begin{table*}[h!]
    \caption{Measurements on the \oiiiseven{} line profile.}
    \label{tab:oiii_specfit}

    \centering
    \resizebox{0.95\linewidth}{!}{
    \begin{tabular}{@{\extracolsep{3pt}} l c c r r r r c c c r c c r r}
    \hline
    \hline
    \noalign{\smallskip}
   & & & \multicolumn{4}{c}{\oiiiseven{} full profile} & &\multicolumn{3}{c}{SBC} & &\multicolumn{3}{c}{NC}\\
   \cline{4-7} \cline{9-11} \cline{13-15}
    \noalign{\smallskip}
    Source & $F_{\textrm{tot}}$ & W & \multicolumn{1}{c}{FWHM} &  \multicolumn{1}{c}{A.I.}   &  \multicolumn{1}{c}{$c(1/2)$} &\multicolumn{1}{c}{ $c(9/10)$} & & \multicolumn{1}{c}{$F/F_{\textrm{tot}}$} & \multicolumn{1}{c}{FWHM}  & \multicolumn{1}{c}{Peak}  &  & \multicolumn{1}{c}{$F/F_{\textrm{tot}}$} & \multicolumn{1}{c}{FWHM}  & \multicolumn{1}{c}{Peak}  \\
    & & \multicolumn{1}{c}{[\AA]} &\multicolumn{1}{c}{ $[\textrm{km s}^{-1}]$}  &  & \multicolumn{1}{c}{$[\textrm{km s}^{-1}]$} & \multicolumn{1}{c}{ $[\textrm{km s}^{-1}]$} & & & \multicolumn{1}{c}{$[\textrm{km s}^{-1}]$} & \multicolumn{1}{c}{$[\textrm{km s}^{-1}]$} & & & \multicolumn{1}{c}{$[\textrm{km s}^{-1}]$} &  \multicolumn{1}{c}{$[\textrm{km s}^{-1}]$} \\
    (1) & (2) &  (3) & \multicolumn{1}{c}{(4)} & \multicolumn{1}{c}{(5)} & \multicolumn{1}{c}{(6)} & \multicolumn{1}{c}{(7)}  &  & \multicolumn{1}{c}{(8)} & \multicolumn{1}{c}{(9)} & \multicolumn{1}{c}{(10)} & & \multicolumn{1}{c}{(11)}  & \multicolumn{1}{c}{(12)} & \multicolumn{1}{c}{(13)}\\
    \noalign{\smallskip}
    \hline
    \noalign{\smallskip}
    \multicolumn{15}{c}{Population A}\\
    \noalign{\smallskip}
    \hline
    \noalign{\smallskip}
BZQJ0544-2241$^{(a)}$ & 0.27 & 6 & 1374 $\pm$ 360: & -0.07 $\pm$ 0.04: &  -369 $\pm$ 40: & -339 $\pm$ 88: & & 0.45 & 1618 & -621 & & 0.55 & 1187 & -247 \\
CTSJ01.03 & 0.44 & 22 & 1709 $\pm$ 123 & -0.18 $\pm$ 0.02 &  -191 $\pm$ 33 & -88 $\pm$ 553 & & 0.25 & 1437 & -1110 & & 0.75 & 1388 & 0 \\
FBQS J2149-0811$^{(a)}$ & 0.14 & 9 & 1477 $\pm$ 440: & -0.20 $\pm$ 0.10: &  -319 $\pm$ 58: & -248 $\pm$ 92: & & 0.25 & 1488 & -1292 & & 0.75 & 1258 & -187 \\
\noalign{\smallskip}
\hline
\noalign{\smallskip}
\multicolumn{15}{c}{Population B}\\
\noalign{\smallskip}
\hline    
\noalign{\smallskip}
PKS0226-038 & 0.31 & 22 & 979 $\pm$ 69 & -0.27 $\pm$ 0.16 & -46 $\pm$ 16 & -18 $\pm$ 31 & & 0.24 & 1169 & -1080 & & 0.76 & 901 & 0 \\
PKS0237-23 & 0.31 & 6 & 1602 $\pm$ 109 & 0.09 $\pm$ 0.01 & -805 $\pm$ 21 & -922 $\pm$ 42 & & 0.83 & 1423 & -899 & & 0.17 & 654:: & -34:: \\
PKS0858-279 & 0.33 & 3 & 1506 $\pm$ 234 &  -0.35 $\pm$ 0.02 & -114 $\pm$ 80 & 27 $\pm$ 42 & & 0.31 & 1273 & -1211 & & 0.69 & 1168 & 65 \\
CTSJ03.14 & 1.34 & 40 & 860 $\pm$ 92 & -0.29 $\pm$ 0.13 & -120 $\pm$ 40 & -38 $\pm$ 21 & & 0.27 & 1774 & -504 & & 0.73 & 693 & -38 \\
PKS1448-232 & 0.36 & 11 & 1154 $\pm$ 74 & -0.08 $\pm$ 0.02 & -271 $\pm$ 17 & -240 $\pm$ 36 & & 0.24 & 1045 & -754 & & 0.76 & 1007 & -161 \\
$[\textrm{HB89}]$1559+088 & 0.31 & 10 & 885 $\pm$ 215 & -0.51 $\pm$ 0.03 & -605 $\pm$ 69 & -538 $\pm$ 52 & & 0.54 & 1581 & -1452 & & 0.46 & 688 & -498 \\
\noalign{\smallskip}
\hline
    \end{tabular}
    } \\ 
    
{\small\raggedright{ {Notes.} {  {(1) Source identification. (2) \oiiiseven{} total flux, in units of $10^{-13}$ erg s$^{-1}$ cm$^{-2}$. (3) Equivalent width, $W$. (4), (5), (6), (7) FWHM, asymmetry index, and the centroid velocities at $\frac{1}{2}$ and $\frac{9}{10}$ intensities of the full profile. (8), (9), (10) Relative intensity, FWHM, and peak shift for the \oiiionly{} semi-broad (in general blueshifted) component. (11), (12), (13) Same for the \oiiionly{} narrow component.} $^{(a)}$ \oiiiseven{} may be affected by the correction of the telluric absorptions present in the spectrum} (see Appendix \ref{appendix_1}).}\par}
    \end{table*}

\begin{table*}[t!]
    \centering
    \caption{Measurements on the \civ{} full broad profile.}
     \centering
    \resizebox{0.8\linewidth}{!}{
    \begin{tabular}{lrrrrrrrccccccccc}
    \hline
    \hline
    \noalign{\smallskip}
    Source & \multicolumn{1}{c}{W} &\multicolumn{1}{c}{FWHM} & \multicolumn{1}{c}{A.I.} & \multicolumn{1}{c}{$c(1/4)$} & \multicolumn{1}{c}{$c(1/2)$} & \multicolumn{1}{c}{$c(3/4)$} & \multicolumn{1}{c}{$c(9/10)$} \\
    & \multicolumn{1}{c}{[\AA]} & \multicolumn{1}{c}{$[\textrm{km\,s}^{-1}]$}  & & \multicolumn{1}{c}{$[\textrm{km\,s}^{-1}]$} & \multicolumn{1}{c}{$[\textrm{km\,s}^{-1}]$} & \multicolumn{1}{c}{$[\textrm{km\,s}^{-1}]$} & \multicolumn{1}{c}{$[\textrm{km\,s}^{-1}]$} \\
    (1) & \multicolumn{1}{c}{(2)} & \multicolumn{1}{c}{(3)} &\multicolumn{1}{c}{(4)} & \multicolumn{1}{c}{(5)} & \multicolumn{1}{c}{(6)} & \multicolumn{1}{c}{(7)} & \multicolumn{1}{c}{(8)} \\
    \noalign{\smallskip}
    \hline
    \noalign{\smallskip}
    \multicolumn{8}{c}{Population A}\\
    \noalign{\smallskip}
    \hline
    \noalign{\smallskip}
WB J0948+0855 &  4.99 $\pm$ 0.21 & 10363 $\pm$ 984 & -0.52 $\pm$ 0.03 & -6901 $\pm$ 642 & -5334 $\pm$ 354 & -3008 $\pm$ 366 & -2207 $\pm$ 235 \\
CTSJ01.03 & 19.35 $\pm$ 0.82 & 4381 $\pm$ 367 & -0.38 $\pm$ 0.06 & -2250 $\pm$ 421 & -1660 $\pm$ 120 & -1100 $\pm$ 126 & -762 $\pm$ 147 \\
\noalign{\smallskip}
\hline
\noalign{\smallskip}
\multicolumn{8}{c}{Population B}\\
\noalign{\smallskip}
\hline    
\noalign{\smallskip}
PKS0226-038 & 32.03 $\pm$ 1.36 & 4898 $\pm$ 508 & -0.31 $\pm$ 0.06 & -1878 $\pm$ 397 & -1148 $\pm$ 153 & -665 $\pm$ 139 & -474 $\pm$ 141 \\
PKS0237-23 & 25.47 $\pm$ 1.08 & 7269 $\pm$ 729 & -0.01 $\pm$ 0.06 &  -1281 $\pm$ 646 & -1461 $\pm$ 168 & -1213 $\pm$ 217 & -1199 $\pm$ 179 \\
CTSJ03.14 & 30.04 $\pm$ 1.27 & 4087 $\pm$ 385 & -0.30 $\pm$ 0.05 & -1632 $\pm$ 369 & -650 $\pm$ 105 & -481 $\pm$ 113 & -471 $\pm$ 113 \\
$[\textrm{HB89}]$1559+088 & 27.87 $\pm$ 1.18 & 4922 $\pm$ 527 & -0.06 $\pm$ 0.05 & -1492 $\pm$ 340 & -1601 $\pm$ 94 & -1420 $\pm$ 144 & -1241 $\pm$ 145 \\
\noalign{\smallskip}
\hline
    \end{tabular}
    }
    \label{tab:civ_full}
    \tablefoot{ {(1) Source identification. (2) Equivalent width. (3) FWHM. (4) Asymmetry index. (5), (6), (7) and (8) Velocity centroids at different fractional intensities in km s$^{-1}$.}}
\end{table*}

\begin{table*}
    \centering
    \caption{Results from \texttt{specfit} analysis on \civ{} and \ion{He}{II}$\lambda1640$.}
     \resizebox{\linewidth}{!}{
    \begin{tabular}{lccccccccccccccccccccccccccc}
    \hline   
    \hline 
\noalign{\smallskip}
&\multicolumn{12}{c}{\civ{}} & \multicolumn{12}{c}{\ion{He}{II}$\lambda$1640} \\
\noalign{\smallskip}
\cline{3-13} \cline{15-25}
\noalign{\smallskip}
& & &\multicolumn{3}{c}{\civ{} BLUE} & & \multicolumn{2}{c}{\civ{} BC} & & \multicolumn{3}{c}{\civ{} VBC} & &&\multicolumn{3}{c}{\ion{He}{II}$\lambda$1640 BLUE} & & \multicolumn{2}{c}{\ion{He}{II}$\lambda$1640 BC} & & \multicolumn{3}{c}{\ion{He}{II}$\lambda$1640 VBC} \\
\noalign{\smallskip}
\cline{4-6} \cline{8-9} \cline{11-13} \cline{16-18} \cline{20-21} \cline{23-25}
\noalign{\smallskip}
Source & & $F_{\textrm{tot}}$ & $F/F_{\textrm{tot}}$ & FWHM & Peak & & $F/F_{\textrm{tot}}$ & FWHM  & & $F/F_{\textrm{tot}}$ & FWHM & Peak & & $F_{\textrm{tot}}$ & $F/F_{\textrm{tot}}$ & FWHM & Peak & & $F/F_{\textrm{tot}}$ & FWHM & & $F/F_{\textrm{tot}}$ & FWHM & Peak\\
& & & & [km s$^{-1}$] & [km s$^{-1}$] & & & [km s$^{-1}$] & & & [km s$^{-1}$] & [km s$^{-1}$] & & & &[km s$^{-1}$] & [km s$^{-1}$] & & & [km s$^{-1}$]  & & & [km s$^{-1}$] & [km s$^{-1}$]\\
(1) & & (2) & (3) & (4) & (5) & & (6) & (7) & & (8) & (9) & (10)  & & (11) & (12) & (13) & (14) & & (15) & (16) & & (17) & (18) & (19)\\
\noalign{\smallskip}
\hline 
\noalign{\smallskip}
\multicolumn{25}{c}{Population A}\\
\noalign{\smallskip}
\hline    
\noalign{\smallskip}
WB J0948+0855 & & 0.75 &0.82 & 22060 & -2129 & & 0.25 & 5558  & & … &  … &  … & & 0.15 & 0.72 & 22060 & -2129 & & 0.28 & 5558 & & … &  … &  … \\
CTSJ01.03 & & 11.87 & 0.68 & 9263 & -530 & & 0.32 & 4562 & & … &  … &  … & & 1.82 & 0.62 & 9263 & -530 & & 0.38 & 4562 & & … &  … &  … \\
\noalign{\smallskip}
\hline 
\noalign{\smallskip}
\multicolumn{25}{c}{Population B}\\
\noalign{\smallskip}
\hline    
\noalign{\smallskip}
PKS0226-038 & & 3.43 & 0.52 & 10535 & -590 & & 0.21 & 3525  & & 0.28 & 10908 & 1842 & & 0.77 & 0.58 & 10535 & -3180 & & 0.22 & 3535 & & 0.20 & 10908 & 1842 \\
PKS0237-23 & & 16.85 & 0.44 & 10778 & -1970 & & 0.20 & 4209 & & 0.36 & 10503 & 2624 & & 3.08 & 0.40 & 10778 & -1970 & & 0.20 & 4209 & & 0.40 & 10503 & 2624 \\
CTSJ03.14 & & 5.89 & 0.48 & 10299 & -1103 & & 0.40 & 3274 & & 0.13 & 10500 & 2035 & & 0.78 & 0.33 & 5500 & -1025 & & 0.07 & 3274 & & 0.61 & 10500 & 2035 \\
$[\textrm{HB89}]$1559+088 & & 4.89 & 0.46 & 7129 & -1394 & & 0.25 & 4226  & & 0.29 & 12931 & 2035 & & 0.66 & 0.34 & 7129 & -1394 & & 0.13 & 4226 & & 0.53 & 12931 & 2035 \\
\noalign{\smallskip}
\hline 
    \end{tabular}
    }
\tablefoot{{ {(1) Source identification. (2) \civ{} total flux, in units of $10^{-13}$ erg s$^{-1}$cm$^{-2}$ . (3), (4), (5) Relative intensity, FWHM, and peak shift for the \civ{} BLUE component. (6), (7) Relative intensity and FWHM of the \civ{} BC, which is centered in the respective rest-frame. (8), (9), (10) Relative intensity, FWHM, and peak shift of the \civ{} VBC. Columns (11)-(19) show the same, but for \ion{He}{II}$\lambda$1640.}}}

    \label{tab:civ_specfit}
\end{table*}

\begin{table}[t!]
    \caption{Measurements on the \siiv{}+\ion{O}{IV}]$\lambda1402$ broad lines.
    }
    \label{tab:siiv_full}

    \centering
    \resizebox{\linewidth}{!}{
    \begin{tabular}{lcrrrrrcccc}
    \hline
    \hline
    \noalign{\smallskip}
    Source & & $F^{(a)}_{\textrm{tot}}$  & \multicolumn{1}{c}{FWHM} &  \multicolumn{1}{c}{A.I.}  & \multicolumn{1}{c}{$c(1/2)$}  &  \multicolumn{1}{c}{$c(9/10)$} \\
    & & & \multicolumn{1}{c}{$[\textrm{km\,s}^{-1}]$}  &   & \multicolumn{1}{c}{$[\textrm{km\,s}^{-1}]$} & \multicolumn{1}{c}{$[\textrm{km\,s}^{-1}]$}\\
    (1) & & (2) &  \multicolumn{1}{c}{(3)} & \multicolumn{1}{c}{(4)} & \multicolumn{1}{c}{(5)} &  \multicolumn{1}{c}{(6)} \\
    \noalign{\smallskip}
    \hline
    \noalign{\smallskip}
    \multicolumn{7}{c}{Population A}\\
    \noalign{\smallskip}
    \hline
    \noalign{\smallskip}
WB J0948+0855 & & 0.92 & 9878 $\pm$ 1013 & 0.07 $\pm$ 0.03 & -2257 $\pm$ 192 & -2633 $\pm$ 411 \\
\noalign{\smallskip}
\hline
\noalign{\smallskip}
\multicolumn{7}{c}{Population B}\\
\noalign{\smallskip}
\hline    
\noalign{\smallskip}
PKS0226-038 & & 1.24 & 6387$\pm$1005 & -0.09$\pm$0.12 & 465$\pm$228 & 756$\pm$415\\
PKS0237-23 & & 2.42 & 5084$\pm$666 & 0.06$\pm$0.03 & 965$\pm$119 & 881$\pm$323\\
$[\textrm{HB89}]$1559+088 & & 1.96 & 7661$\pm$1114 & 0.00$\pm$0.06 & -566$\pm$284 & -752$\pm$474\\
\noalign{\smallskip}
\hline
    \end{tabular}
    }
\tablefoot{ {(1) Source identification. (2) Total flux, in units of $10^{-13}$ erg s$^{-1}$cm$^{-2}$. (3) FWHM. (4) Asymmetry index. (5) and (6) Velocity centroids at $\frac{1}{2}$ and $\frac{9}{10}$ fractional intensities.} }
    \end{table}

\begin{table*}[]
    \centering
    \caption{Results from \texttt{specfit} analysis on the 1900\AA\ blend.}
     \resizebox{\linewidth}{!}{
    \begin{tabular}{lcccccccccccccccccccccccccccccccccccccc}
    \hline   
    \hline 
\noalign{\smallskip}
& & \multicolumn{8}{c}{\aliii{}} & &\multicolumn{2}{c}{\ion{Si}{III}$\lambda$1892} & &\multicolumn{16}{c}{\ion{C}{III}]$\lambda$1909+\ion{Fe}{II}$\lambda$1914}\\
\noalign{\smallskip}
\cline{2-10} \cline{12-13} \cline{15-31}
\noalign{\smallskip}
& & & &\multicolumn{3}{c}{\aliii{} BLUE} & & \multicolumn{2}{c}{\aliii{} BC$^{(b)}$} & & & & & & & & \multicolumn{2}{c}{\ion{C}{III}]$\lambda1909$ BC$^{(b)}$} & & \multicolumn{3}{c}{\ion{C}{III}]$\lambda1909$ VBC} & & \multicolumn{2}{c}{\ion{C}{III}]$\lambda1909$ NC$^{(b)}$} & & \multicolumn{3}{c}{\ion{Fe}{III}$\lambda1914$}\\
\noalign{\smallskip}
\cline{5-7} \cline{9-10}  \cline{18-19} \cline{21-23} \cline{25-26} \cline{28-30} 
\noalign{\smallskip}
Source & & $F_{\textrm{tot}}$ & W & $F/F_{\textrm{tot}}$ & FWHM & Peak & & $F/F_{\textrm{tot}}$ & FWHM &  & $F_{\textrm{tot}}$ & W & & $F_{\textrm{tot}}$ & W & & $F/F_{\textrm{tot}}$ & FWHM  & & $F/F_{\textrm{tot}}$ & FWHM & Peak & & $F/F_{\textrm{tot}}$ & FWHM  & & $F_{\textrm{tot}}$ & FWHM & Peak\\
& & & [\AA] & & [km s$^{-1}$] & [km s$^{-1}$] & & & [km s$^{-1}$]   & & & [\AA] & & & [\AA] & & & [km s$^{-1}$]  & & & [km s$^{-1}$] & [km s$^{-1}$] & & & [km s$^{-1}$]  & & & [km s$^{-1}$] & [km s$^{-1}$]\\
(1) & & (2) & (3) & (4) & (5) & (6) & & (7) & (8) & & (9)  & (10) & & (11) & (12) & & (13)  & (14) & & (15) & (16) & (17) & & (18) & (19)  & &(20) &  (21) & (22) &\\
\noalign{\smallskip}
\hline 
\noalign{\smallskip}
\multicolumn{30}{c}{Population A}\\
\noalign{\smallskip}
\hline    
\noalign{\smallskip}
BZQJ0544-2241 & & 12.61 & 5.44 & … &  … &  … & & 1.00 & 5559  & & 9.17 & 4.03 & & 1.23 & 5.46 & & 1.00 & 5382 & & … &  … &  … & & … &  … &  & 9.17 & 5105 & 3169 \\
WB J0948+0855 & & 8.55 & 5.31 & 0.33 & 3184 & -2011 & & 0.67 & 5549  & & 5.57 & 4.38  & & 1.28 & 1.19 & & 1.00 &  5549  & & … &  … &  … & & … &  … &   & 1.46 &  5177 &  1618 \\
\noalign{\smallskip}
\hline 
\noalign{\smallskip}
\multicolumn{30}{c}{Population B}\\
\noalign{\smallskip}
\hline    
\noalign{\smallskip}
PKS0226-038 & & 3.38 & 3.52 & … &  … &  … & & 1.00 & 4722  & & 5.26 & 5.60 & & 0.87 & 9.97 & & 0.65  & 4722  & & 0.35 & 7018 & 2467 & & 0.06 & 999 &  & … &  … &  … \\
PKS0858-279 & & 32.88 & 2.96 & … &  … &  … & & 1.00 & 4973  & & 44.3 & 4.05 & &  7.75 & 7.15 & & 0.66 & 4973  & & 0.34 & 8588 & 4908 & & … &  … &  & … &  … &  … \\
CTSJ03.14 & & 2.12 & 1.64 & … & … & … & & 1.00 & 3044  & & 2.97 & 2.46 & & 1.10 & 9.78 & & 0.83 & 3044  & & 0.17 & 6954 & 2948 & & 0.03 & 970 &  & … &  … &  … \\
$[\textrm{HB89}]$1559+088 & & 3.42 & 2.55 & … &  … &  … & & 1.00 & 4226  & & 9.15 & 7.02 & & 0.96 & 7.65 & &  0.42 & 4226  & & 0.58 & 7338 & 2207 & & 0.01 & 999 &  & … &  … &  … \\
\noalign{\smallskip}
\hline 

    \end{tabular}
    } \\
\tablefoot{{{ {(1) Source identification. (2) \aliii{} total flux, in units of $10^{-14}$ erg s$^{-1}$cm$^{-2}$. (3) \aliii{} Equivalent width $W$. (4), (5) and (6) Relative intensity, FWHM, and peak shift respectively for the \aliii{} BLUE component. (7) and (8) Relative intensity and FWHM for the \aliii{} BC, centered in the respective rest-frame. (9) and (10) \ion{Si}{III}$\lambda$1892 total flux (in units of $10^{-14}$ erg s$^{-1}$ cm$^{-2}$) and $W$. (11), (12) Same for \ion{C}{III}]$\lambda$1909. (13) and (14) Relative intensity and FWHM for \ion{C}{III}]$\lambda$1909 BC, centered in the respective rest-frame. (15), (16) and (17) Relative intensity, FWHM, and peak shift for the \ion{C}{III}]$\lambda$1909 VBC. (18) and (19) Relative intensity and FWHM for the \ion{C}{III}]$\lambda$1909 NC, centered in the respective rest-frame. (20), (21) and (22) \ion{Fe}{III}$\lambda$1914 total flux (in units of $10^{-14}$ erg s$^{-1}$ cm$^{-2}$), FWHM and peak shift.} }  } }
    \label{tab:1900A_specfit}
\end{table*}

\par  {The measurements resulting from the \texttt{specfit} analysis of the 1900\AA\ blend are presented in Table \ref{tab:1900A_specfit}.} 
{As \ion{Si}{III}]$\lambda1892$ has the  same peak rest-frame velocity and FWHM of \aliii{}, we do not report these values for the \ion{Si}{III}] line in the table.} 

\begin{table}[t!]
    \caption{Radio properties of the full ISAAC sample.}
    \centering
    \resizebox{\linewidth}{!}{
    \begin{tabular}{l c rrrcc}
    \hline
    \hline
    \noalign{\smallskip}
    Source & & \multicolumn{1}{c}{$f_{\rm 1.4GHz}$} & \multicolumn{1}{c}{$P_{\rm{1.4GHz}}$} & \multicolumn{1}{c}{$\log R_{\textrm{K}}$} & Survey & Radio\\
    & & $[\textrm{mJy}]$ &  & & & Class.\\
    (1) & & (2) & (3) & (4) & (5) & (6)\\
    \noalign{\smallskip}
    \hline
    \noalign{\smallskip}
    \multicolumn{6}{c}{{ISAAC1}}\\
    \noalign{\smallskip}
    \hline
    \noalign{\smallskip}
HE 0001-2340 & & < 0.90	& < 0.15 & < 0.48 & NVSS & RQ \\
$[\rm HB89]$0029+073 & & < 0.24	& < 0.08 & < -0.12 & FIRST & RQ \\
CTQ 0408 & & < 2.49	& < 0.85 & < 0.43 & SUMSS & RQ \\
SDSSJ005700.18+143737.7	& &	< 0.90 & < 0.21	& < 0.35 & NVSS & RQ \\
H 0055-2659	& &	< 0.90 & < 0.38	& < 0.60 & NVSS & RQ \\
SDSSJ114358.52+052444.9	& & < 0.28 & < 0.06	& < 0.17 & FIRST & RQ \\
SDSSJ115954.33+201921.1	& & < 0.29 & < 0.11	& < -0.04 & FIRST & RQ \\
SDSSJ120147.90+120630.2	& & < 0.29 & < 0.11	& < -0.26 & FIRST & RQ \\
SDSSJ132012.33+142037.1	& & < 0.29 & < 0.06	& < 0.15 & FIRST & RQ \\
SDSSJ135831.78+050522.8	& & < 0.28 & < 0.06 &	< -0.15 & FIRST & RQ \\
Q 1410+096 & & < 0.30 & < 0.11 & < -0.13 & FIRST & RQ \\
SDSSJ141546.24+112943.4	& & 6.50 & 1.41	& 1.27 & NVSS & RI \\
B1422+231 & & 284.67 & 119 & 2.78 & FIRST & RL \\
SDSSJ153830.55+085517.0	& & < 0.31 & < 0.12	& < 0.36 & FIRST & RQ \\
SDSSJ161458.33+144836.9	& & < 0.28 & < 0.06	& < -0.13 & FIRST & RQ \\
PKS 1937-101 & & 838 & 381 & 2.99	& NVSS & RL	\\
PKS 2000-330 & & 446 & 202 & 3.14 &	NVSS & RL \\
SDSSJ210524.49+000407.3	& & 2.50 & 0.46	& 0.54 & NVSS & RQ	\\
SDSSJ210831.56-063022.5	& & < 0.37 & < 0.07	& < 0.18 & FIRST & RQ \\
SDSSJ212329.46-005052.9	& & < 0.29 & < 0.05	& < -0.45 &	FIRST & RQ \\
PKS 2126-15	& & 589	& 207 & 2.97 & NVSS & RL \\
SDSSJ235808.54+012507.2	& & < 0.28 & < 0.11	& < -0.32 &	FIRST & RQ \\
\noalign{\smallskip}
\hline
\noalign{\smallskip}
\multicolumn{6}{c}{ISAAC2}\\
\noalign{\smallskip}
\hline
\noalign{\smallskip}
PKS0226-038	& & 939	& 135 & 3.50 & NVSS & RL \\
PKS0237-23 & & 6256	& 1043 & 3.80 & NVSS & RL \\
BZQJ0544-2241 & & 133 & 10.78 & 2.06 & NVSS & RL \\
PKS0858-279	& & 1474 & 233 & 2.77 &	NVSS & RL \\
CTSJ01.03 & & 2.90 & 0.55 & 0.92 & NVSS & RQ \\
WB J0948+0855	& & 36.50 & 4.88 & 1.78 & NVSS & RI	\\
CTSJ03.14 & & 151	& 30.31	& 2.40 & NVSS & RL \\
PKS1448-232	& & 439	& 73.03	& 2.83 & NVSS & RL \\
$[\textrm{HB89}]$1559+088 & & 4.50 & 0.79 & 0.87 & NVSS & RQ	\\
FBQS J2149-0811	& & < 0.36 & < 0.05	& < 0.02 & FIRST & RQ \\
\noalign{\smallskip}
\hline
    \end{tabular}
     }
    \tablefoot{ {(1) Source identification. (2) Radio flux at 1.4 GHz. (3) Radio power $P$ at 1.4 GHz, in units of $10^{26}$ W Hz$^{-1}$.  (4) Radioloudness parameter. (5) The survey from where we obtain the radio data. (6) Radio classification.}  }

    \label{tab:radio_isaac}
\end{table}

\subsection{Radio data}
\label{sec:radiodata}

\par  {The radio properties {for the complete ISAAC sample} are reported in Table \ref{tab:radio_isaac}.} The radio fluxes used for estimate these parameters were obtained from the 1.4-GHz NRAO VLA Sky Survey \citep[NVSS,][]{Condon_1998} and from the VLA Faint Images of the Radio Sky at Twenty-Centimeters survey \citep[FIRST,][]{Gregg_1996,Becker_1995} catalogues. In the case of CTQ\,0408, not covered by these two surveys, the upper limit for the flux has been estimated from the SUMSS catalogue \citep{Mauch_2003}.   The $R_{\textrm{K}}$ parameter is determined as the ratio between the rest-frame specific fluxes at 1.4 GHz and at the $g$-band effective wavelength \citep[following][] {Ganci_2019}. We then consider three different ranges of radio-loudness: radio-quiet (RQ; $R_{\textrm{K}}< 10$), radio-intermediate (RI; $10 \leq R_{\textrm{K}} \leq 70$), and radio-loud (RL; $R_{\textrm{K}}\geq 70$). 

\par If  the object is not detected (for the 50\% of the sample: 16 sources, 15 from ISAAC1, and one from ISAAC2), an upper limit on the radio flux is set equal to a detection threshold ($\sim 2$ times the rms, in both FIRST and NVSS catalogs) at the position of the source.  Our preference is to rely on the FIRST detection limit as its maps have a higher sensitivity than NVSS and allows for a better restrictive upper limit. However, in cases where the object's position does not fall within the current coverage of the FIRST catalog, we then utilize the NVSS detection limit. 
When considering the complete ISAAC sample, we have 10 RL, two RI, and 20 RQ sources. 

For objects from both HEMS and FOS sample, we re-estimated the values of $R_\textrm{K}$ using the same methodology employed for the ISAAC data. In the HEMS sample, from the 28 sources four are RL and 24 RQ. As mentioned in section \ref{sect:comparison_samples}, in the low luminosity FOS sample there are 34 RL and 50 RQ.

By default, we have defined that the sources initially classified as radio-intermediate are considered together as radio-quiet objects unless we have evidences that the source is jetted.  The origin of enhancement of radio emission in RIs is possibly related to star formation \citep{Condon_2013,bonzinietal15,Caccianiga_2015, Ganci_2019}, powerful broad line region and extended ionized outflows \citep[][and references therein]{panessaetal19}, or to intrinsic reddening in the optical fluxes.
\section{Results}

\label{section:results}


\subsection{ ISAAC1 and ISAAC2 basic results: a synopsis}
\label{sec:synopsis}
The optical and UV main spectral properties of ISAAC1 and ISAAC2 are fully consistent. 

\hb\ -- The dichotomy between Pop. A and Pop. B is  preserved in the complete ISAAC sample, with Pop. A quasars usually presenting lower values of \hb{} $W$\ and FWHM. However, it is noteworthy that the lowest FWHM(\hb{}) values in ISAAC2 are observed in two Pop. B quasars (CTSJ03.14 and PKS1448-232), both of which are blazar candidates and therefore the FWHM may be strongly influenced by the source orientation. Among the ISAAC2 \hb\, Pop. A profiles, all are symmetric except for BZQJ0544-2241, which requires an additional (usually weak) BLUE component to fully represent the \hb{} line. In contrast, the \hb\ profiles in Pop. B sources are all redward asymmetric with the exception of $[\textrm{HB89}]$1559+088, which is nearly symmetric and also requires a BLUE component.  This additional component is added when the blueshift observed in the \hb\ profile does not correspond to the blueshifted SBC observed in \oiiiseven{}.  The full ISAAC sample has eight quasars (five Pop. A and three Pop. B) with a BLUE component, which contribution to the profile has an average value of $\approx$ 8\%. The mean asymmetry index for the whole ISAAC sample Pop. B is 0.26 (same for only ISAAC2) and they have the full broad profile with an averaged velocity centroid $c(1/4)$ of 1740\,\kms{}.  This is attributed to the fact that in Pop. B the most prominent broad component is the VBC, which accounts for 54\% of the \hb{} profile in the ISAAC2 sample (57\% for the whole ISAAC), with median FWHM and shift values of $\approx$ 10420 and 3149\,\kms, respectively. Consistent values are found for the full ISAAC sample, with median values of the FWHM and shift of the VBC of 11760 and 3000 \kms , respectively, in good agreement with the values obtained in HEMS sources \citep[e.g.,][see also \citealt{vietrietal18,vietrietal20,wolf_2020} for extremely broad \hb\ profiles]{Marziani_2009,sulentic_2017}.

\paragraph{\oiii\ --}  In the full ISAAC sample, both populations (A \& B) exhibit significant blue shifts, with the most pronounced ones observed in Pop. A  sources. In most of the full ISAAC sample both the \oiiionly\ SBC and NC are often not in the rest-frame. Additionally, for 18 out of the 32 sources from the whole ISAAC sample (16 from ISAAC1 and two from ISAAC2), the SBC is the strongest component, accounting for $\sim$ 64\% of the flux in the case of Pop. A and $\sim$ 52\% for Pop. B. Consequently, the two populations A and B also differ in terms of FWHM, with the Pop. A on average showing slightly broader profiles due to the presence of a stronger SBC. Moreover, in contrast to the low-$z$ scenario where the majority of the \oiiiseven{} profiles exhibit blueshift velocity centroids at half intensity, $c(1/2)$, lower than 250 \kms \citep{zamanovetal02, marziani_2016a}, in our complete high-$z$ ISAAC sample the 62\% of the sources have blueshifts in \oiiionly{} greater than 250 \kms\, and are considered outliers. However, in ISAAC2, we do not observe the few very large \oiiionly\ profiles (FWHM $\sim 4000$ \kms) measured in ISAAC1.  This may be a consequence of the fact that ISAAC1 is primarily composed of radio-quiet sources (which typically exhibit broader \oiii{} profiles than radio-loud ones, see \S \ref{sec:oiii}), while a significant fraction of ISAAC2 is classified as radio-loud.  {Significant \oiiionly\ outflows appear to be a common feature at high redshift and have also been observed in other samples \citep[see e.g.,][]{Kakkad_2020}.} As discussed in \citetalias{Deconto-Machado_2023}, the \oiiionly\ profile almost always appear peculiar if compared to the ones of optically selected samples at low redshift \citep{zamanovetal02,marzianietal03b,bianetal05,zhangetal11,craccoetal16}, with higher shift amplitude and broader widths.

\paragraph{\civ\ -- } Consistent results with ISAAC1  are also observed for ISAAC2 \civ{} profiles. In general, the Pop. B sources from ISAAC2 tend to exhibit slightly larger values of $W$\ than Pop. A. This is also observed in the full ISAAC sample (median $W$ $\sim$ 27\ \AA\ for Pop. B vs.  $\sim 21$ \AA\ for Pop. A). Conversely, Pop. A sources tend to show greater FWHM, asymmetry, and centroid velocities, which is in agreement with the HEMS sample \citep{sulentic_2017}. 
The  BLUE is the strongest component in the full ISAAC sample, and   contributes to $\sim$ 72\%\ and $\sim$ 51\% of the full \civ{} flux for Pop. A and Pop. B, respectively. Consistently,  the largest blueshifts are found for the BLUE components of Pop. A sources, which achieve a median value of $c(1/2)$ of the full \civ{} profile $\sim$ -2590 $\pm$ 1890 km s$^{-1}$ for the entire ISAAC sample. In the case of Pop. B, this value is $\sim$ -1530 $\pm$ 760 km s$^{-1}$. Centroid shifts are comparable to the ones derived in recent works that followed a similar analysis strategy \citep[and especially \citet{sulentic_2017} which analyses the HEMS sample]{coatman_2016,vietrietal18,templeetal23}. 
Both \oiii\ and \civ\ suggest the presence of extremely powerful winds in the NLR and BLR \citep{marzianietal16a,marzianietal16,Fiore_2017,bischettietal17}. At variance with low-$z$, low-luminosity sources, the blueshifted emission associated with the winds are prominent in both Pop. A and B, as further discussed in \S \ref{sect:dom_outflows} and \S \ref{sect:param_outflows}.

\paragraph{1900\AA\ blend -- } In ISAAC2 sample we have 1900\AA\, blend data for six of the 10 objects (two Pop. A and four Pop. B). The 1900\AA\ blend properties observed in the ISAAC2 sample also appear consistent with the full ISAAC sample, as well as with the expectations from recent works involving intermediate redshift sources \citep{templeetal20,marzianietal22,Buendia-rios}.  Considering the whole ISAAC sample (11 Pop. A and nine Pop. B quasars), we find that the BLUE component in \aliii{} is detected in only five Pop. A with an average blueshift of $\sim -1720$ km s$^{-1}$ and account for $\sim$ 40\% of the full profile. For Pop. B, all nine sources show \aliii{} profiles well-represented by a single rest-frame BC with averaged FWHM $\sim 4240$ km s$^{-1}$. For \ion{Si}{III}]$\lambda$1892, only three Pop. A exhibit a BLUE component, with an averaged shift of $\sim -1880$ km s$^{-1}$. Regarding the \ion{C}{III}]$\lambda$1909 profiles, Pop. A sources are suitably represented by a rest-frame BC with FWHM $\sim 3370$ km s$^{-1}$. Nine out of the 11 Pop. A sources needed an additional \ion{Fe}{III} component at $\sim$ 1914\AA. The Pop. B \ion{C}{III}]$\lambda$1909 profiles are fitted by the combination BC+VBC, where the VBC is weaker than that of H$\beta$, as it represents only 37\% of the full profile with FWHM of $\sim$ 7240 km s$^{-1}$. 


\begin{figure}[t!]
    \centering
    \includegraphics[width=0.95\columnwidth]{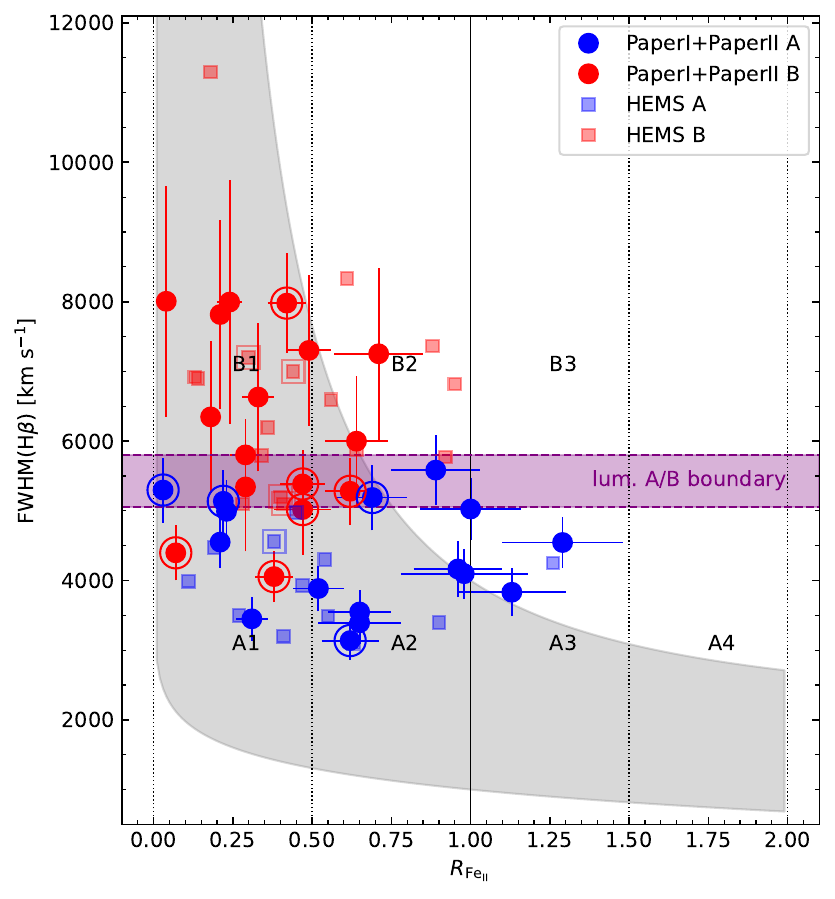}
    \caption{Location on the MS of the 10 new sources  for this work together with the ones from \citetalias{Deconto-Machado_2023} (circles) and the HEMS sample (squares). Pop. A quasars are represented by blue symbols and Pop. B by red symbols. The radio-loud sources are surrounded by an open circle/square in the plot. Grey region indicates the distribution of the main sequence of quasars at low-$z$. Purple-shaded area shows the luminosity-dependent boundary between Pops. A and B for high luminosity sources, as described in detail in \citetalias{Deconto-Machado_2023}.}
    \label{fig:optical_plane}
\end{figure}

\subsection{The optical plane of the 4DE1 parameter space}
\label{sec:optical_plane}

{The classification as A/B population in the quasar MS is based on two parameters that can be measured only on the NIR spectra: the FWHM of \hb{} full broad profile and the strength of the \ion{Fe}{II} blend at 4570\AA,} {defined as the ratio between the intensities of the blue blend of \feii\, at 4750\AA\, and \hb{}, $R_{\textrm{\ion{Fe}{II}}}$.} {The Pop. A/Pop. B  limit was set according to the luminosity-dependent FWHM value, as explained in \citetalias{Deconto-Machado_2023}. We measured the parameter $R_{\textrm{\ion{Fe}{II}}}$ parameter to assign a number from 1 to 4 to the Spectral Type (ST) covering the range $0 \le R_{\textrm{\ion{Fe}{II}}} \le 2$ with a step $\Delta R_{\textrm{\ion{Fe}{II}}}=0.5$ \citep{Sulentic_2002}.} 

The location of the whole ISAAC sample in the MS optical plane is shown in  Fig. \ref{fig:optical_plane}, where are also represented the 28 sources belonging to the HEMS sample.  As discussed in \citetalias{Deconto-Machado_2023}, the sources of high luminosity, such as the 10 new quasars presented here, show a  displacement in the direction of increasing FWHM(H$\beta$) by 1000 -- 1500 \kms\ respect to the low-luminosity, low-\textit{z} samples \citep[see also][]{sulentic_2017}.
{Of the 10 new sources, six are classified as Pop. B and four as Pop. A. CTSJ03.14 and PKS1448-232 (see the models in Figs. \ref{fig:CTSJ0314} and \ref{fig:PKS1448}, respectively), two RLs classified as B1, are located bellow the Pop A/B boundary for their luminosity. Both sources are blazar candidates (see \citet{Massaro_2014} for CTSJ03.14 and \citet{Massaro_2015} for PKS1448-232), implying that the FWHM of \hb{} may be significantly lowered by a pole-on  orientation of the line-emitting region.} In the complete sample considered in this work (ISAAC1+ISAAC2+HEMS), there are 30 sources classified as Pop. A (including five RL) and 30 classified as Pop. B (with eight of them being RL).



\begin{figure*}
    \centering
    \includegraphics[width=0.35\linewidth]{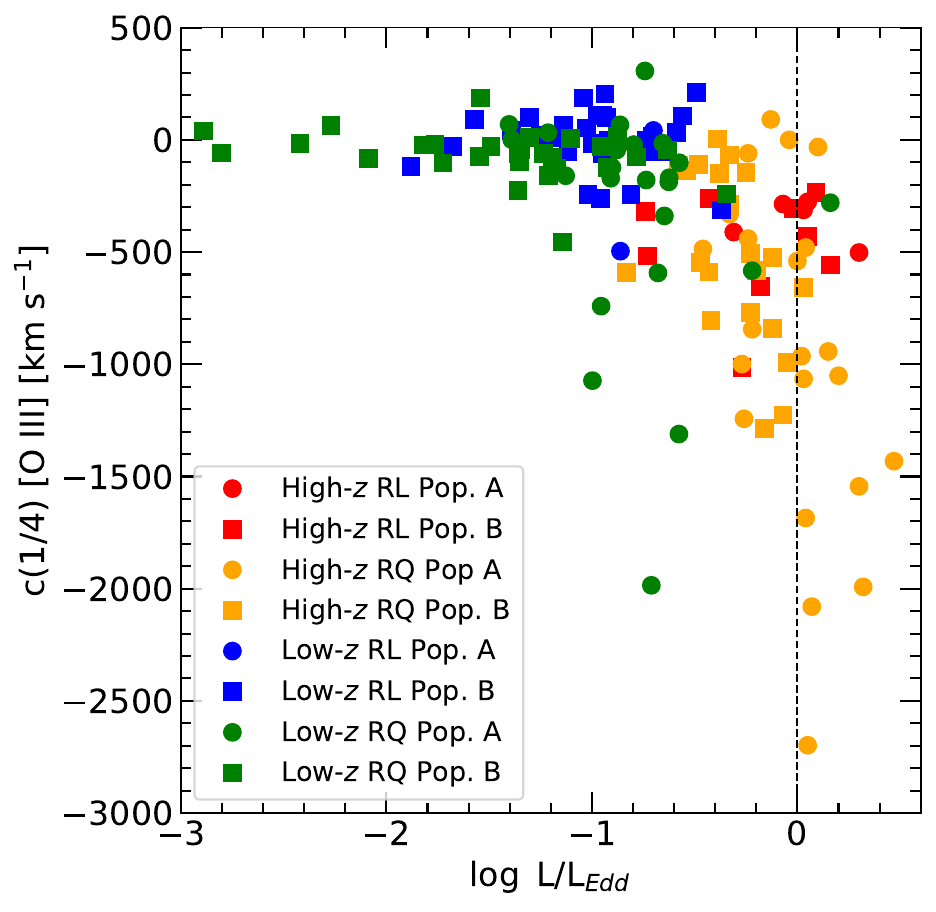}
    \includegraphics[width=0.35\linewidth]{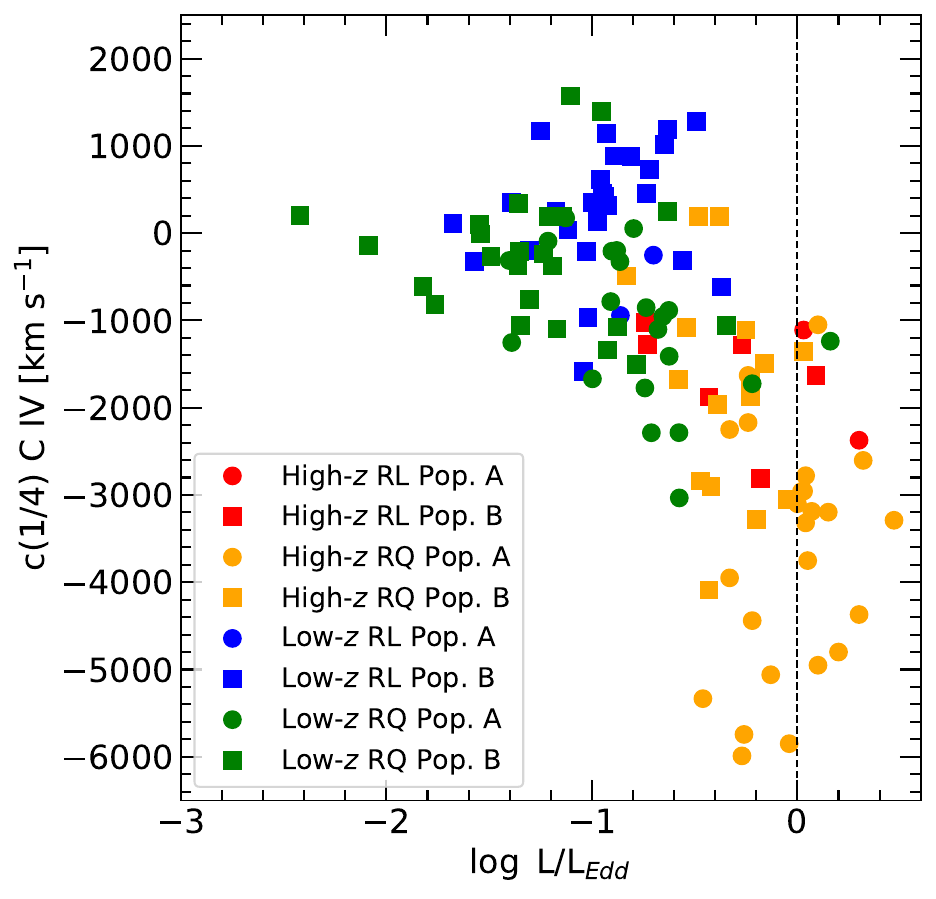}
    \caption{Velocity centroid $c$(1/4) \textit{versus} $L/L_{\textrm{Edd}}$ for  \oiiiseven{} (\textit{left}) and \civ{} (\textit{right}). {Dark green and blue symbols correspond respectively to RQ and RL sources at low-z, while orange (RQ) and red (RL) symbols to high-z quasars. Within each colour, filled squares represent  Pop. B sources and the bullets Pop. A quasars.}} 
    \label{fig:lledd_oiii_civ}
\end{figure*}

   
 

\subsection{The role of physical parameters}
\label{physical_parameters}

\par  {In \citetalias{Deconto-Machado_2023}, we show that while the bolometric luminosity, $L_{\mathrm{bol}}$, and the black hole mass ($M_{\mathrm{BH}}$)  exhibit trends with the \civonly{} blueshifts, the primary correlation is observed with the Eddington ratio, with the higher \civonly{} blueshift in the higher accretion rate \citepalias[see e.g. \S 6.3 of][and references therein]{Deconto-Machado_2023}. These results are reconfirmed in the present work. By including the new data from ISAAC2 in the sample composed by both the high- and low-$z$ data, we find a relation between \civonly{} $c(1/4)$ blueshift { {(< -250 km s$^{-1}$)}} and $L_{\mathrm{bol}}$  with a slope of 0.21$\pm$0.07 (with a Pearson correlation coefficient c.c. of 0.61, $\rho \sim 10^{-9}$). \textcolor{blue}{} The slope  0.14 $\pm$ 0.06, and Pearson c.c. $\sim 0.48$ ($\rho \sim 10^{-6}$) suggest an {even weaker} relation    between \civonly{} $c(1/4)$ and the black hole mass $M_{\mathrm{BH}}$.} Regarding the Eddington ratio, we find the following relation  (with a Pearson c.c. $\sim$ 0.66 and a $\rho$-value for the null hypothesis $\sim 10^{-11}$):}
\begin{equation*}
    \log |c(1/4)_{\mathrm{\ion{C}{IV}}}|= (0.43 \pm 0.05)\log L/L_{\mathrm{Edd}}+(3.39 \pm 0.04).
\end{equation*}

\par { {Similar results, albeit less significant, are observed for \oiiionly{}. When analyzing the 45 sources (34 high-$z$ and 11 low-$z$) that exhibit significant \oiiionly{} outflows, we find a slope of $0.12 \pm 0.07$ between \oiiionly{} $c(1/4)$ and $L_{\rm bol}$ (Pearson c.c. $\sim 0.31$, $\rho \sim 10^{-2}$). Concerning $L/L_{\rm Edd}$, we obtain (Pearson c.c. $\sim$ 0.48, $\rho \sim 10^{-4}$): }}
\begin{equation*}
    \log |c(1/4)_{\mathrm{[\ion{O}{III}]}}|= (0.33 \pm 0.09)\log L/L_{\mathrm{Edd}}+(2.91 \pm 0.04).
\end{equation*}



\par In tight agreement with these relations, Fig. \ref{fig:lledd_oiii_civ} shows the dependence of $c$(1/4) on $L/L_{\textrm{Edd}}$ for the \oiiiseven{} and \civ{} emission lines, {where we have included all the measures of c(1/4) velocity centroids of both lines (blue-shifted or not) and for both low- and high-$z$ samples.} For the two emission lines, strong outflows (i.e., $\gtrsim$ 500 km \,$^{-1}$ for \oiiiseven{} and $\gtrsim$ 2000 km\,s$^{-1}$ for \civ{}) are found at Eddington ratios $\log L/L_{\textrm{Edd}} \gtrsim -0.8$ {(${ L/L_{\rm Edd} \sim 0.2}$)}, consistent with {previous results \citep[see e.g.][]{marzianietal03b,bianetal05, Komossa_2008, marzianietal16a,craccoetal16,coatman_2016,Wang_2018,Ayubinia_2023}}. 

 {{The effect of high Eddington ratio is strengthened at high luminosity.} { The high-$z$\ -- low-$z$\ comparison is consistent with the weaker luminosity effect (outflow velocity $\propto L^\frac{1}{4}$) on both the RL and RQ samples that is expected for radiation driven winds and that becomes appreciable if samples spread over a wide range of luminosity are considered \citep[][c.f.  \citealt{sulentic_2017}]{laorbrandt02}}.}

\par  {The radioloudness  appears to play a role in reducing the outflow velocities in both \oiiiseven{} and \civ{}.   {The sources that present the most significant blueshifts in both \oiiionly{} and \civonly{} in Fig. \ref{fig:lledd_oiii_civ} are all radio-quiet. Conversely, the radio-loud sources (blue and red symbols)  exhibit smaller outflow velocities. } This result is in a good agreement with previous findings  \citep[e.g.,][]{marzianietal03b, Laor2018} and is discussed in more details in \S \ref{sec:intercomparison} and \ref{sec:rk_trends}.}

\begin{figure}[t!]
    \centering
    \includegraphics[width=\linewidth]{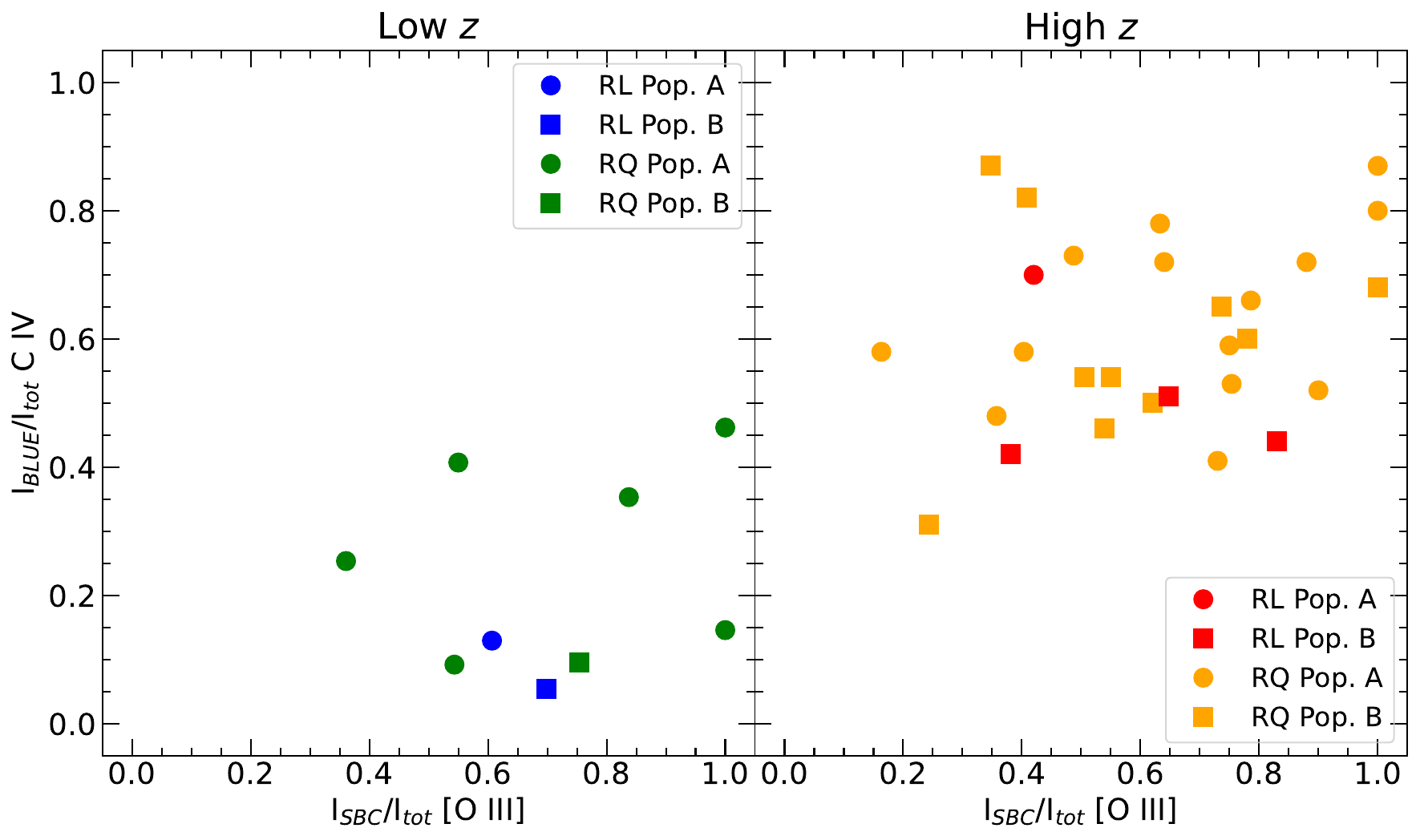}
    \caption{Comparison between the relative intensities of the blueshifted component for both \oiiiseven{} and \civ{}. Low redshift ($z$) sources represented in the $left$ panel and high $z$ in the $right$ one.} 
    \label{fig:intensities}
\end{figure}

\begin{figure*}[t!]
    \centering
    \includegraphics[width=\linewidth]{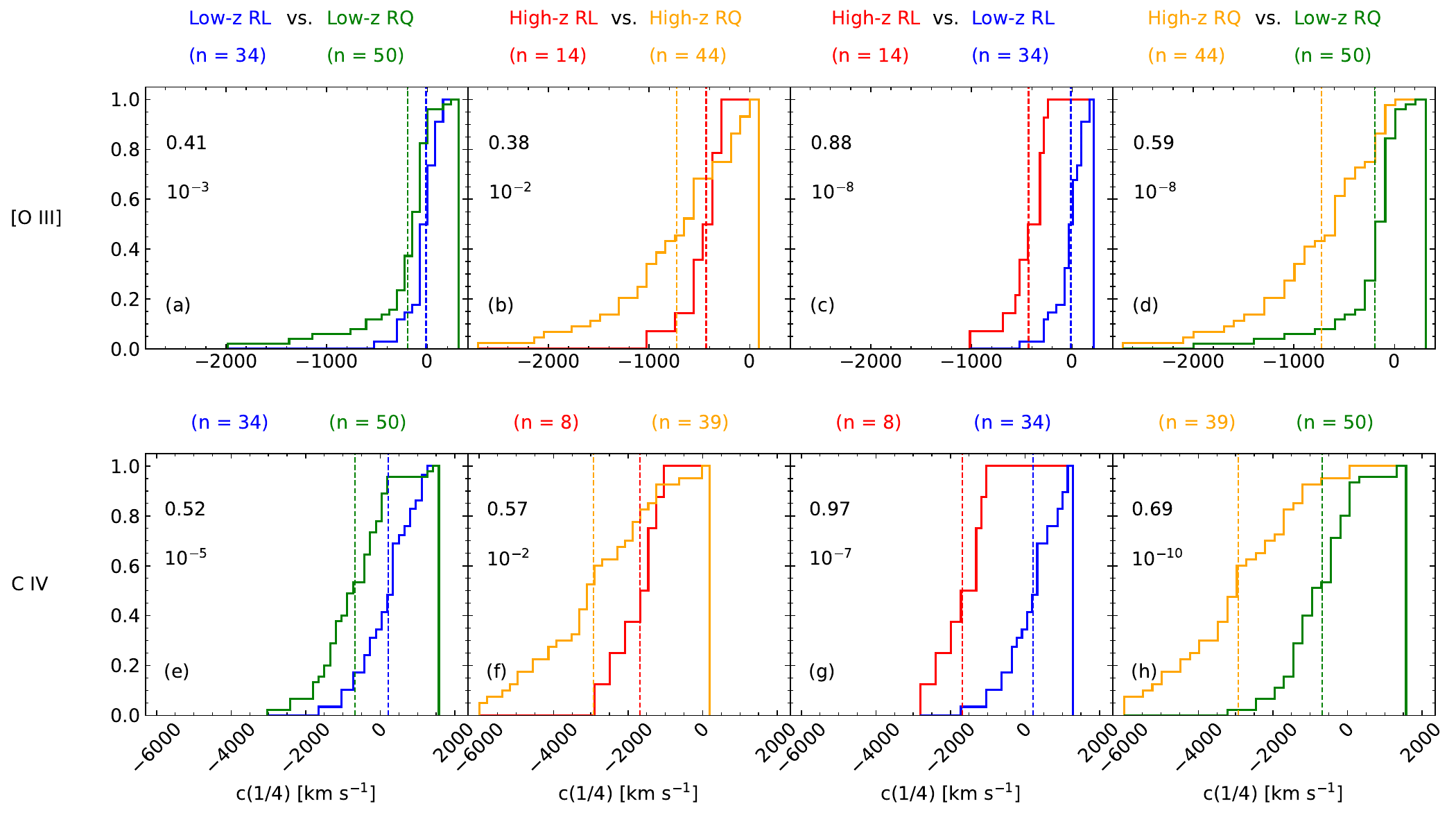}
    \caption{Centroid velocities at 1/4 intensities ($c$(1/4)) cumulative distributions of the samples separated by radioloudness and redshift ranges for \oiiiseven{} (\textit{top} panels) and \civ{} (\textit{bottom} panels). Kolmogórov-Smirnov (KS) tests were performed comparing different subsamples identified at the top of the panels. KS statistics for each comparison are shown on the top left of each plot together with the respective $\rho$-value. Vertical lines indicate the mean value for each subsample.}
    \label{fig:distr_acumuladas}
\end{figure*}

\subsection{Inter-comparison between RL and RQ at low and high $z$: \oiiiseven{} \textit{vs.} \civ{}}
\label{sec:intercomparison}

{Fig. \ref{fig:intensities} presents a comparison between \oiiionly{} and \civonly{}} in terms of the relative intensity of the outflowing component (SBC for \oiiiseven{} and BLUE for \civ{}) with respect to the full profiles, $I_{\rm BLUE}/I_{\rm tot}$. In this case, we exclusively consider sources that are identified as blue-outliers in \oiiionly{} and which also present blueshifts in \ion{C}{IV}. { {Therefore, the sample shown in Fig. \ref{fig:intensities} has 28 high-$z$ sources (14 from ISAAC and 14 from HEMS) and 9 low-$z$ FOS sample.}} {{ {In addition, in the FOS sample there are seven objects in which only a blueshifted SBC \oiiionly{} component (with no clear NC) is detected, which nevertheless present shifts in the $c(1/2)$ lower than -250. Clearly identifying any outflow in these cases would require higher resolution and an improved S/N than the data available in the present work.}}  

At low redshift, the behaviour of the BLUE component appears to be different in \oiiionly{} and \civonly{} in terms of relative intensity:  {while the \oiiionly{} outflowing component can achieve a $I_{\textrm{SBC}}/I_{\textrm{tot}}$ up to a nearly $\sim$ 100\% of the total intensity of the full  profile, the \civonly{}  outflowing component never surpasses $\sim$ 40\% of the entire profile, with the lower values found for the RL sources}


\par The right panel of Fig. \ref{fig:intensities} shows the same analysis but for the high-redshift range. At variance with the low-$z$ sample, both the \oiiiseven{} and \civ{} emission lines exhibit very similar behaviour with respect to their outflowing component, and their ratio $I_{\textrm{BLUE}}/I_{\textrm{tot}}$ can vary from $\sim$ 20\% to 100\% of the full profile, with the largest contributions found in radio-quiet sources. {The radio-loud/radio-quiet difference at high redshift is consistent with the one found at low redshift, with the RL sources presenting clearly fainter \oiiiseven{} SBC and \civ{} BLUE.}

\par {We have analyzed the RL and RQ blueshift distributions in an attempt to characterize other parameters affected by radioloudness.  } Fig. \ref{fig:distr_acumuladas} shows the cumulative distribution functions of centroid velocities at 1/4 flux intensity ($c(1/4)$) of both \oiiiseven{} and \civ{} emission lines for the samples by separating them according to the radio classification and the redshift range. { All sources with \oiiionly{} or \civonly{} spectral information have been taken into account regardless of the shift}. 
The panel (a) shows a comparison between RL and RQ at low redshift for the \oiiiseven{} $c(1/4)$. In this case, the Kolmogorov-Smirnov (KS) test reveals that the RL and RQ subsamples are statistically indistinguishable, although the RQ in general present larger values of $c(1/4)$ towards the blue, which indicates stronger outflows than for RL sources. A similar result is found when comparing RL and RQ $c(1/4)$ at high-$z$ (panel (b) of Fig. \ref{fig:distr_acumuladas}).

\par Panel (c) of Fig. \ref{fig:distr_acumuladas} compares the RL sources at high- and low-$z$ ranges. The KS test indicates that the distributions are significantly different ($\sim$ 0.88, $\rho \sim 10^{-8}$), with the RL sources at high redshift presenting larger outflows velocities (mean value $\sim$ $-$420 km\,s$^{-1}$) than the ones at lower redshifts ($\sim$ $-$40 km\,s$^{-1}$). The same is found when comparing RQ sources at high and low redshift (panel (d)). The difference in this case is still significant due to more prominent outflow at high $z$, with $c(1/4)$ mean values  $\approx -740$ and $\approx -200$ \kms\ for high and low $z$, respectively.

\par The bottom panels of Fig. \ref{fig:distr_acumuladas} show the results for the \civ{} emission line. In general, the $c(1/4)$ of this line follows the same behaviour of \oiiionly{} albeit with stronger outflows. In both RQs and RLs there is a statistically very significant difference between the distributions of \civonly{} c(1/4) at high and low redshift. The more extreme difference is found between RL sources at high and low $z$, where the distributions are completely different (KS coefficient of 0.97, and a probability of coming from the same distribution $\rho \sim 10^{-7}$). The significance is extremely high, even if the RL subsample in the case of \civ{} at high $z$ counts with only seven sources. 

 
 {The cumulative distributions of both \oiiionly{} and \civonly{} centroid shifts demonstrate that the main difference between samples is when they widely differ in luminosity, not radioloudness.  In the two redshift domains considered  (corresponding on average to a difference in luminosity of a factor $\approx 60 $), Fig. 
 \ref{fig:lledd_oiii_civ} and Fig. \ref{fig:distr_acumuladas} show that there is second order effect related to radioloudness, in both \civ\ and \oiii. The effect is more significant at low-$z$ (with a $\rho \sim 10^{-5}$), although a similar difference in shift amplitudes is detected also at high-redshift. In the latter case, the small number of sources makes the effect not statistically significant. Summing up, the effect of radioloudness is consistent in the low-$z$\ and high-$z$\ samples, {with lower blueshifted velocities and outflows in the RLs,} albeit weaker than the one due to luminosity {and the accretion rate (see section \ref{physical_parameters})}.}



\subsection{Trends with radio loudness}
\label{sec:rk_trends}
    
\subsubsection{\ion{Fe}{II} }

\begin{figure}[t!]
    \centering
   \includegraphics[width=\linewidth]{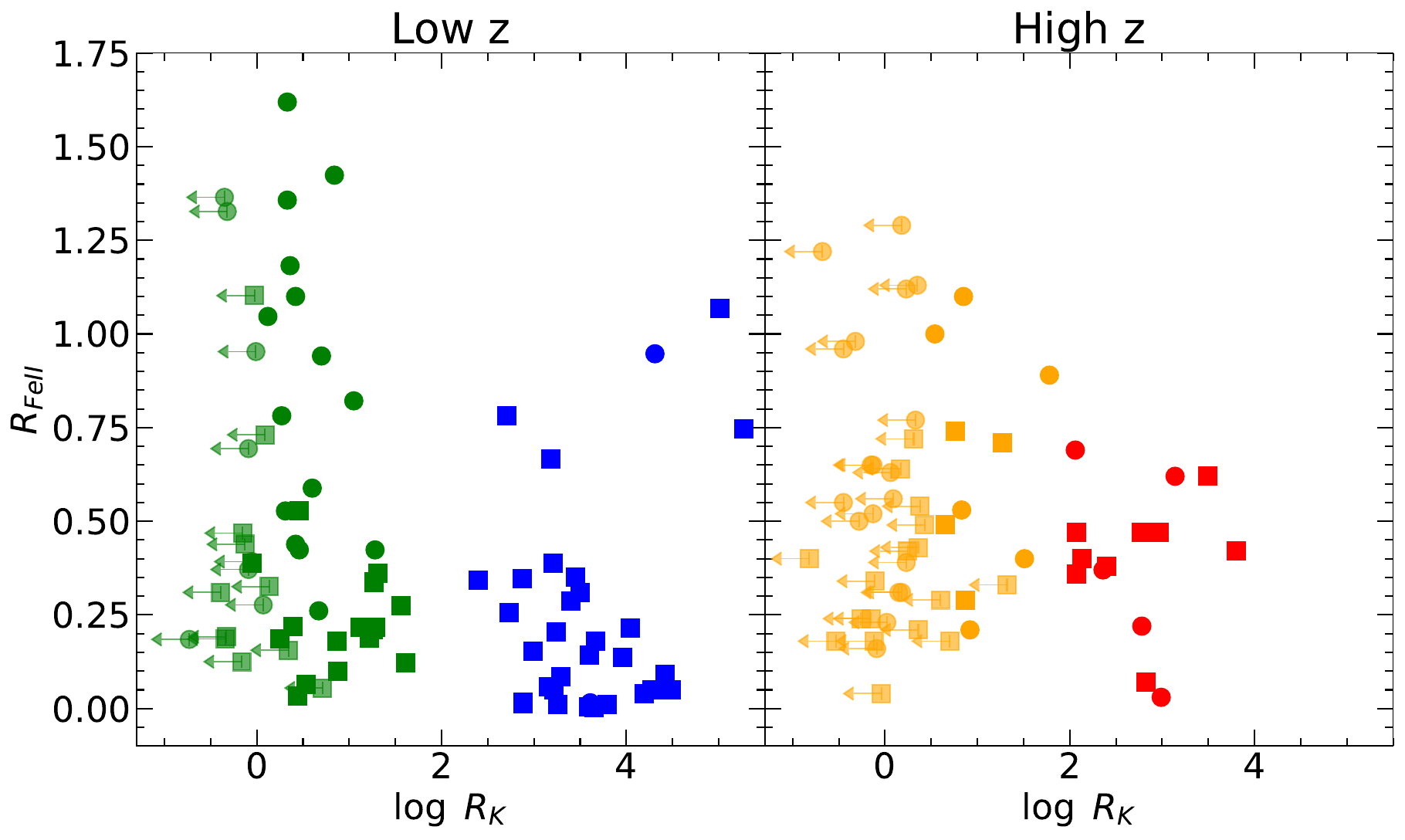}
    \caption{Relation between radioloudness $R_{\textrm{K}}$ and $R_{\textrm{\ion{Fe}{II}}}$ for low-$z$ (\textit{left}) and high-$z$ (\textit{right}) ranges. $R_{\textrm{K}}$ upper limits are indicated by arrows alongside the symbol. Color scheme as in Fig. \ref{fig:lledd_oiii_civ}. }
    \label{fig:Rfeii_rk}
\end{figure}

\par The relation between the $R_{\textrm{\ion{Fe}{II}}}$ with the radioloudness parameter $R_{\textrm{K}}$ is shown in Fig. \ref{fig:Rfeii_rk}. In both low- and high-$z$ ranges, the RQ sources are the ones that can present $R_{\textrm{\ion{Fe}{II}}}$ values from $\sim 0$ up to more than 1, with the most extreme cases found for the low-redshift radio-quiet Pop. A quasars. With the exception of the these extreme sources, the RQ at high and low $z$ present a very similar distribution, with the mean $R_{\textrm{\ion{Fe}{II}}}$ value for low-$z$ RQ is $\sim$ 0.52, while for high-$z$ RQ it is $\sim$ 0.55. 

\par The radio-loud quasars tend to present lower values of $R_{\textrm{\ion{Fe}{II}}}$, which rarely exceeds the threshold of $R_{\textrm{\ion{Fe}{II}}}=1$. The fact that the RL sources are weaker \ion{Fe}{II} emitters than the RQ ones has already been observed in previous studies \citep[e.g.,][]{Yuan_2003,Netzer_2004,Sulentic_2004}. 
In general, the sources that present strong \ion{Fe}{II} emission are found to also present very high Eddington ratios, which are  {more frequently}  {typically} found in radio-quiet Pop. A.  
Hence, this difference in $R_{\textrm{\ion{Fe}{II}}}$ between RL and RQ sources may also be linked to the accretion rate, which somehow can contribute to enhance the \ion{Fe}{II} emission in strong accretors.

The mean $R_{\textrm{\ion{Fe}{II}}}$ values of low- and high-redshift RL sources are $\sim 0.25$\ and $\sim$ 0.40, respectively. 
This result is intriguing, although it may be most easily related to sample differences in Eddington ratio at high- and low-$z$, with high-$z$\ sources at the lowest Eddington ratios missing because they are too faint to be detected in major surveys such as the SDSS \citep{sulentic_2014}.

\begin{figure}[t!]
    \centering
    \includegraphics[width=\linewidth]{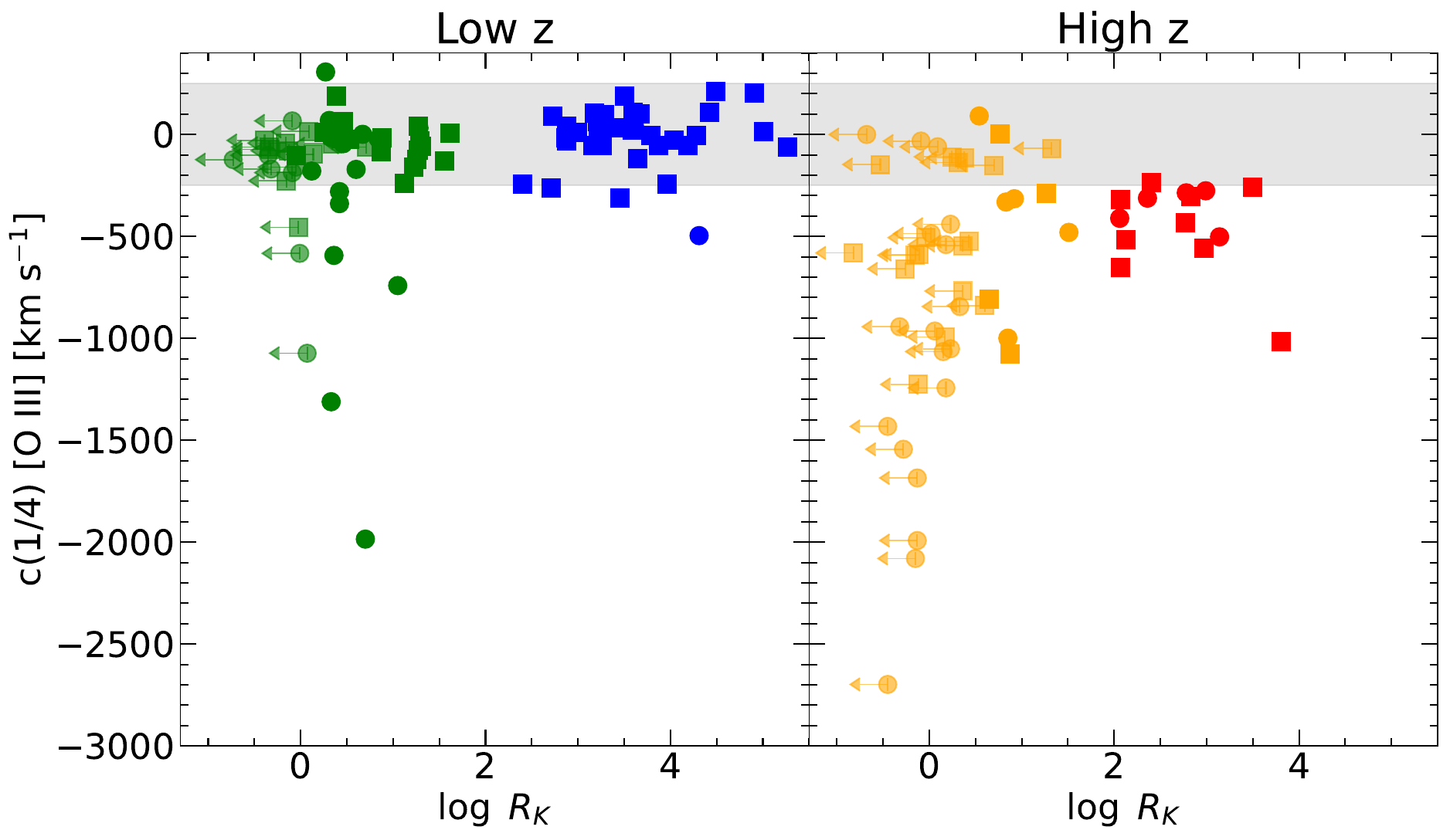}
    \caption{Relation between radioloudness and velocity centroid at 1/4 intensity of the \oiiiseven{} emission line for low-$z$ (\textit{left}) and high-$z$ (\textit{right}) ranges. Grey shaded areas in both plots indicate the $c$(1/4) range between 250 and -250 km s$^{-1}$. $R_{\textrm{K}}$ upper limits are indicated by arrows alongside the symbol. Color scheme as in Fig. \ref{fig:lledd_oiii_civ}.}
    \label{fig:c14_rk_oiii}
\end{figure}

\subsubsection{\oiiiseven{} }

\label{sec:oiii}
\par Fig. \ref{fig:c14_rk_oiii} shows the relation between $R_{\textrm{K}}$ and $c$(1/4) of \oiiiseven{} for low- and high-$z$ ranges separately. In both ranges, RQ sources are the ones with larger outflows velocities compared to RL, with the RQ achieving values of $\sim -2000$ at low and $\sim -2700$\,\kms at high $z$. This difference in outflow velocity between RQ and RL quasars has been already reported in the recent literature \citep[e.g.,][and references therein]{marzianietal03b, marziani_2016a, Ganci_2019} and may be consequence of the presence of relativistic jets in RL sources that potentially can reduce the impact of the outflows on the optical emission. In contrast, RQ lack such strong radio jets, allowing the outflows to have a more significant influence on the optical emission \citep[e.g.,][]{Padovani_2016}.  

\par At low redshifts, the majority of the sources (including both RQ and RL) do not exhibit \oiiiseven{} $c$(1/4) values that exceed 250 km s$^{-1}$, and therefore do not present significant \oiiiseven{} outflows. { {Only eight objects (seven RQ and one RL) within this redshift range in our FOS sample present relevant \oiiionly{} $c(1/4)$ blueshifts of $\sim -500$\,\kms.}} 

\par The right panel of Fig. \ref{fig:c14_rk_oiii} shows the same analysis but focused on high-$z$ sources. In this redshift range, $\sim$ 75\% of the radio-quiet sources present blueshifts larger than 250 \kms, and in some cases can achieve very strong velocities as is the case of SDSS J212329.46-005052.9 ($\sim$ 2700\,\kms). { { All the high-redshift radio-loud sources from our sample present significant blueshifts, with a \oiiiseven{} $c$(1/4) range between $\sim$ 230  and 1000 km\,s$^{-1}$. }}

\begin{figure}[t!]
    \centering
    \includegraphics[width=0.98\linewidth]{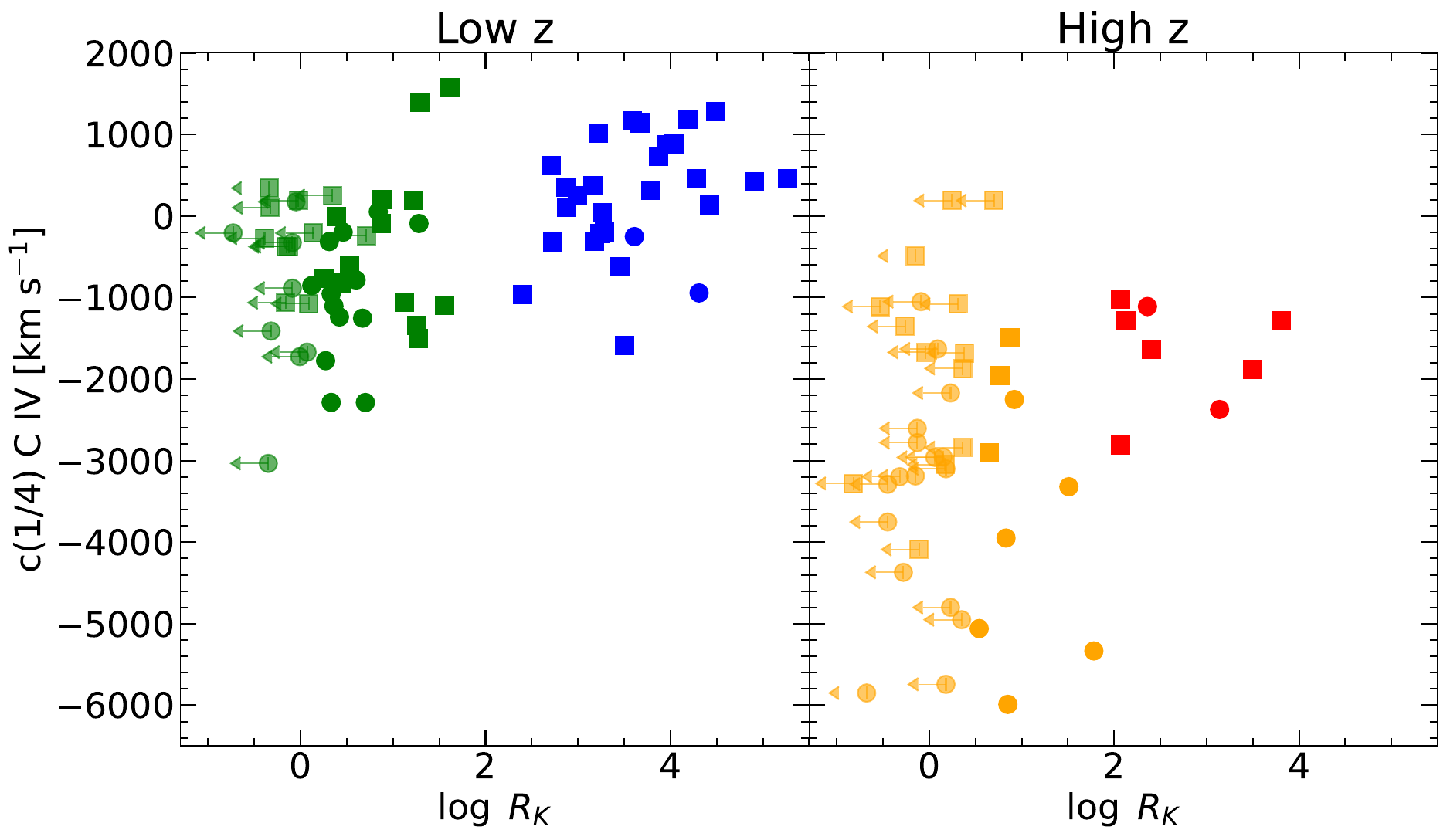}
    \caption{Relation between radioloudness and velocity centroid at 1/4 intensity of the \civ{} emission line for low-$z$ (\textit{left}) and high-$z$ (\textit{right}) ranges. $R_{\textrm{K}}$ upper limits are indicated by arrows alongside the symbol. Color scheme as in Fig. \ref{fig:lledd_oiii_civ}.}
    \label{fig:c14_rk_civ}
\end{figure}

\subsubsection{\civ{} }

\par It has been known since long that \oiiiseven{} and \civ{} show similarities in their phenomenology \citep[e.g.,][]{zamanovetal02,Coatman_2019,Deconto-Machado_2023}. In Fig. \ref{fig:c14_rk_civ} we repeat the \oiii\ analysis described in \S \ref{sec:oiii} for \civ{}. As for \oiiiseven{}, the majority of RQ sources at low $z$ present negative values of \civ{} $c$(1/4) while the wide majority of RL have positive values or values very close to 0-500 km s$^{-1}$ to the blue. Regarding the high-$z$ range,  {it is already known that the largest shifts are also found in RQ sources \citep[see e.g.,][and references therein]{richards_2011,Richards_2021}. This is not different in our high-$z$ sample, case in which} practically all sources (including both RQ and RL) present very significant outflow velocities. {The average \civonly{} $c(1/4)$ blueshift for RL sources is $\sim 1900$ km s$^{-1}$.}} The relation between the \oiiiseven{} and \civ{} emission lines does not seem to be affected by the radio emission, however the comparison between RQ and RL at high $z$ for the \civ{} emission line is limited by the fact that we have a very small sample of RL in this redshift range (only seven objects, two Pop. A and five Pop. B) and more data are definitely required.

\begin{figure*}[ht!]
    \centering
    \includegraphics[width=0.35\linewidth]{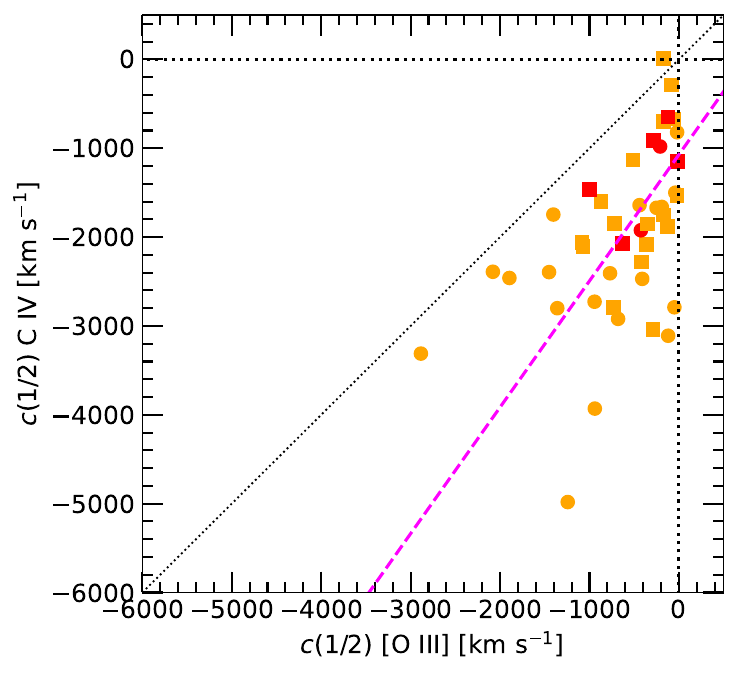}
    \hspace{1cm}
    \includegraphics[width=0.35\linewidth]{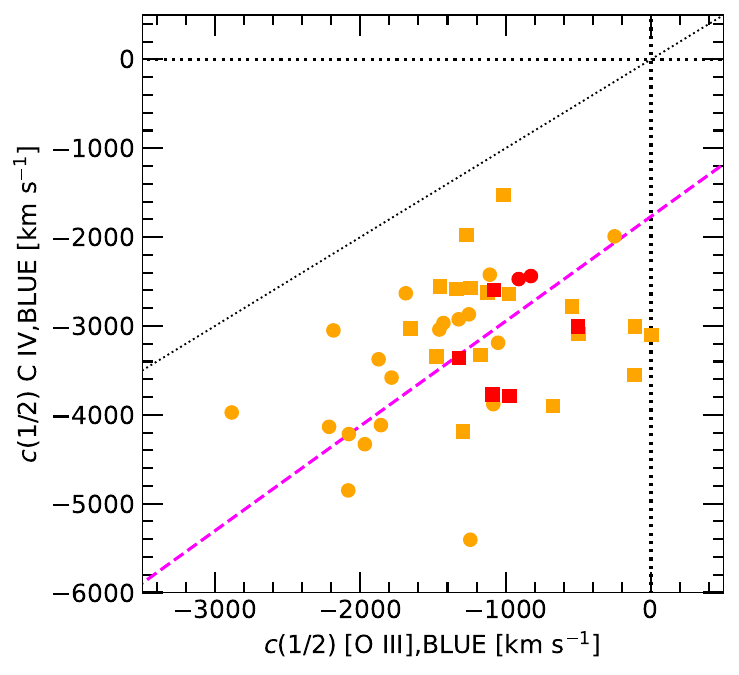}
    \caption{ {Centroid velocity at 1/2 flux intensity ($c(1/2)$) of \civ{} \textit{vs.} $c(1/2)$ of \oiiiseven{} for the full (\textit{left}) and outflow (\textit{right}) profiles of the high-$z$ sample. The magenta lines indicate the linear regression between $c(1/2)$ of \civonly{} and \oiiionly{} for both cases obtained through the bisector method. Dotted black lines represent the 1:1 relation. Color scheme is the same as in Fig. \ref{fig:lledd_oiii_civ}.}}
    \label{fig:oiii_civ_c12}
\end{figure*}




\section{Discussion}
\label{section:discussion}

\subsection{Dominance of outflows}
\label{sect:dom_outflows}
\par \civ{} somehow seems to be a magnified version of \oiiiseven{}, presenting very similar trends however with stronger outflow velocities. 
The left plot of Fig. \ref{fig:oiii_civ_c12} shows the relation between the centroid velocity at half intensity ($c(1/2)$) for the full profiles of \oiiiseven{} and \civ{} at high redshift, including the sources from HEMS, \citetalias{Deconto-Machado_2023}, and from the present paper. The result reported in \citetalias{Deconto-Machado_2023} is strengthened: ISAAC2 objects that present strong shifts in the \oiii{} emission line profiles do present them also in \civ{}.  {The bisector linear relation after including all ISAAC
data is}
\begin{equation*}
    c(1/2)_{\textrm{\ion{C}{IV}}}=(1.42 \pm 0.23)\times c(1/2)_{\textrm{[\ion{O}{III}]}}+(-1071\pm 148)
\end{equation*}
 {with a  correlation coefficient $\approx$ 0.51.  This trend is consistent with the results reported by  \citet[][]{Coatman_2019} and \citet{vietrietal20} at high redshift. }

\par {In both  \oiiionly{} and \civonly{}, the largest blueshifts are found in Pop. A sources (represented by orange bullets in the figures), which usually present the largest accretion rates. Fig. \ref{fig:oiii_civ_c12} also shows that the largest outflow velocities in \oiiionly{} and in \civonly{} are found in radio-quiet sources, indicating that the accretion may be the main driver of these outflows \citep{nesvadbaetal07,kukretietal23}.} A similar relation is found when considering the $c$(1/2) of only the outflowing (BLUE) components of both \oiiiseven{} and \civ{} emission lines, as shown in the right plot of Fig. \ref{fig:oiii_civ_c12}. In this case, the orthogonal linear relation (with a c.c. of 0.40) is given by:
\begin{equation*}
c(1/2)_{\textrm{\ion{C}{IV},BLUE}}=(1.27 \pm 0.99)\times c(1/2)_{\textrm{[\ion{O}{III}],BLUE}}+(-1432\pm 1323)
\end{equation*}

At high $z$, the outflows appear to be more prominent than at low $z$ (see also Fig. \ref{fig:lledd_oiii_civ}), consolidating the idea of accretion rate as the main driver of the outflows. The accretion rate in most of the sources at high $z$ is higher {($-0.6 \lesssim \log$  \lledd $\lesssim 0.4$) than in the case of sources at low $z$\   \citep[$-2 \lesssim \log$ \lledd $\lesssim 0$, see also e.g.,][]{cavalierevittorini00,hopkinsetal06}, even if the difference could be in part due to a selection effect, as recalled in  Section \ref{sec:rk_trends}. }

\begin{figure}[t!]
    \centering
    \includegraphics[width=0.8\linewidth]{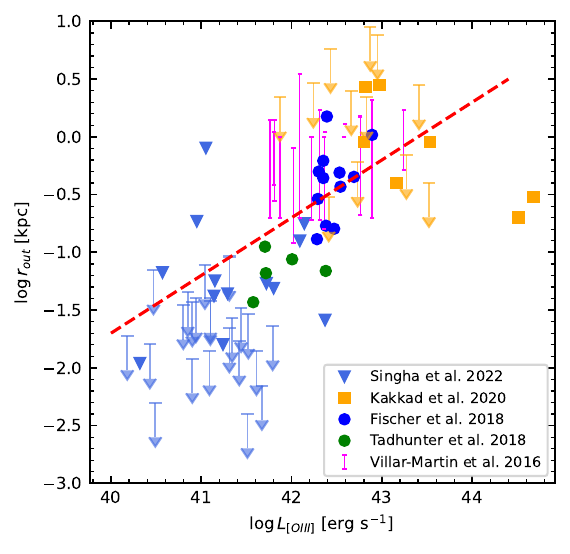}
    \caption{ {Relation between the outflow radio in kpc and the luminosity of the \oiiiseven{} emission line for the five plotted samples. Red dashed line represents the adopted least squares linear regression. {Magenta vertical lines symbolise the minimum and maximum estimated radii in the Villar-Martin sample. Arrows correspond to upper limits.}}}
    \label{fig:estimando_raio}
\end{figure}

\begin{figure*}[t!]
    \centering
    \includegraphics[width=0.33\linewidth]{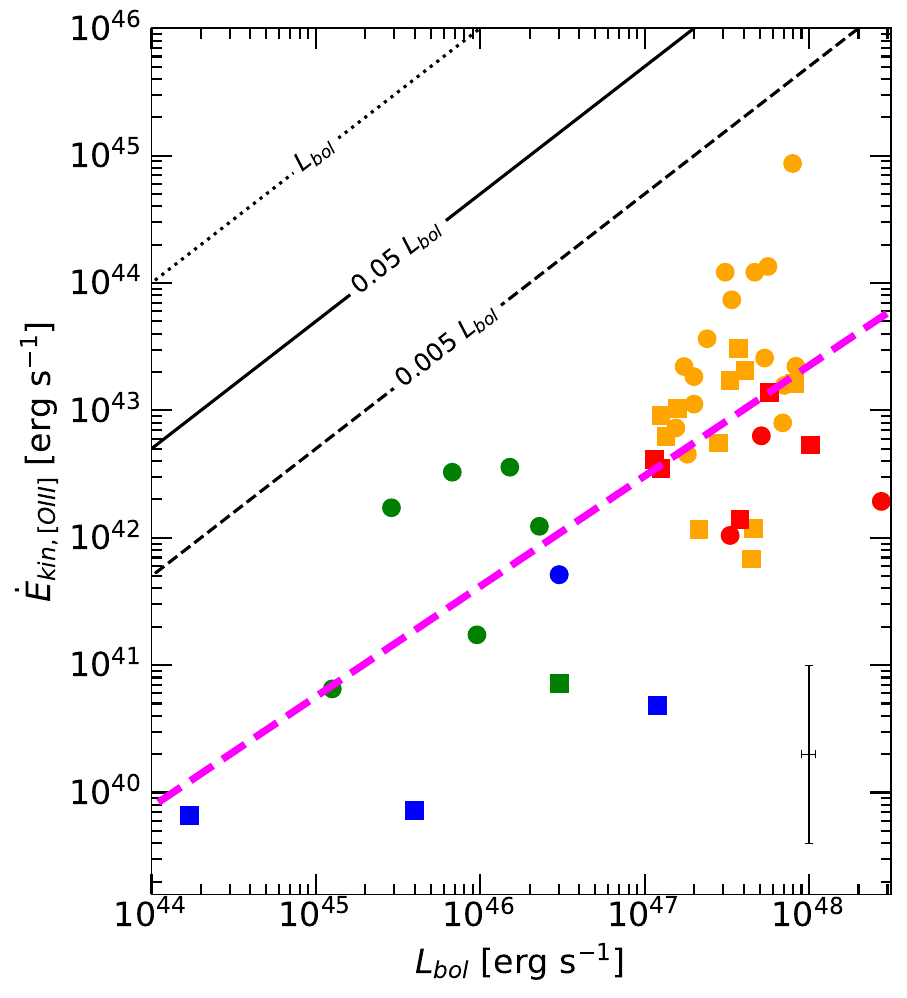}
    \hspace{0.5cm}
    \includegraphics[width=0.33\linewidth]{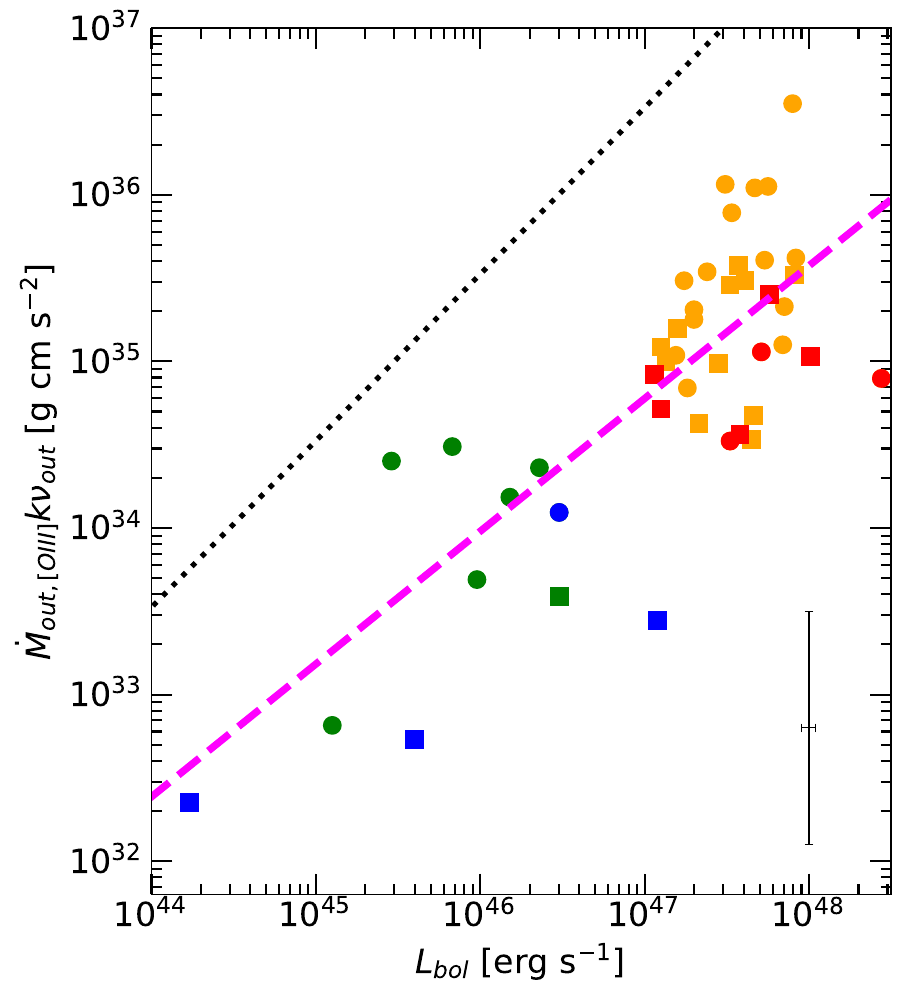}

    \caption{Kinetic power $\dot{E}_{\textrm{kin}}$ (left) and thrust (\textit{right}) \textit{vs.} bolometric luminosity for the \oiiiseven{} outflow. The two outflow parameters were estimated varying the outflow radius according with Eq. \ref{eq:radio_oiii} and assuming $v_{\textrm{out}}=c(1/2)$. Magenta dashed lines indicate the least-squares linear regression. Black dotted, continuous, and dashed lines on the left plot show $\dot{E}_{\textrm{kin}}=L_{\textrm{bol}}$, $\dot{E}_{\textrm{kin}}=0.05L_{\textrm{bol}}$, and $\dot{E}_{\textrm{kin}}=0.005L_{\textrm{bol}}$ respectively. Dotted line in the right panel indicates $\dot{M}_{\textrm{out}}v_{\textrm{o}}=L_{\textrm{bol}}/c$.  {Typical uncertainties are displayed in the bottom-right corner of the plots.} Color scheme as in Fig. \ref{fig:lledd_oiii_civ}. }
    \label{fig:outflow_oiii}
\end{figure*}

\subsection{Dynamical parameters of the outflows}
\label{sect:param_outflows}

\par The mass outflow rate {($\dot{M}$)}, the kinetic power {($\dot{E}_{\mathrm{kin}}$)}, and the thrust {($\dot{M}$\,$v_{o}$)} were estimated adopting a simple biconical outflow, and by using an  {analogous}  methodology  {to the one of } \citet{marzianietal16,Marziani_2017}.  {It is}  summarized in Appendix \ref{app:outflows}. {In this framework, the relations are formally identical for \oiiionly\ and \civ, and namely:}
 \begin{equation*}
    M_{\textrm{ion}} \propto  L^\mathrm{out} Z^{-1} n_\mathrm{H}^{-1},
\end{equation*}
\noindent where $L$ is the outflow-emitted line luminosity, $n_\mathrm{H}$\ the electron density and $Z$\ the metallicity.  The mass outflow rate ($\dot{M}^{\textrm{out}}_\mathrm{ion}$) and the thrust ($\dot{M}_{\mathrm{ion}}^{\textrm{out}}$\,$v_{\mathrm{out}}$) at a radius $r$ and with an outflow velocity $v_{\mathrm{out}}$, might be written as:
\begin{equation*}
\dot{M}_\textrm{ion}^\textrm{out} \propto L^\textrm{out}\ v_{\mathrm{out}}\ r^{-1} Z^{-1} n_\textrm{H}^{-1}
\end{equation*}
\begin{equation*}
    \dot{M}_{\mathrm{ion}}^{\textrm{out}}v_{\mathrm{out}}\propto L^{\mathrm{out}}\ v^2_{\mathrm{out}}\ r^{-1} Z^{-1} n_\mathrm{H}^{-1}
\end{equation*}
  
The kinetic power, ${\dot{E}_{\mathrm{kin}}}$, is then given by $\dot{E}_{\mathrm{out}}\sim \frac{1}{2}\dot{M}_{\mathrm{out}}v_{\mathrm{out}}^2$, which leads to
\begin{equation*}
    \dot{E}_{\mathrm{kin}} \propto L^{\mathrm{out}} v_{\mathrm{out}}^3\ r^{-1} Z^{-1}n_\mathrm{H}^{-1}
\end{equation*}

The parameters entering the previous equations are estimated utilizing different relations for \oiiionly\ and \civonly, and will be briefly discussed in next sections and Appendix \ref{app:outflows}.  { The adopted parameters and relevant scaling relations  are summarized in Table \ref{tab:summary}. } 

{ When estimating wind parameters in quasars, such as mass outflow rate, thrust, and kinetic power, several sources of error can affect the accuracy of these measurements. These errors arise from various observational and methodological issues, including the complexity of the line profiles, spatial resolution limitations, and assumptions inherent in the models used for interpretation,  {{where the last two are expected to be the dominating sources of uncertainties in the estimation of the wind parameters.}} 
Specifically for \civ,\ the uncertainty in the conversion between line luminosity and ionized gas could be $\sim 3$\ at a $3 \sigma$ confidence level. We  have considered a range of densities of 0.5 dex around $\log n_\mathrm{H} $ radius of the emitting region; $Z$ estimates in quasars suggest very high metallicity, and a range around $5 Z_\odot$, where between 2 and 20 times solar seems possible both at high and low redshifts  \citep{hamannferland93,garnicaetal22,florisetal24}. The geometry of the outflow would contribute a factor $3$, via the comparison between the spherically symmetric case and a flat layer.  The 1$\sigma$ uncertainty in the emitting radius is $\pm 30$\%\ if ascribed only to the uncertainty of the scaling law parameter, while uncertainty on the outflow velocity is  typically $\sim 30$\%. Propagating quadratically these uncertainties would result in a typical   factor $\approx 5 $ at 1$\sigma$\ confidence level.}
\par  {Similar considerations were applied to \oiiionly{}: a factor of $\approx 3$ at 1$\sigma$ in density, a factor of $\approx 2$ in $Z$ and in the luminosity-to-ionized mass gas conversion, and a factor of $\approx 2$ in the zero point of the radius-luminosity ($L$(\oiiionly)) relation. These  factors contribute to a comparable uncertainty estimate, a factor of $\approx 5$ at 1$\sigma$ confidence level, and should be taken into account in the interpretation of the analysis that follows.}

\begin{table}[]
    \centering
   \caption{{Summary of assumed scaling relations and parameter values  for wind dynamics. }   \label{tab:summary}}
    \begin{tabular}{lccccc}\hline\hline
    \noalign{\smallskip}
\multirow{2}{*}{Parameter}	&	\multicolumn{2}{c}{Low $z$}  	&&	\multicolumn{2}{c}{High $z$}			\\ \cline{2-3}\cline{5-6}
\noalign{\smallskip}
	&	\civonly\	&	\oiiionly\	&&	\civonly\	&	\oiiionly\	\\ \hline
 \noalign{\smallskip}
$M_\mathrm{ion}$ 	&	$\propto L_{\rm \ion{C}{iv}}^{\rm out}$	&	$\propto  L_{\rm [\ion{O}{iii}]}^{\rm out}$	&&	$\propto L_{\rm \ion{C}{iv}}^{\rm out}$ &	$\propto  L_{\rm [\ion{O}{iii}]}^{\rm out}$	\\
Z [$Z_\odot$]	&	5	&	5	&	& 5	&	2	\\
$n_\mathrm{H}$\ [cm$^{-3}$]	&	$10^{9.5}$	&	 $10^3$	&&	$10^{9.5}$	&	$10^3$	\\
$r$ [cm]	&  $\propto$$\lambda\,L_\lambda$(1350) 	&  $\propto$$L$(\oiiionly)	&&  $\propto$$\lambda\,L_\lambda$(1350) 	&  $\propto$$L$(\oiiionly)	\\
$v$ [\kms]	&	$c(\frac{1}{2})+ 2\sigma$	&	$c(\frac{1}{2})$	&&	$c(\frac{1}{2})+ 2\sigma$	&	$c(\frac{1}{2})$	\\
\noalign{\smallskip}
\hline		
         \end{tabular}
\end{table}

\subsubsection{\oiiiseven{}}
\label{section:outflow_oiii}

{The considered situation is the one in which the outflow radius $r$ is allowed to vary depending on the \oiiiseven{} luminosity of each source. In the case of sources spanning four orders of magnitude in luminosity, it is nonphysical to assume that the size of the emitting region is constant, as there is evidence of compact emission at low-$z$ \ \citep{zamanovetal02}, and of kpc-sized outflows at high-luminosity \citep[e.g.,][]{harrisonetal14}. Conventional scaling laws of narrow-line regions with luminosity are not expected to exclusively trace the outflowing \oiiionly{} component \citep{bennertetal02,bennertetal06}.  }


In Fig. \ref{fig:estimando_raio} is shown the outflow radius as a function of the \oiiiseven{} luminosity for the integral field spectroscopy (IFU) data analyzed by \citet{Villar-martin_2016}, \citet{Fischer_2018}, \citet{Tadhunter_2018}, \citet{Kakkad_2020}, and \citet{singha_2022}  {where is also represented a least squares linear relation between these two quantities, imposing a slope of 0.5:} 
\vspace{-0.2cm}
\begin{equation}
\label{eq:radio_oiii}
   \log \left(r\right)=(0.5)\times \log \left(L_{\textrm{[O III],full}}\right) + (-21.7\pm 0.1).
\end{equation}
{The slope 0.5, {which gives a reasonable representation of the radius-L$\mathrm{_{[OIII]}}$ relation in Fig. \ref{fig:estimando_raio},} is imposed to ensure consistent ionization conditions as a function of luminosity. This scaling relation was then applied to our sources, considering the luminosity of the entire \oiiionly{} emission line profile. {Our low-$z$ sample exhibits  \oiiionly{} luminosities (full profile) ranging from approximately $10^{40.6}$ up to $10^{43.1}$ erg\,s$^{-1}$, while the high-$z$ sources have luminosities between $10^{43.4}$ and $10^{44.9}$ erg\,s$^{-1}$. Consequently, we find \oiiionly{} outflows radius $r$ ranging from $\sim$ 0.04 to 0.71 kpc at low-$z$, {in agreement with recent values obtained for IFU resolved \oiiionly{} in low-$z$ AGN \citep[see e.g.][]{Deconto-Machado_Manga_2022}, and from 0.97 to 5.81 kpc at high-$z$.}}}

 \par  {The parameters used to estimate the outflow properties associated with the \oiiiseven{} emission line are listed in Table \ref{tab:parametros_outflows_oiii_civ}.} 
 { {The estimated values of {the mass rate} $\dot{M}_{\textrm{out}}$, derived from equation \ref{eq:mout_oiii} in the Appendix, span from $\sim$ 0.08 to 3.43 $\textrm{M}_{\odot}$ yr$^{-1}$ for low luminosities ($\lesssim 10^{46}$ erg s$^{-1}$) and from  {$\sim$ 6.2 to 114.0} $\textrm{M}_{\odot}$ yr$^{-1}$ for higher luminosities.}} 

\par  {The relations between the \oiiionly{} outflow kinetic power $\dot{E}_{\textrm{kin}}$, and thrust $\dot{M}_{\textrm{out}}v_o$ with the bolometric luminosity $L_{\textrm{bol}}$ are shown in Fig. \ref{fig:outflow_oiii}.  {We have assumed a density $n_{\rm H}=10^3$ cm$^{-3}$ for both high- and low-$z$ contexts. For metallicity, we have adopted $Z=5Z_{\odot}$ for low-$z$, considering the compactness of the outflow, and $Z=2Z_{\odot}$ for high-$z$, aligning with the typical values used in the computations for the NLR metallicity at high $z$ \citep[see e.g., ][]{xuetal18}. Additionally, as detailed in Section \ref{app:oiii_outflow}, we assume the outflow velocity to be $c(1/2)$ of the BLUE component. If we instead follow the assumption of $c(1/2)+2\sigma$ as done by \citet{Fiore_2017}, it would result in an outflow velocity $\sim 2.5$ larger in our estimates of the outflow parameters. Typical errors are represented in the errorbars of Fig. \ref{fig:outflow_oiii}.}  {The results of the linear correlation analysis between these outflow parameters for \oiiiseven{} and \civ{} are reported in Table \ref{tab:outflow_relations}. }
 {We find correlation coefficients of 0.66 and 0.74 (with $\rho$-value of $\sim 10^{-7}$ and $\sim 10^{-9}$, respectively) between the kinetic power and the thrust with the bolometric luminosity.} Similar correlations have already been extensively discussed by other authors \citep[e.g.][and references therein]{Carniani_2015,Feruglio_2015,Fiore_2017}.

\par In our analysis of \oiiionly{} $\dot{E}_{\textrm{kin}}$ with luminosity-dependent $r$, none of the sources display a ratio reaching at least $5\%$ of the bolometric luminosity (filled line in Fig. \ref{fig:outflow_oiii}), a threshold needed for a significant impact on the host galaxy dynamics, to account for the black hole mass - velocity dispersion correlation and  host-spheroid co-evolution \citep[e.g.,][]{dimatteoetal05}. 
Star formation quenching might be easier, if an AGN  outflow  induces a wind in the  diffuse interstellar medium that in turns induces a flattening and shredding of molecular  clouds \citep{Hopkins_2010}. A threshold for this effect occurs when the kinetic efficiency is much lower than the 5\%\ limit,  $\dot{E}_{\textrm{kin}}/L_{\textrm{bol}}\sim 5\times 10^{-3}$\ (short dashed line in Fig. \ref{fig:outflow_oiii}).  Based on the criteria of both \citet{Hopkins_2010} and \citet{dimatteoetal05}, the \oiiiseven{} outflows parameters accepted at phase value are   not providing an efficient feedback mechanism neither at low- nor at high$-z$ (although see the discussion of Section \ref{comparison_oiii_civ}).  {Similar results have recently been found at low redshift \citep[and references therein]{Kim_2023}. They derived $\dot{E}_{\mathrm{kin}}\lesssim$ 0.1\% $L_{\rm bol}$ for a sample of low-$z$ type-1 AGN. It is likely that part of the \oiiionly\ emitting gas has already dissipated part of its energy and momentum, as the emitting regions can be extended over several kpc. By the same token, the AGN outflows traced by the \oiiionly\ emission  are likely to impact on the central kiloparsecs scales, but their effects on galactic scale is more debatable.  }

\par 
 { {The thrust ranges from $\sim2.2 \times 10^{32}$ to $3.1 \times 10^{34}$ g cm s$^{-2}$, in the case of the low-luminosity sample;}} at high luminosity, the thrust increases significantly to the larger outflow masses and higher velocities and the  values are  found between  $\sim 3.3 \times 10^{34}$ to $3.5 \times 10^{36}$ g cm s$^{-2}$.} In all cases, however, the \oiiionly{} thrust  is  $\ll L/c$ {, in agreement with \citet{vietrietal20}. 


\setcounter{table}{13}
\begin{table}[t!]
    \centering
     \caption{{ {Least-squares linear relations ($y=a+b*x$) between different outflows properties and the bolometric luminosity.}}}
    \begin{tabular}{lccccccr}
    \hline
    \hline
    \noalign{\smallskip}
        y & x & Sources &  $a\pm \delta a$ & $b \pm \delta b$ & RMSE & CC & $\rho$-value\\
        (1) & (2) & (3) & (4) & (5) & (6) & (7) & (8)  \\
    \noalign{\smallskip}
    \hline 
    \noalign{\smallskip}
     \multicolumn{8}{c}{\oiiiseven{}}\\
    \noalign{\smallskip}
    \hline
    \noalign{\smallskip}
    $\log \dot{E}_{\textrm{kin,44}}$ & $\log L_{ \textrm{bol,44}}$  &  {45} &  {-4.09$\pm$0.44} &  {0.86$\pm$0.14} &  {0.009} &  {0.69} &  {1.7$\times 10^{-6}$\ \ }\\
    $\log \dot{M}_{\textrm{out}}v_{\textrm{o,34}}$ & $\log L_{ \textrm{bol,44}}$  &  {45} &   {-1.62$\pm$0.31} &  {0.79$\pm$0.09} &  {0.017} &  {0.78} &  {2.7$\times 10^{-10}$}\\
    \noalign{\smallskip}
    \hline
    \noalign{\smallskip}
    \multicolumn{8}{c}{\civ{}}\\
    \noalign{\smallskip}
    \hline
    \noalign{\smallskip}
    $\log \dot{E}_{\textrm{kin,44}}$ & $\log L_{ \textrm{bol,44}}$  &  {68} &  {-3.59$\pm$0.25} &  {1.24$\pm$0.08} &  {0.075} &  {0.87} &  {2.1}$\times10^{-22}$\\
    $\log \dot{M}_{\textrm{out}}v_{\textrm{o,34}}$ & $\log L_{ \textrm{bol,44}}$  &  {68} &   {-2.08$\pm$0.22} &  {1.19$\pm$0.07} &  {0.069} &  {0.90} &  {3.7}$\times10^{-25}$\\
    \noalign{\smallskip}
  
    \hline
    \end{tabular}
    \label{tab:outflow_relations}
    \tablefoot{ {(1), (2) Fitted parameters. (3) Number of sources. (4), (5) Linear correlation coefficients. (6) Root mean square root. (7), (8) Pearson $r$ score and its associated null hypothesis probability value.} }
\end{table}

\begin{figure*}[ht!]
    \centering

\includegraphics[width=0.33\linewidth]{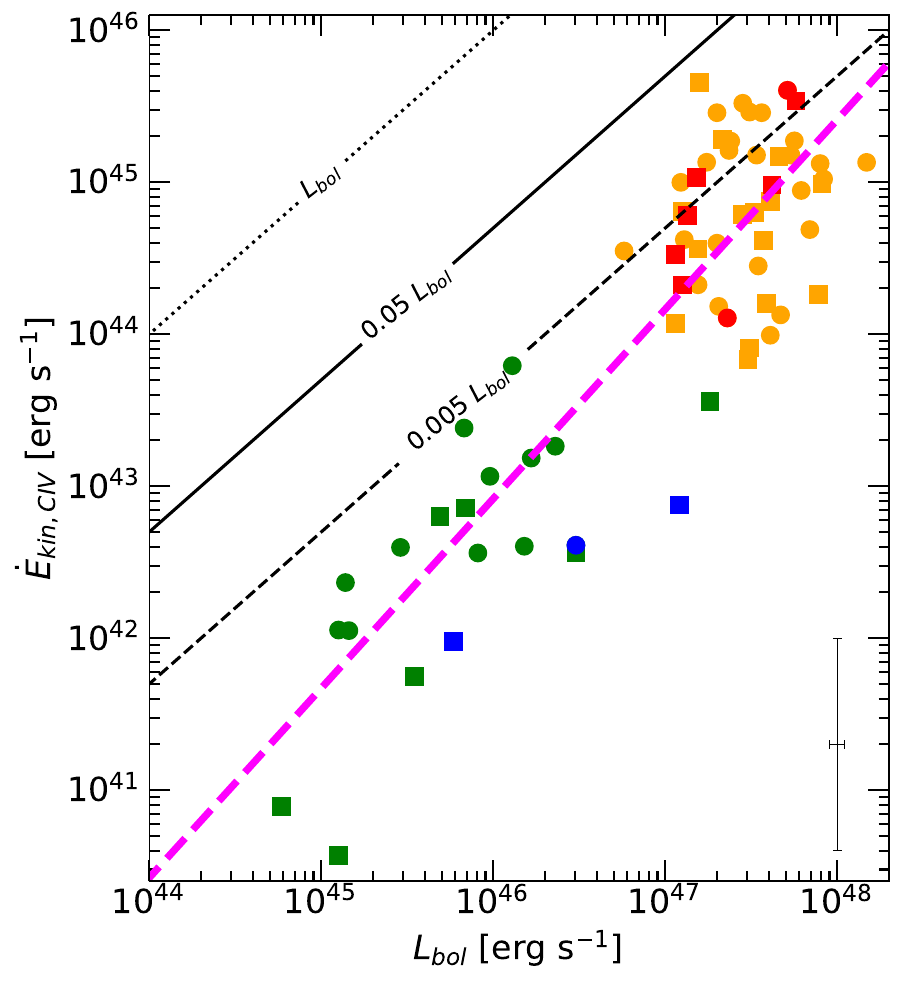}
\hspace{0.5cm}
\includegraphics[width=0.33\linewidth]{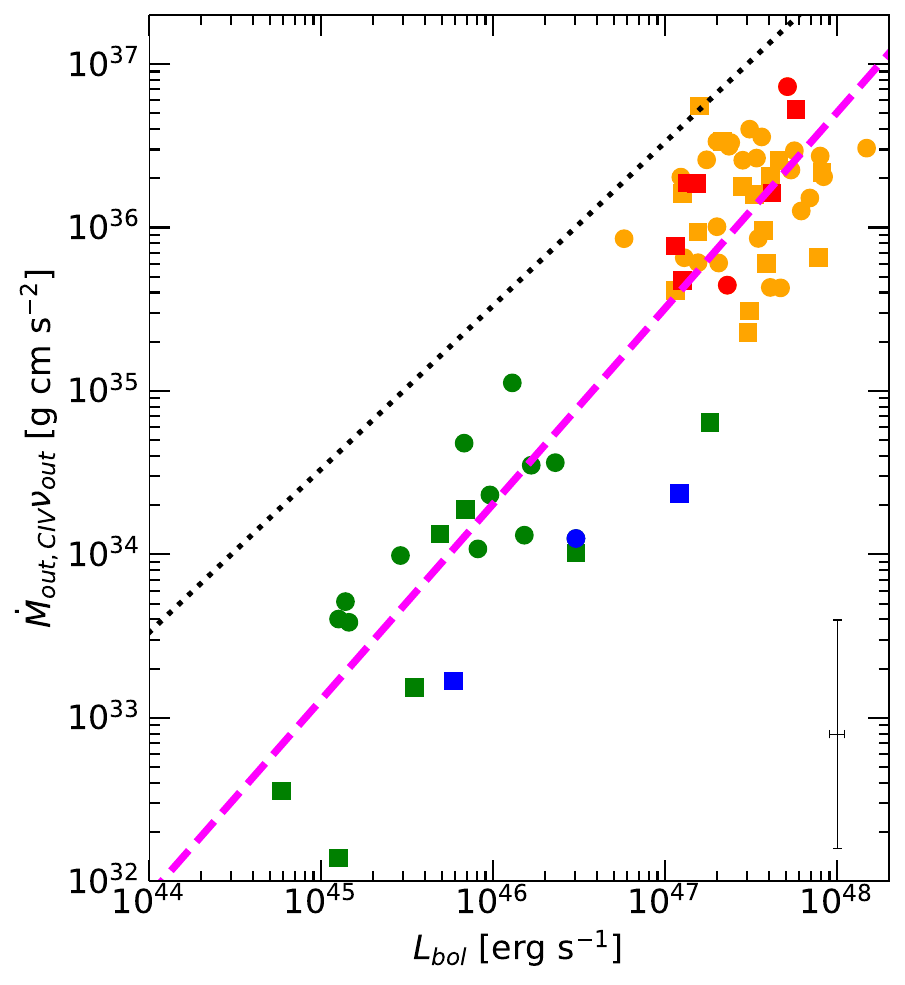}


   \caption{Kinetic power $\dot{E}_{\textrm{kin}}$ (\textit{left plot}), and thrust (\textit{right}) \textit{vs.} bolometric luminosity for the \civ{} emission line. The two outflow parameters estimated varying outflow radius depending on $\lambda L_{\lambda}$ as defined in \citet{Kaspi_2021}. Magenta dashed lines indicate the least squares linear regression. Black dotted, continuous, and dashed lines show $\dot{E}_{\textrm{kin}}=L_{\textrm{bol}}$, $\dot{E}_{\textrm{kin}}=0.05L_{\textrm{bol}}$, and $\dot{E}_{\textrm{kin}}=0.005L_{\textrm{bol}}$ respectively. Dotted line in the right panel indicates $\dot{M}_{\textrm{out}}v_{\textrm{o}}=L_{\textrm{bol}}/c$.  {Typical uncertainties are displayed in the bottom-right corner of the plots.} Color scheme as in Fig. \ref{fig:lledd_oiii_civ}. }
    \label{fig:outflow_civ}
\end{figure*}

\begin{table*}[t!]
 \caption{Average values and standard deviations of the \oiiiseven{} and \civ{} outflow parameters for the different samples.} 
    \centering
    \begin{tabular}{lcccccccccccc}
    \hline
    \hline
    \noalign{\smallskip}
    & \multicolumn{3}{c}{\oiiiseven{}} & & \multicolumn{3}{c}{\civ{}}\\
    \cline{2-4} \cline{6-8}
    \noalign{\smallskip}
    Sample & $\dot{M}_{\rm out}$ & $\dot{M}_{\rm out}v_{\rm o}$ & $\dot{E}_{\textrm{kin}}$  & &  $\dot{M}_{\rm out}$ & $\dot{M}_{\rm out}v_{\rm o}$ & $\dot{E}_{\textrm{kin}}$\\
    & [M$_{\odot}$ yr$^{-1}$] & [$10^{35}$ g cm s$^{-2}$] & [$10^{43}$ erg s$^{-1}$] & & [M$_{\odot}$ yr$^{-1}$] & [$10^{36}$ g cm s$^{-2}$] & [$10^{45}$ erg s$^{-1}$]\\
    (1) & (2) & (3) & (4) & & (5) & (6) & (7)\\
    \noalign{\smallskip}
    \hline
    \noalign{\smallskip}
    ISAAC &  {34.35 $\pm$ 24.73} &  {3.12 $\pm$ 3.48} &  {3.50 $\pm$ 5.30} & &  {38.59 $\pm$ 20.25} &  {3.05 $\pm$ 1.53} &  {2.03 $\pm$ 1.16} \\
    HEMS  &  {28.71 $\pm$ 30.01} &  {4.07 $\pm$ 9.03} &  {10.11 $\pm$ 29.33} & &  {24.32 $\pm$ 14.51} &  {1.18 $\pm$ 0.87} &  {0.48 $\pm$ 0.45}\\
    FOS  &  {1.58 $\pm$ 1.12} & 0.11 $\pm$ 0.11 &  {0.24 $\pm$ 0.34} & &  {0.37 $\pm$ 0.39} &  {0.01 $\pm$ 0.01} &  {0.01 $\pm$ 0.01}\\
    \noalign{\smallskip}
    \hline
    \end{tabular}
    \label{tab:promedios_outflow}
    \tablefoot{ {(1) Sample identification. (2), (3), (4) Outflow mass rate, thrust, and kinetic power of \oiiionly{} outflows. (5), (6), (7) Same for \civonly{} outflows. }}
\end{table*}

\subsubsection{\civ{}}
\label{section:outflow_civ}
\par \
{The ionized gas mass producing the \civ{}\ line was 
estimated following the   results obtained from CLOUDY photoionization  computations by using two different spectral energy distributions (SEDs) representing the two main samples of sources considered in this work: a SED from \citet{mathewsferland87}  for the low-$z$\ FOS sample, and the more appropriate  SED from \citet{krawczyketal13} for high luminosity high-$z$\ ISAAC and HEMS samples, and taking also into account the constrains imposed by the measures of observed quantities in the spectra as the W(\civonly) or the \civ/\hb\ ratio. All the calculations and assumptions for the  \civonly{} outflow  parameters are reported in  Appendix  \ref{appendix:civ_outflows}}.   For \civ, we   consider {that (at variance with the \oiiionly), the \civonly{} outflow is accelerated to a final} outflow velocity of $c(1/2)+2\sigma$ of the blueshifted  component.  

{As for  \oiiionly{},  we estimated the radius $r$ independently for each source, and based on the luminosity at 1350\AA. In the case of \civonly, we use} the consolidated scaling relation of \citet{Kaspi_2021}:
\begin{equation}
\label{eq:radio_civ}
  \frac{r_\mathrm{BLR}}{10 \textrm{ lt days}}=(0.34 \pm 0.11)\left( \frac{\lambda L_{\lambda}(1350\AA)}{10^{43} \textrm{erg s}^{-1}} \right)^{0.45 \pm 0.05},
\end{equation}
{where $r_\mathrm{BLR}$ is the radius of the BLR.}
\par {{For our sources, the specific luminosities at 1350\AA\ range from approximately $10^{44.5}$ up to $10^{46.5}$ erg\,s$^{-1}$ at low $z$ }} and from $10^{45.9}$ to $10^{47.6}$ erg\,s$^{-1}$ at high $z$. With these luminosity values, we find \civ{} outflow radii varying from $0.06$ to $0.37$ pc at high $z$ and $0.01$ to 0.11 pc at low $z$.
\par  {In Table \ref{tab:parametros_outflows_oiii_civ} we list the properties of the \civ{} used in estimating the outflow parameters.} 
The relations between the \civonly{} outflow parameters and the bolometric luminosity are shown in Fig. \ref{fig:outflow_civ} and the respective linear correlations are reported in Table \ref{tab:outflow_relations}. The outflow parameters considered in this work {(kinetic power $\dot{E}_{\textrm{kin}}$ and thrust $\dot{M}_{\textrm{out}}v_{\textrm{o}}$)} show very good correlations with the bolometric luminosity (as expected), reaching correlation coefficients $\sim$ 0.9 and $P \ll 10^{-20}$ {in both cases}.

\par The left plot of Fig. \ref{fig:outflow_civ} shows the kinetic power of \civonly{} outflows. We find that 15 {from 42} high-$z$ sources present a $\dot{E}_{\textrm{kin}}/L_{\textrm{bol}}$ ratio  $\gtrsim 5 \times 10^{-3}$, {8 of them have $\dot{E}_{\textrm{kin}}/L_{\textrm{bol}}$ around 0.01 and 3 sources exhibit  a ratio close to 0.05 (between 1 to 3\%).}
The situation is  different when we consider the low-$z$ sample. %
{In this case, only two sources, $[\textrm{HB89}]$1259+593 and $[\textrm{HB89}]$1543+489, have a $\dot{E}_{\textrm{kin}}/L_{\textrm{bol}}$ ratio $\gtrsim 5 \times 10^{-3}$. }  This is hardly surprising, considering that the luminosity of the BLUE component is  highly correlated with 1350\AA\ luminosity with a slope $\approx 1.8 \gg 1$, and that outflow velocities are systematically higher at high $L$. At high $L$, the kinetic power reach values for which a substantial feedback effect might be possible: {a considerable fraction of sources present $\dot{E}_{\rm kin} \gtrsim 0.005\, L_{\rm bol}$ and some of them are close to the limit 0.05 $L_\mathrm{bol}$\ that is the minimum energetic requirement for feedback to lead to the black hole mass - host velocity dispersion relation according to \citet{dimatteoetal05}.}


\par The right panel of Fig. \ref{fig:outflow_civ} shows the behavior of the thrust in different luminosity ranges. 
Thrust values from  {$\sim\,10^{35}$ to $10^{36}$}\,g\,cm\,s$^{-2}$ at high luminosity and from $\sim\,10^{32}$ to $10^{35}$\,g\,cm\,$^{-2}$ at low L. {At high luminosity, the thrust values are slightly closer to their AGN $L/c$\ momentum rate than at low luminosity.} It seems reasonable that the \civonly{} emitting gas in the inner BLR may not have yet suffered losses in energy and momentum (unlike \oiiionly).  In simple word, this would imply that, at least at high $L$, the outflow is somehow able to exploit the full luminosity of the continuum for its acceleration.  Radiative acceleration is apparently less efficient at low-$z$. The reason for the difference --  at low luminosity most sources remain substantially below the threshold lines at 0.005 $L_\mathrm{bol}$\ and $L_\mathrm{bol}/c$ -- is not entirely clear. A simple explanation is that the acceleration might be involving a smaller fraction of the gas mass at low $L$. Lowering the gas density by an order of magnitude would reconcile the low-$z$ values with the critical limits. A lower density (or a higher degree of ionization, see Appendix \ref{appendix:civ_outflows}) would increase the gas mass needed to explain the observed line luminosity and thus increase the outflowing gas mass proportionally. Lower density gas might be associated with lower column density gas that can be more efficiently accelerated \citep{Netzer_2010}. 

We conclude that there is the possibility  of a significant feedback effect due  to the mildly ionized BLR outflow from the high ionization broad line gas observed in luminous quasars at the cosmic noon, and that this possibility appears more remote for lower-luminosity, low-$z$\ quasars.  
This conclusion appears to be valid for both RQ and RL sources: the slightly lower velocities measured on the \civ\ profile of  the RLs are not enough to significantly affect the outflow dynamical parameters with respect to the RQ population (Fig. \ref{fig:outflow_civ}).  



\begin{figure*}[ht!]
\centering
    \includegraphics[width=0.42\linewidth]{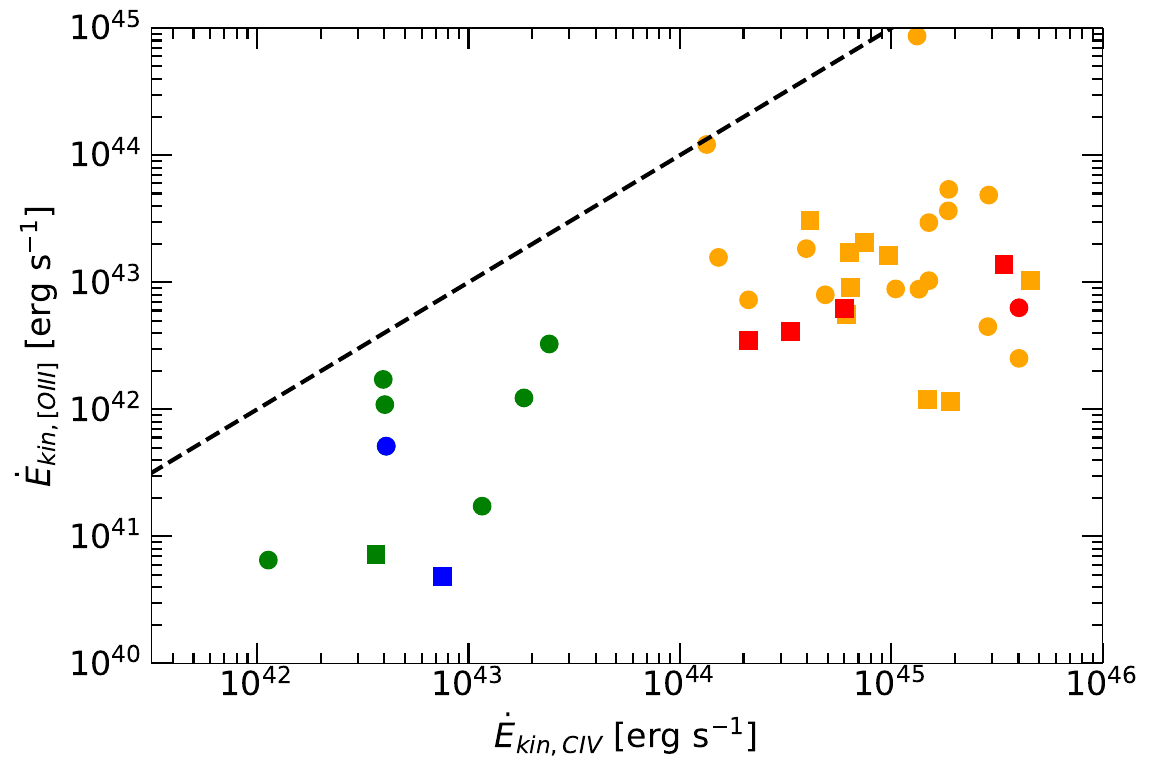}
    \hspace{1cm}
   \includegraphics[width=0.423\linewidth]{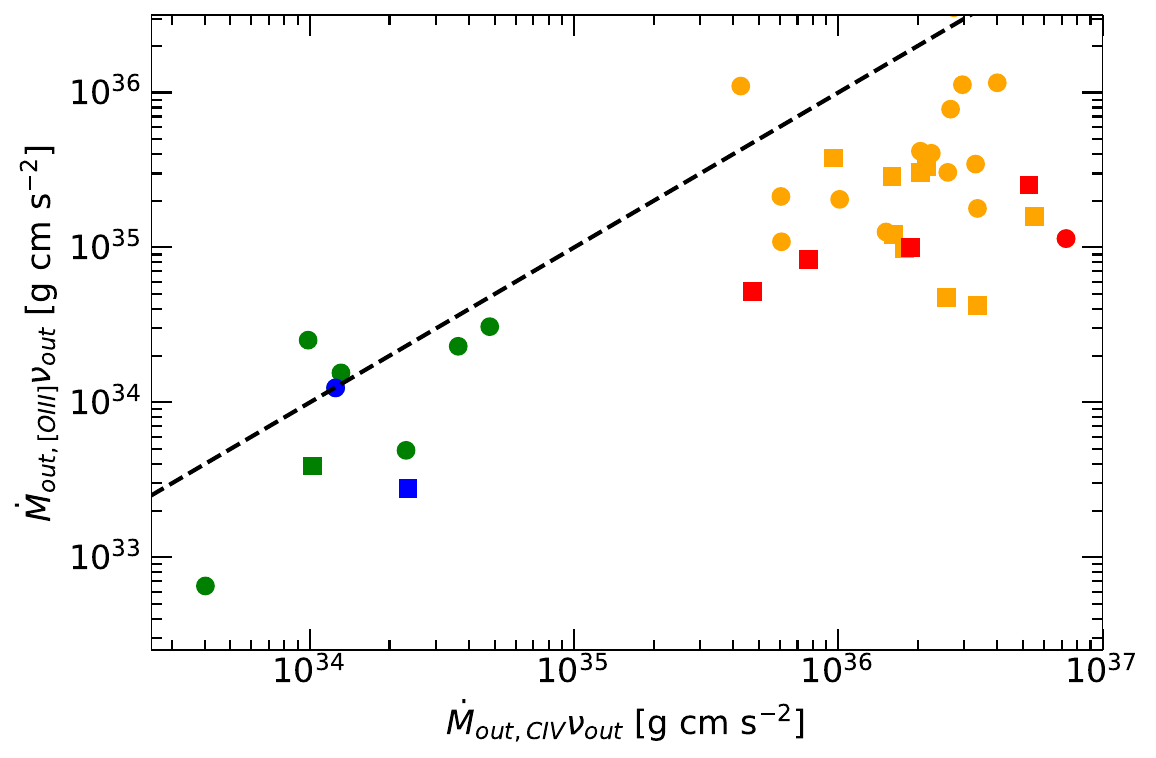}

    \hspace{1cm}
    \caption{Comparison between the  kinetic power $\dot{E}_{kin}$ (\textit{left}), and thrust (\textit{right}) of \oiiiseven{} and \civ{} outflows, for high-z (orange and red symbols) and low-z (blue and green symbols) sources. Both parameters have been estimated by evaluating the outflow radius in each source according Eqs.\ref{eq:radio_oiii} and \ref{eq:radio_civ} for \oiiionly{} and \civonly{} respectively. Black dashed lines indicate the 1:1 relation. Color scheme as in Fig. \ref{fig:lledd_oiii_civ}. }
    \label{fig:comp_outflow_oiii_civ}
\end{figure*}



\subsubsection{Comparison between the outflow parameters of \oiiiseven{} and \civ{}}
\label{comparison_oiii_civ}

\par { {The average values and standard deviations of \oiiiseven{} and \civ{} outflow parameters for the three analysed samples (ISAAC and HEMS at high-z, and FOS at low-z) are listed in Table \ref{tab:parametros_outflows_oiii_civ}. }
Fig. \ref{fig:comp_outflow_oiii_civ} presents a comparison between the estimated kinetic power and thrust for the \oiiiseven{} (left panel) and \civ{} (right panel) emission lines.  {From the results obtained independently for \oiiionly{} and \civonly{} and at high- and low-z, in the previous sections, as well as from the comparison of the outflow parameters in the BLR and NLR through the analysis of both lines, we infer that:} 



\begin{itemize}

\item{The ISAAC and HEMS (high-redshift) samples share similar values for the outflow parameters, mass outflow rate, thrust, and kinetic power. This is true for both \oiiionly{} and \civonly{}, with the \civonly\ consistently stronger wind parameter than those estimated from \oiiionly. The FOS (low-redshift) sample exhibits substantially lower values for all outflow parameters  {($\approx$ one and two orders of magnitude in \oiiionly{} and \civonly{} respectively)}. }

\item {At high redshift, the \civonly{} outflow parameters indicate a significant influence: {16 out of 42 have $\dot{E}_{\rm kin} > 0.005 L_{\rm bol}$, and in some cases, it approaches $\sim$ 3\% of $L_{\rm bol}$}. Regarding \oiiionly{}, our results suggest that although the outflowing gas at high redshift seems to be induced by the AGN, its impact is possibly restricted to the central kiloparsecs, due to its low efficiency.}

\item  {In contrast to the outflows at low redshift,  {some of} the \civonly{} outflows at high redshift exhibit thrust values on the order of $L_{\rm bol}/c$, suggesting that the outflowing gas in this scenario may be momentum-conserving. This characteristic is not observed in \oiiionly{} for both high- and low-redshift ranges.}

\item It is  intriguing that at low-$z$, \civonly\ and \oiiionly{} have consistent thrust values   at variance with high-luminosity sources  (Fig. \ref{fig:comp_outflow_oiii_civ}), even if both of them fall short of the $L/c$\ line. Accepted at face value, this might be  a consequence of the compactness of the emitting regions in low-$z$\ quasars: $\dot{M}_{\textrm{out}}v_{\textrm{o}}$\ is inversely proportional to the radius and in the case of the low-luminosity NLSy1s, the \oiiionly\ emitting regions might be very compact ($\lesssim 10$ pc), therefore preserving a   signature of the AGN radiative acceleration.

\end{itemize}


\par

\subsection{A tentative physical explanation for the RL/RQ difference in wind profiles}

\label{sect:high_low_z}
\par Observations of the \civ\ line in the ISAAC sample suggest systematically lower blueshifts for radio-loud than for radio-quiet quasars {in agreement with previous studies both at high and low-$z$ \citep[see e.g.][]{willsetal95, sulenticetal95,corbinboroson96,marziani_1996,richards_2011}.}  According to \citet[][and references therein]{Sulentic_2015}, two phenomena might be relevant here.



First, the pressure exerted by the jet in its propagation   surpasses the thermal pressure of the BLR gas and the hydrostatic pressure of the expected emitting gas for the blueshifted component attributed to the accretion disc wind. This suggests the existence of an ``avoidance zone'' near the radio axis, that may be wider  in case of powerful relativistic jets, potentially suppressing emission along radial lines close to the jet axis. In addition,  the impact on the inner BLR due to the cocoon associated with the relativistic ejection.   The wind might start farther from the central BH due to cocoon pressure, reaching a lower terminal velocity. 

\section{Conclusions}
\label{section:conclusions}
\par We presented an additional sample of 10 high-redshift, high-luminosity quasars observed with the VLT/ISAAC spectrograph that complete our ISAAC sample, discussed in \citetalias{Deconto-Machado_2023}. These data cover the \hb{} spectral range,  shifted to the near-infrared due to its redshift,  and are combined with rest-frame UV spectra, observed in the optical, available in the literature. The analysis at high redshift involves 32 sources from our ISAAC sample and 28 sources from the HEMS survey, {including both RL and RQ and with bolometric luminosities in the range from $\approx$ 10$^{47}$ to 10$^{48.5}$[erg\,s$^{-1}$].} A comparison sample is made with 84 low-redshift, low-luminosity sources selected from the FOS data.  We performed a spectroscopic exploration of UV and optical emission line diagnostics along the quasar main sequence, including a dedicated analysis of the differences between RL and RQ sources at both high and low redshifts. Additionally, we investigated the main feedback properties from both the BLR and NLR using high-ionization lines (\civonly{} and \oiiionly{}, respectively). Our main conclusions are as follows.

\begin{enumerate}
   \item The shift in the Main Sequence towards broader \hb\ profiles and higher R$_{\ion{Fe}{II}}$ observed {at high redshift} in \citetalias{Deconto-Machado_2023}, is reconfirmed with the inclusion of the 10 new sources.   RL quasars from our high-$z$ sample consistently exhibit a tendency towards lower R$_{\ion{Fe}{II}}$ compared to the RQ ones.
   
   \item High-$z$ RL/RQ have stronger outflows than low-$z$ RL/RQ as inferred from both \oiiiseven{} and \civ, which indicates that the accretion rate may be the primary driver of these outflows. 
   \item RL have smaller outflows compared to RQ in both high- and low-$z$ ranges. This suggests that the radio-loudness has a second order   effect on the outflow. However, more data on high-$z$ RL is needed.
   
    \item Both RQ and RL outflows may induce a significant feedback effect at the high luminosity encountered in the high-$z$ sample.  
    
    \item  In the end, the outflow ``power" is proportional to the AGN luminosity, via the line luminosity that enters linearly, and the increase of outflow velocity with luminosity that is expected to be weaker ($\propto L^\frac{1}{4}$ for a radiation driven wind), but to become significant   if a   luminosity range as large as $\sim 10^3- 10^4$\ is considered, as in the case of the samples presented in this paper.  Within limited luminosity ranges, the dominant effect is associated with the radiative output per unit mass (i.e., the Eddington ratio).  
    
\item The \civ{} line follows the same trends of \oiiiseven{}, however with larger outflow velocities and stronger outflow parameters. The relations between these two emission lines does not seem to be affected by the radio emission. 
\end{enumerate}

\par Our findings suggest that, regardless of the influence of accretion and radio emission, the outflow is likely to significantly affect the host galaxy at high   luminosity. Less clear is the role of outflows from the BLR and NLR at low-$z$\  due to the much lower luminosity, and smaller outflow velocity. 



\begin{acknowledgements}
 The authors thank the anonymous referee for his/her valuable suggestions that helped us to improve the present paper. A.D.M. and A.d.O. acknowledge financial support from the Spanish MCIU through projects PID2019–106027GB–C41, PID2022-140871NB-C21 by “ERDF A way of making Europe”, and the Severo Ochoa grant CEX2021- 515001131-S funded by MCIN/AEI/10.13039/501100011033. A.D.M. acknowledges the support of the INPhINIT fellowship from ``la Caixa'' Foundation (ID 100010434). The fellowship code is LCF/BQ/DI19/11730018. A.D.M. is very thankful for the kind hospitality at the Padova Astronomical Observatory. This research has made use of the NASA/IPAC Extragalactic Database (NED) which is operated by the Jet Propulsion Laboratory, California Institute of Technology, under contract with the National Aeronautics and Space Administration. In this work, we made use of astronomical tool IRAF, which is distributed by the National Optical Astronomy Observatories, and archival data from FIRST, NVSS, and 2MASS. Funding for SDSS-III has been provided by the Alfred P. Sloan Foundation, the Participating Institutions, the National Science Foundation, and the U.S. Department of Energy Office of Science. The SDSS-III web site is http://www.sdss3.org/. 
\end{acknowledgements}

\bibliographystyle{aa} 
\bibliography{bib} 
%
%
\onecolumn
\setcounter{table}{12}

\LTcapwidth=16cm
\begin{small}
\begin{longtable}{p{4.1cm}cccccccccccc}

\caption{Properties of the \oiiiseven{} and \civ{} outflows for the objects of each sample considered in this work  {that present clear outflows}.} \\
\hline\hline
\hline\hline
\noalign{\smallskip}
& & \multicolumn{5}{c}{[\ion{O}{III}]} & & \multicolumn{5}{c}{\ion{C}{IV}}\\
\cline{3-7} \cline{9-13}
\noalign{\smallskip}
Source & $L_{\rm bol}$ & $L_{\textrm{full}}$ & $L_{\textrm{SBC}}$ & $r_{\textrm{out}}$ & $I_{\rm SBC}/I_{\rm tot}$ & c(1/2)$_{\textrm{SBC}}$ & & $L_{\textrm{BLUE}}$ & $L(\textrm{1350\AA})$ & $r_{\textrm{out}}$ & $c(1/2)_{\textrm{BLUE}}$ & FWHM$_{\textrm{BLUE}}$  \\
\noalign{\smallskip}
& [erg s$^{-1}$] & [erg s$^{-1}$] & [erg s$^{-1}$] & [kpc] & & [km s$^{-1}$] & & [erg s$^{-1}$] & [erg s$^{-1}$] & [pc] & [km s$^{-1}$] & [km s$^{-1}$]\\
(1) & (2) & (3) & (4) & (5) & (6) &  (7) & & (8) & (9) & (10) & (11) & (12)\\
\noalign{\smallskip}
\hline
\noalign{\smallskip}
\endfirsthead
\caption{Properties of the \oiiiseven{} and \civ{} outflows for the objects of each sample considered in this work. (cont.)}\\
\hline\hline
\noalign{\smallskip}
& & \multicolumn{5}{c}{[\ion{O}{III}]} & & \multicolumn{5}{c}{\ion{C}{IV}}\\
\cline{3-7} \cline{9-13}
\noalign{\smallskip}
Source & $L_{\rm bol}$ & $L_{\textrm{full}}$ & $L_{\textrm{SBC}}$ & $r_{\textrm{out}}$ & $I_{\rm SBC}/I_{\rm tot}$ & $c(1/2)_{\textrm{SBC}}$ & & $L_{\textrm{BLUE}}$ & $L(\textrm{1350\AA})$ & $r_{\textrm{out}}$ & $c(1/2)_{\textrm{BLUE}}$ & FWHM$_{\textrm{BLUE}}$  \\
& [erg s$^{-1}$] & [erg s$^{-1}$] & [erg s$^{-1}$] & [kpc] & & [km s$^{-1}$] & & [erg s$^{-1}$] & [erg s$^{-1}$] & [pc] & [km s$^{-1}$] & [km s$^{-1}$]\\
(1) & (2) & (3) & (4) & (5) & (6) & (7) & & (8) & (9) & (10) & (11) & (12) \\
\noalign{\smallskip}
\hline
\noalign{\smallskip}
\endhead
\hline
\endfoot
\noalign{\smallskip}
     \multicolumn{13}{c}{ISAAC1+ISAAC2}\\
     \noalign{\smallskip}
     \hline
     \noalign{\smallskip}
    $[\rm HB89]$ 0029+073	&	47.65	&	44.02	&	43.97	&	2.03	&	1.00 &-396	&	&	\multicolumn{5}{c}{no \ion{C}{IV} spectrum available}	\\
SDSSJ005700.18+143737.7	&	47.56	&	\multicolumn{5}{c}{[\ion{O}{III}] not covered by the spectrum}	&	&	44.74	&	46.73	&	0.14	&	-4765	&	13187	\\
PKS0226-038	&	47.18	&	\multicolumn{5}{c}{full [\ion{O}{III}] $c(1/2) > -250$ km s$^{-1}$ }	&	&	44.77	&	46.81	&	0.15	&	-2593	&	10535	\\
PKS0237-23	&	47.76	&	44.06	&	43.98	&	2.13	& 0.83	& -1091	&	&	45.43	&	47.48	&	0.30	&	-3773	&	10778	\\
BZQJ0544-2241	&	47.52	&	43.81	&	43.47	&	1.60	&	0.45 & -620	&	&	\multicolumn{5}{c}{\ion{C}{IV} at the edge of the spectrum}	\\
CTSJ01.03	&	47.37	&	\multicolumn{5}{c}{full [\ion{O}{III}] $c(1/2) > -250$ km s$^{-1}$ }	&	&	45.49	&	47.68	&	0.37	&	-2423	&	9263	\\
WB J0948+0855	&	47.45	&	\multicolumn{5}{c}{no [\ion{O}{III}] outflow detected}	&	&	44.29	&	46.97	&	0.17	&	-6917	&	22060	\\
CTSJ03.14	&	47.62	&	\multicolumn{5}{c}{full [\ion{O}{III}] $c(1/2) > -250$ km s$^{-1}$ }	&	&	45.05	&	47.59	&	0.33	&	-3003	&	10299	\\
SDSSJ114358.52+052444.9	&	47.20	&	43.41	&	43.31	&	1.02	&	0.78 & -1294	&	&	45.17	&	47.30	&	0.25	&	-4185	&	14345	\\
SDSSJ115954.33+201921.1	&	47.66	&	44.32	&	44.06	&	2.89	&	0.55 & -498	&	&	45.21	&	47.47	&	0.29	&	-3090	&	9933	\\
SDSSJ120147.90+120630.2	&	47.91	&	44.65	&	44.49	&	4.20	&	0.62 & -977	&	&	45.32	&	47.40	&	0.27	&	-2639	&	7464	\\
SDSSJ132012.33+142037.1	&	47.24	&	43.98	&	43.79	&	1.95	&	0.64 & -1428	&	&	45.04	&	46.87	&	0.16	&	-2967	&	8739	\\
SDSSJ135831.78+050522.8	&	47.49	&	44.16	&	44.13	&	2.39	&	0.90 & -2079	&	&	45.05	&	47.12	&	0.20	&	-4217	&	11983	\\
Q1410+096	&	47.73	&	44.34	&	44.20	&	2.95	&	0.73 & -1254	&	&	44.88	&	47.15	&	0.21	&	-4135	&	10881	\\
PKS1448-232	&	47.58	&	44.11	&	43.49	&	2.27	&	0.24 & -753	&	&	\multicolumn{5}{c}{no \ion{C}{IV} spectrum available}	\\
SDSSJ153830.55+085517.0	&	47.33	&	43.56	&	43.56	&	1.20	&	1.00 & -544	&	&	45.31	&	47.39	&	0.27	&	-2780	&	10024	\\
$[\textrm{HB89}]$1559+088	&	47.57	&	44.06	&	43.82	&	2.15	&	0.56 & -1611	&	&	44.92	&	47.21	&	0.22	&	-2554	&	7129	\\
SDSSJ161458.33+144836.9	&	47.53	&	44.18	&	44.06	&	2.45	&	0.75 & -1874	&	&	45.15	&	47.26	&	0.24	&	-3375	&	9333	\\
PKS1937-101	&	48.44	&	44.79	&	44.55	&	4.97	&	0.57 & -483	&	&	\multicolumn{5}{c}{no \ion{C}{IV} spectrum available}	\\
PKS2000-330 & 47.71 & 44.43 & 44.06 & 3.27 & 0.42 & -826 & & 45.27 & 46.50 & 0.11 & -4325 & 7835\\
SDSSJ210524.49+000407.3	&	47.79	&	\multicolumn{5}{c}{full [\ion{O}{III}] $c(1/2) > -250$ km s$^{-1}$ }	&	&	44.71	&	47.40	&	0.27	&	-5071	&	10418	\\
SDSSJ210831.56-063022.5	&	47.30	&	43.37	&	43.37	&	0.97	&	1.00 & -1243	&	&	44.88	&	47.22	&	0.23	&	-5406	&	13618	\\
SDSSJ212329.46-005052.9	&	47.75	&	43.95	&	43.90	&	1.88	&	0.88 & -2376	&	&	45.21	&	47.51	&	0.31	&	-3974	&	10217	\\
PKS2126-15	&	48.01	&	44.38	&	43.85	&	3.08	&	0.30 & -988	&	&	\multicolumn{5}{c}{no \ion{C}{IV} spectrum available} \\
FBQSJ2149-0811	&	47.26	&	43.68	&	43.08	&	1.38	&	0.25 & -1291	&	&	\multicolumn{5}{c}{no \ion{C}{IV} spectrum available}	\\
SDSSJ235808.54+012507.2	&	47.92	&	44.40	&	44.40	&	3.15	&	1.00 & -1052	&	&	45.04	&	47.08	&	0.20	&	-3189	&	8302	\\

\noalign{\smallskip}
\hline
\noalign{\smallskip}
\multicolumn{13}{c}{HEMS}\\
\noalign{\smallskip}
\hline
\noalign{\smallskip}
HE0035-2853	&	47.19	&	\multicolumn{5}{c}{full [\ion{O}{III}] $c(1/2) > -250$ km s$^{-1}$ }	&	&	44.79	&	46.73	&	0.14	&	-3028	&	5530	\\
HE0043-2300	&	47.36	&	\multicolumn{5}{c}{full [\ion{O}{III}] $c(1/2) > -250$ km s$^{-1}$ }	&	&	44.73	&	46.75	&	0.14	&	-2472	&	3840	\\
HE0058-3231 & 47.49 & \multicolumn{5}{c}{full [\ion{O}{III}] $c(1/2) > -250$ km s$^{-1}$ }	&	&	44.63 & 46.70 & 0.13 & -3004 & 5200\\
HE0109-3518	&	47.61	&	\multicolumn{5}{c}{full [\ion{O}{III}] $c(1/2) > -250$ km s$^{-1}$ }	&	&	45.31	&	47.63	&	0.35	&	-1991	&	3020	\\
HE0122-3759	&	47.38	&	43.65	&	43.35	&	1.34	&	0.63 & -2083	&	&	45.17	&	47.08	&	0.20	&	-4851	&	7520	\\
HE0203-4627	&	47.10	&	43.41	&	42.80	&	1.02	&	0.24 & -1322	&	&	44.25	&	46.46	&	0.10	&	-3362	&	6450	\\
HE0205-3756 & 47.85 & 46.84 & 43.56 & 1.73 & 0.48 &  -1456 & & 45.14 & 47.11 & 0.20 & -3040 & 4640\\
HE0248-3628	&	46.76	&	\multicolumn{5}{c}{full [\ion{O}{III}] $c(1/2) > -250$ km s$^{-1}$ }	&	&	44.80	&	46.98	&	0.18	&	-3880	&	5140	\\
HE0251-5550	&	47.84	&	44.42	&	43.74	&	3.22	&	0.16 & -1253	&	&	45.45	&	47.38	&	0.27	&	-2869	&	4170	\\
HE0349-5249 & 47.89 &  \multicolumn{5}{c}{full [\ion{O}{III}] $c(1/2) > -250$ km s$^{-1}$ }	& & 45.07 & 47.06 & 0.19 & -3104 & 5525\\
HE0359-3959	&	47.11	&	\multicolumn{5}{c}{[\ion{O}{III}] region dominated by \ion{Fe}{II}}	&	&	44.11	&	46.54	&	0.11	&	-6042	&	7940	\\
HE0436-3709 & 47.48 &  \multicolumn{5}{c}{full [\ion{O}{III}] $c(1/2) > -250$ km s$^{-1}$ }	& & 44.05 & 45.97 & 0.06 & -3551 & 5670\\
HE0507-3236	&	47.19	&	43.82	&	43.32	&	1.61	&	0.36 & -1323	&	&	44.68	&	46.69	&	0.13	&	-2925	&	4720	\\
HE0512-3329	&	47.30	&	43.77	&	43.31	&	1.52	&	0.40 & -1785	&	&	44.63	&	46.32	&	0.09	&	-3581	&	5000	\\
HE0926-0201	&	47.61	&	44.31	&	44.01	&	2.84	&	0.41 & -1338	&	&	45.36	&	47.13	&	0.21	&	-2584	&	5560	\\
HE0940-1050	&	47.90	&	44.93	&	44.26	&	5.81	&	0.75 & -4857	&	&	45.40	&	47.48	&	0.30	&	-4331	&	6270	\\
HE1039-0724	&	47.06	&	\multicolumn{5}{c}{full [\ion{O}{III}] $c(1/2) > -250$ km s$^{-1}$ }	&	&	44.70	&	46.73	&	0.14	&	-1976	&	4350	\\
HE1104-1805	&	47.54	&	\multicolumn{5}{c}{full [\ion{O}{III}] $c(1/2) > -250$ km s$^{-1}$ }	&	&	45.08	&	47.15	&	0.21	&	-2632	&	4610	\\
HE1120+0154	&	47.59	&	\multicolumn{5}{c}{full [\ion{O}{III}] $c(1/2) > -250$ km s$^{-1}$ }	&	&	45.08	&	47.06	&	0.19	&	-1526	&	4400	\\
HE1347-2457 & 48.17 &\multicolumn{5}{c}{[\ion{O}{III}] region dominated by \ion{Fe}{II}}	& & 45.54 & 47.53 & 0.31 & -5538 & 7720 \\
HE1349+0007	&	47.10	&	43.42	&	43.08	&	1.03	&	0.35 & -1477	&	&	45.06	&	46.85	&	0.16	&	-3347 &	5377	\\
HE1409+0101	&	47.52	&	44.18	&	44.03	&	2.46	&	0.74 & -1173	&	&	45.11	&	46.97	&	0.17	&	-3328	&	5390	\\
HE2147-3212	&	47.09	&	\multicolumn{5}{c}{full [\ion{O}{III}] $c(1/2) > -250$ km s$^{-1}$ }	&	&	44.96	&	46.82	&	0.15	&	-4116	&	6650	\\
HE2156-4020	&	47.45	&	43.66	&	43.34	&	1.35	&	0.51 & -1126	&	&	45.46	&	47.38	&	0.27	&	-2625	&	5000	\\
HE2202-2557	&	47.13	&	43.53	&	43.20	&	1.16	&	0.38 & -1240	&	&	45.22	&	46.66	&	0.13	&	-2573	&	4490	\\
HE2349-3800	&	47.06	&	43.69	&	43.41	&	1.40	&	0.65 & -975	&	&	44.48	&	46.45	&	0.10	&	-3788	&	5710	\\
HE2352-4010	&	47.67	&	44.25	&	44.11	&	2.67	&	0.79 & -2185	&	&	44.82	&	47.14	&	0.21	&	-3050	&	3760	\\
\noalign{\smallskip}
\hline
\noalign{\smallskip}
\multicolumn{13}{c}{FOS}\\
\noalign{\smallskip}
\hline
\noalign{\smallskip}
PG 0044+030	&	46.48	&	43.05	&	43.00	&	0.66	&	0.75 & -366	&	&	42.01	&	46.01	&	0.06	&	-3251	&	4580	\\
Mrk 1502	&	45.46	&	41.14	&	40.73	&	0.07	&	0.55 & -1352	&	&	41.25	&	44.58	&	0.01	&	-2270	&	6750	\\
3C 057	&	46.48	&	43.10	&	42.84	&	0.71	&	0.61 & -817	&	&	42.38	&	46.45	&	0.10	&	-3355	&	3697	\\
{3C 84} & {44.23} & {42.32}& {41.25} & {0.28} & {0.91} & {-437} && \multicolumn{5}{c}{{\ion{C}{IV} strongly affected by noise}}\\
{LEDA 75249} & {45.84} &  \multicolumn{5}{c}{{no [\ion{O}{III}] outflow detected}} & & {42.01} & {45.53} &  {0.04} & {-2206} & {6318}\\
$[\textrm{HB89}]$ 0850+440	&	46.18	&	42.37 &	42.10	&	0.31	&	0.54 & -1396	&	&	42.12	&	45.71	&	0.05	&	-2926	&	3739	\\
SDSS J100402.61+285535.3	&	46.36	&	42.88	&	42.77	&	0.55	& 0.70 &	-1058	&	&	42.18	&	45.81	&	0.05	&	-3198	&	8043	\\
PG 1116+215	&	45.98	&	42.83	&	42.44	&	0.52	&	0.36 & -700	&	&	41.93	&	45.68	&	0.05	&	-2882	&	8348	\\
{LBQS 1138+0204} & 45.69 & \multicolumn{5}{c}{{no [\ion{O}{III}] outflow detected}} & & {41.57} & {45.32} & {0.03} & {-2553} & {8122}\\
3C 273	&	47.08	&	43.10	&	42.94	&	0.71	&	0.69 & -344	&	&	42.62	&	46.35	&	0.09	&	-1645	&	5587	\\
SBS 1259+593	&	46.11	&	\multicolumn{5}{c}{no [\ion{O}{III}] outflow detected}	&	&	42.74	&	46.15	&	0.07	&	-3866	&	8394	\\
FBQS J131217.7+351521	&	45.60	&	42.15	&	41.98	&	0.24	&	0.62 & -264	&	&	\multicolumn{5}{c}{full \ion{C}{IV} c(1/2) > -250, strongly affected by absorption}	\\
FBQS J1405+2555	&	45.14	&	\multicolumn{5}{c}{no [\ion{O}{III}] outflow detected} &	&	41.23	&	45.37	&	0.03	&	-3841	&	6074	\\
$[\textrm{HB89}]$ 1415+451	&	45.10	&	{40.62} & {40.60} & {0.04} & {1.00} & {-593}&	&	41.21	&	44.65	&	0.02	&	-1472	&	4862	\\
$[\textrm{HB89}]$ 1425+267	&	45.54	&	\multicolumn{5}{c}{no [\ion{O}{III}] outflow detected}	&	&	40.64	&	44.82	&	0.02	&	-4283	&	3543	\\
Mrk 478	&	45.16	&	\multicolumn{5}{c}{no [\ion{O}{III}] outflow detected}	&	&	41.26	&	44.89	&	0.02	&	-1526	&	5032	\\
$[\textrm{HB89}]$ 1444+407	&	45.91	&	\multicolumn{5}{c}{no [\ion{O}{III}] outflow detected}	&	&	41.92	&	45.64	&	0.04	&	-3110	&	4232	\\
$[\textrm{HB89}]$ 1538+477	&	47.26	&	\multicolumn{5}{c}{{no [\ion{O}{III}] outflow detected}}	&	&	42.64	&	46.49	&	0.11	&	-3492	&	9104	\\
$[\textrm{HB89}]$ 1543+489	&	45.83	&	41.89	&	41.81	&	0.18	&	0.84 & -2105	&	&	42.30	&	45.81	&	0.05	&	-3301	&	7901	\\
Mrk 509	&	45.10	&	\multicolumn{5}{c}{no [\ion{O}{III}] outflow detected}	&	&	39.73	&	44.52	&	0.01	&	-2713	&	3069	\\
PG 2112+059	&	46.22	&	\multicolumn{5}{c}{no [\ion{O}{III}] outflow detected}	&	&	42.33	&	45.90	&	0.06	&	-3120	&	6573	\\
MR 2251-178	&	44.77	&	\multicolumn{5}{c}{no [\ion{O}{III}] outflow detected}	&	&	40.32	&	44.54	&	0.01	&	-2363	&	2334	\\
$[\textrm{HB89}]$ 2349-014	&	45.77	&	\multicolumn{5}{c}{{no [\ion{O}{III}] outflow detected}}	&	&	40.17	&	44.53	&	0.01	&	-5746	&	6476	\\

\label{tab:parametros_outflows_oiii_civ}
\end{longtable}
\end{small}
{\small{ {Notes.} {  {(1) Source identification. (2) Bolometric luminosity. (3) Luminosity of the \oiiionly{} full profile applied in the scaling law for estimating the outflow radius $r_{\rm out}$ (Eq. \ref{eq:radio_oiii}). (4) Luminosity of only the \oiiionly{} outflowing component, named AS SBC for \oiiionly. (5) \oiiionly{} outflow radius estimated using the Eq. \ref{eq:radio_oiii}. (6) \oiiionly{} SBC relative intensity. (7) Outflow velocity that we assume to be the centroid at $\frac{1}{2}$ intensity of the \oiiionly{} SBC component. (8) Luminosity of the \civonly{} outflowing BLUE component. (9) Specific luminosity at 1350\AA, entering in Eq. \ref{eq:radio_civ}. (10) Estimated \civonly{} outflow radius. (11) \civonly{} outflow velocity. (12) FWHM of the \civ{} BLUE component. }\par}

\begin{appendix}
\onecolumn

\section{Multicomponent fits in the optical and UV ranges and individual notes}
\label{appendix_1}
\subsection{PKS0226-038}

\begin{figure}[h!]
    \centering
    \includegraphics[width=0.69\linewidth]{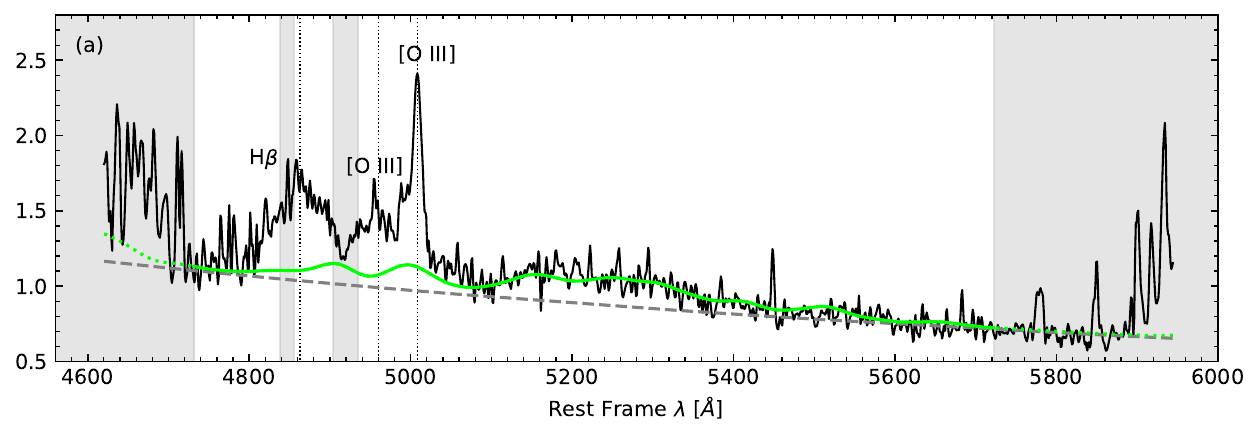}
    \includegraphics[width=0.305\linewidth]{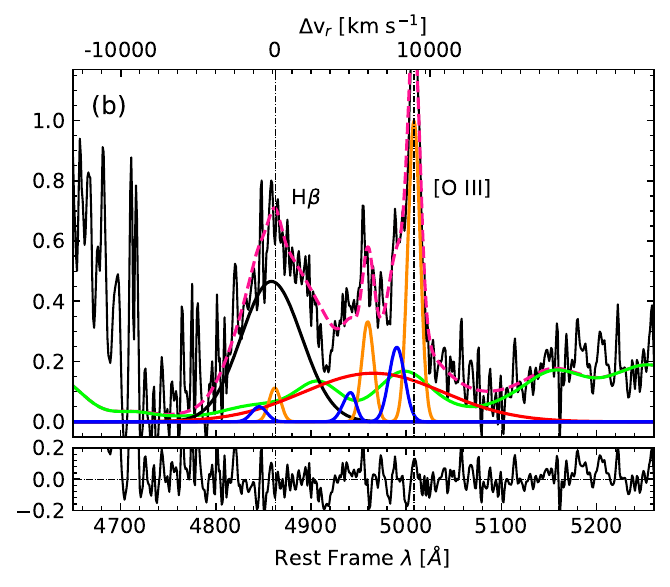}
    \\
    \includegraphics[width=\linewidth]{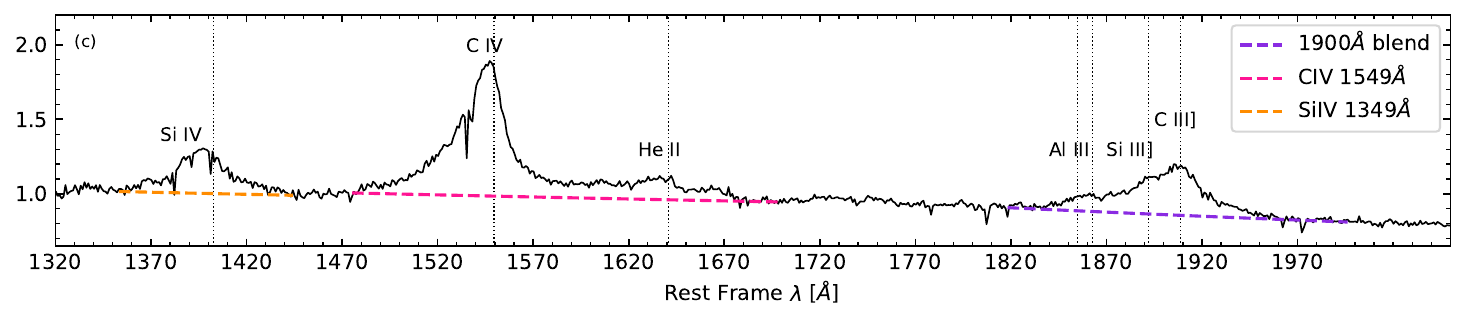}
    \\
    \includegraphics[width=0.33\linewidth]{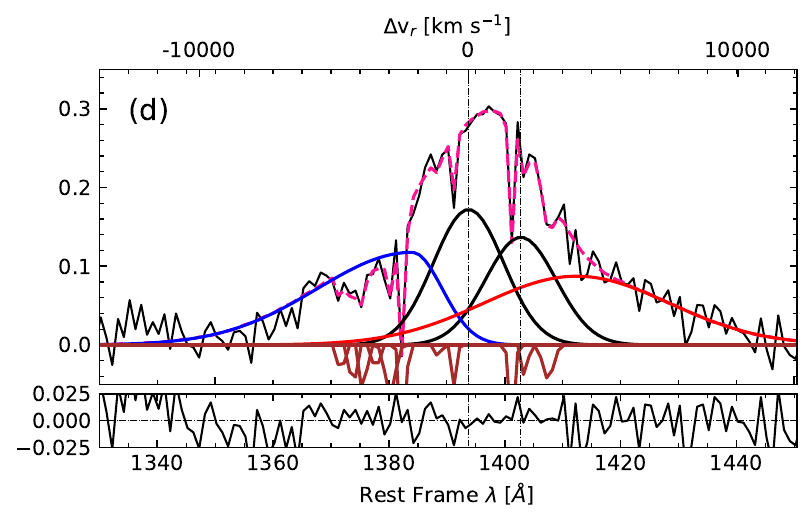}
    \includegraphics[width=0.32\linewidth]{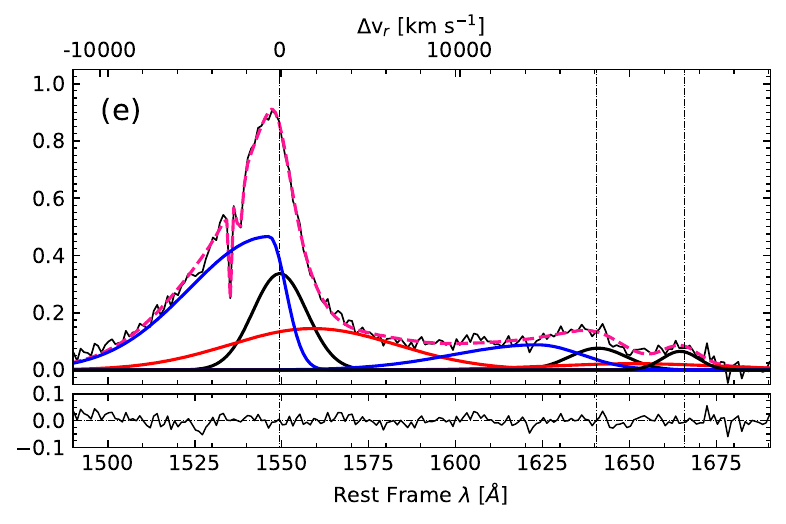}
    \includegraphics[width=0.336\linewidth]{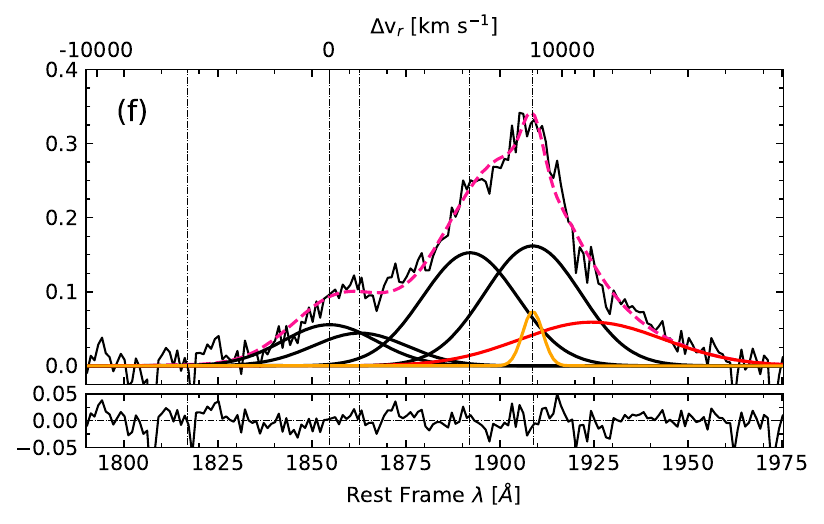}
    \caption{PKS0226-038. \textit{(a)} Rest-frame spectrum covering the \hb{} spectral range obtained with VLT/ISAAC. The spectrum is normalised by the continuum at 5100 \AA . The grey dashed line traces the power law that represent the continuum level as obtained with the \texttt{specfit} multicomponent analysis. The green line shows the \ion{Fe}{II} contribution.  Dotted green line indicates the expected \ion{Fe}{II} contribution for the other parts of the spectra that were not considered in the fitting. The vertical dotted lines indicate the rest-frame of the main emission lines in the \hb{} spectral range and the grey-shaded area indicate the regions that were not considered in the fittings and/or are affected by tellurics. The white area indicates the region used to anchor both the continuum and the \ion{Fe}{II} template. \textit{(b)} Result of the fitting after continuum subtraction (upper panel) and the respective residuals (bottom panel) for the \hb{} region. Pink dashed line shows the final fit. \textit{(c)} Continuum-normalised UV spectrum with the adopted continuum marked in different colours depending on the spectral region. \textit{(d)} Model of the \ion{Si}{IV}$\lambda$1397+\ion{O}{IV}]$\lambda$1402. \textit{(e)} Model of the \civ{}+\ion{He}{II}$\lambda$1640. \textit{(f)} Model of the 1900\AA\ blend. Broad components (BC) are represented by a black line meanwhile red line shows the VBC. Orange lines represent narrow components and the blue ones correspond to the blueshifted components. The region in which the \ion{Fe}{II} or \ion{Fe}{III} template was fitted is represented by the solid green lines. Brown lines represent the absorptions seen in the spectrum and were modelled as negative-flux Gaussians.}
    \label{fig:PKS0226}
\end{figure}

VBC shift affected by telluric residual.

\clearpage
\subsection{PKS0237-23}

\begin{figure}[h!]
    \centering
    \includegraphics[width=0.69\linewidth]{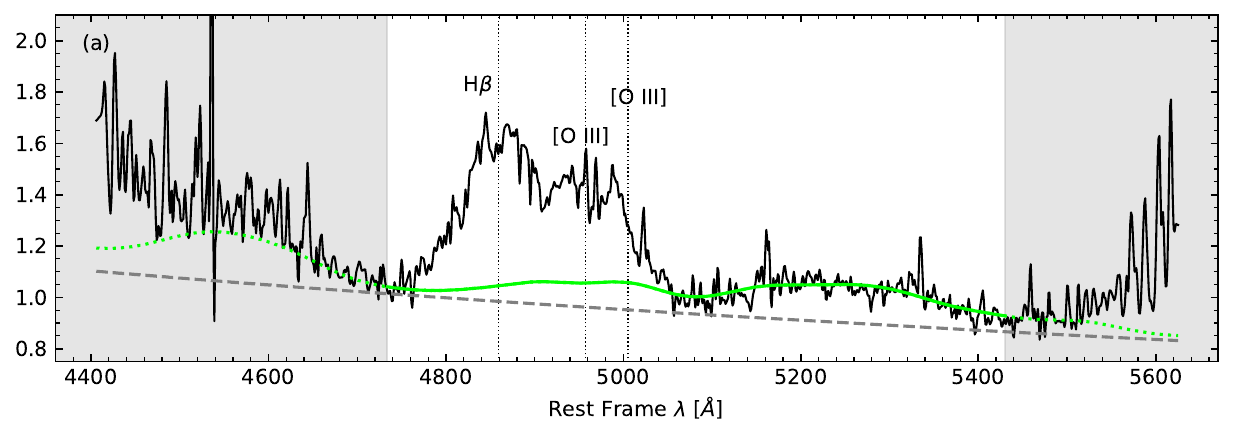}
    \includegraphics[width=0.305\linewidth]{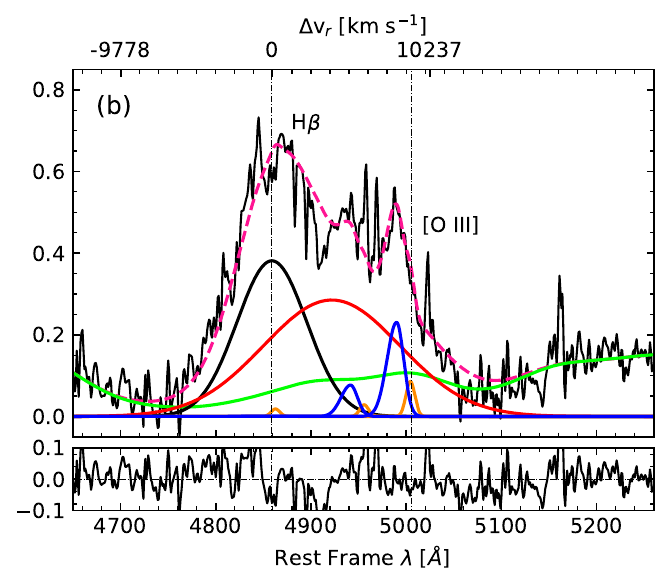}
    \\
    \includegraphics[width=\linewidth]{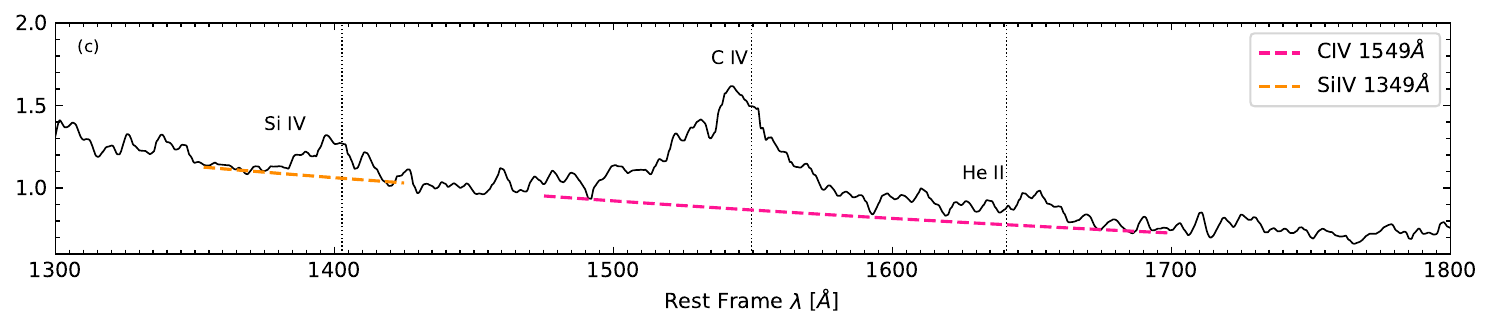}
    \raggedright
    \includegraphics[width=0.33\linewidth]{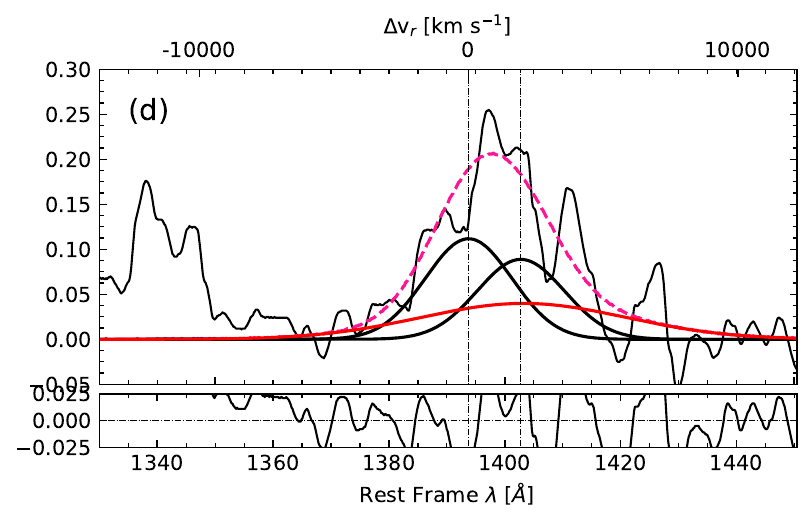}
    \includegraphics[width=0.32\linewidth]{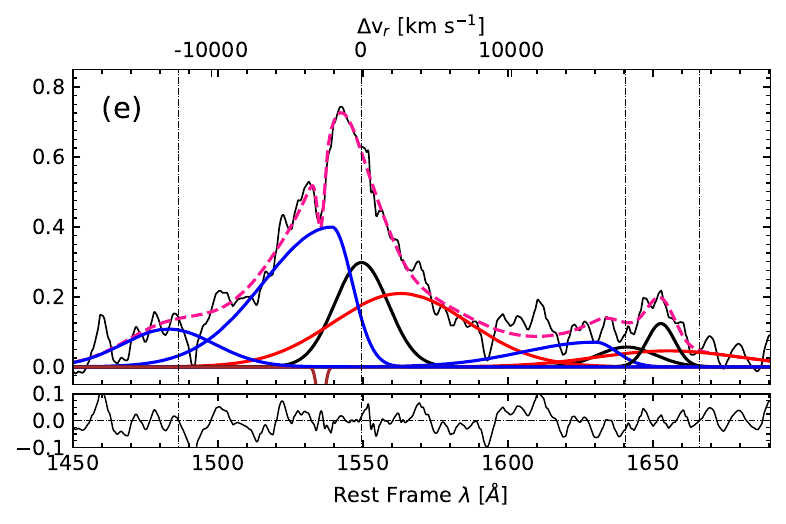}
    \caption{PKS0237-23. Colours and lines as Figure \ref{fig:PKS0226}.}
    \label{fig:PKS0237}
\end{figure}

\clearpage
\subsection{BZQJ0544-2241}
\begin{figure}[h!]
    \centering
    \includegraphics[width=0.69\linewidth]{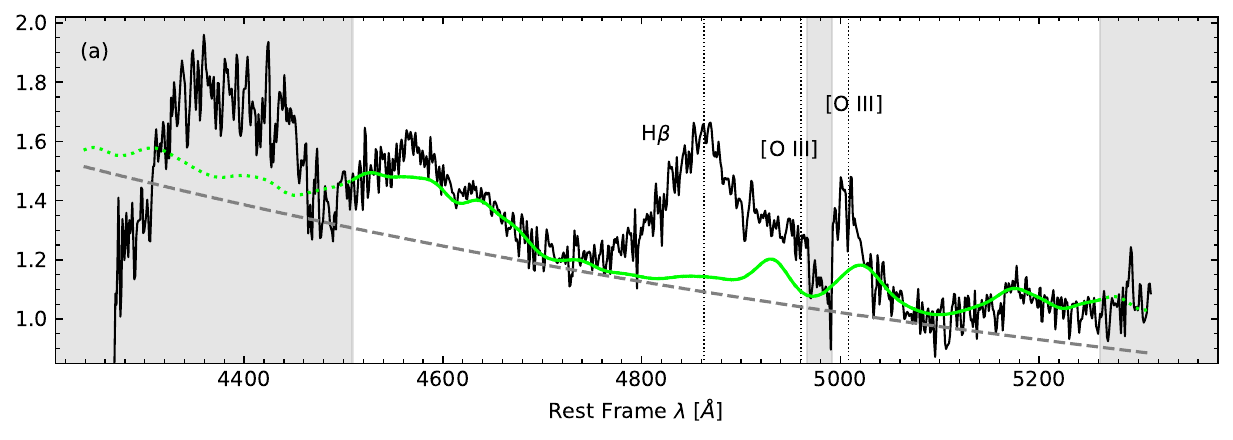}
    \includegraphics[width=0.305\linewidth]{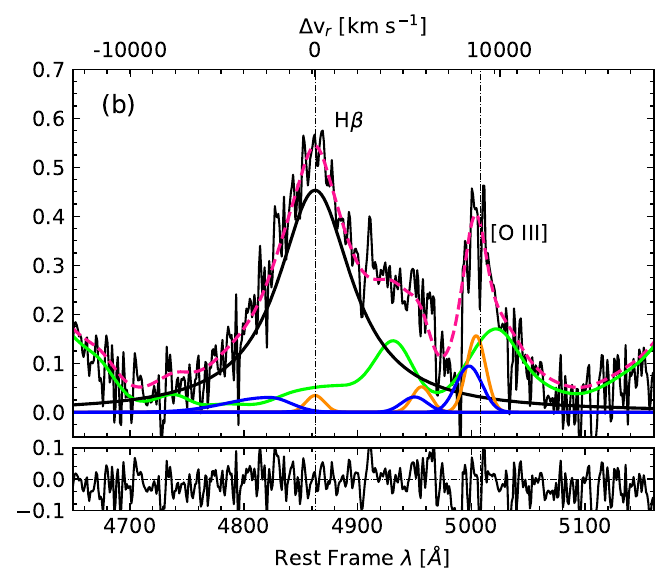}
    \\
    \includegraphics[width=\linewidth]{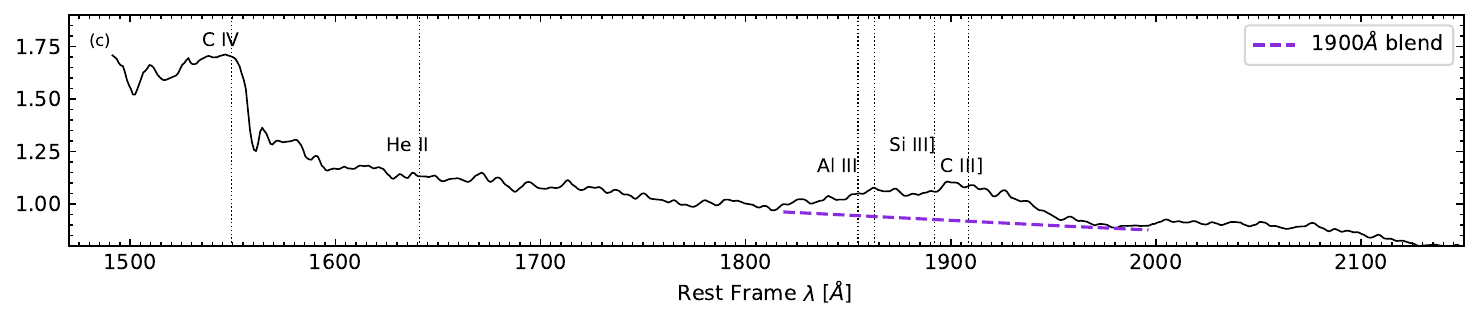}
    \raggedleft
    \includegraphics[width=0.33\linewidth]{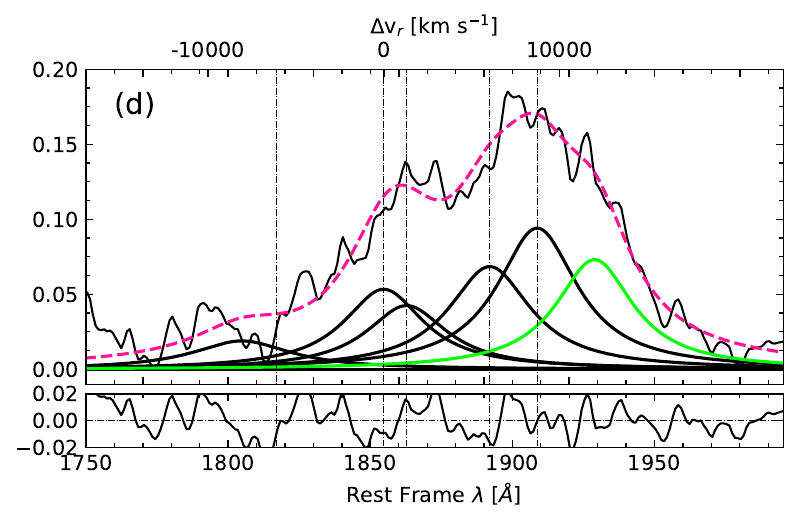}
    \caption{BZQJ0544-2241. Colours and lines as Figure \ref{fig:PKS0226}.}
    \label{fig:0544}
\end{figure}

\clearpage

\subsection{PKS0858-279}

\begin{figure}[h!]
    \centering
    \includegraphics[width=0.695\linewidth]{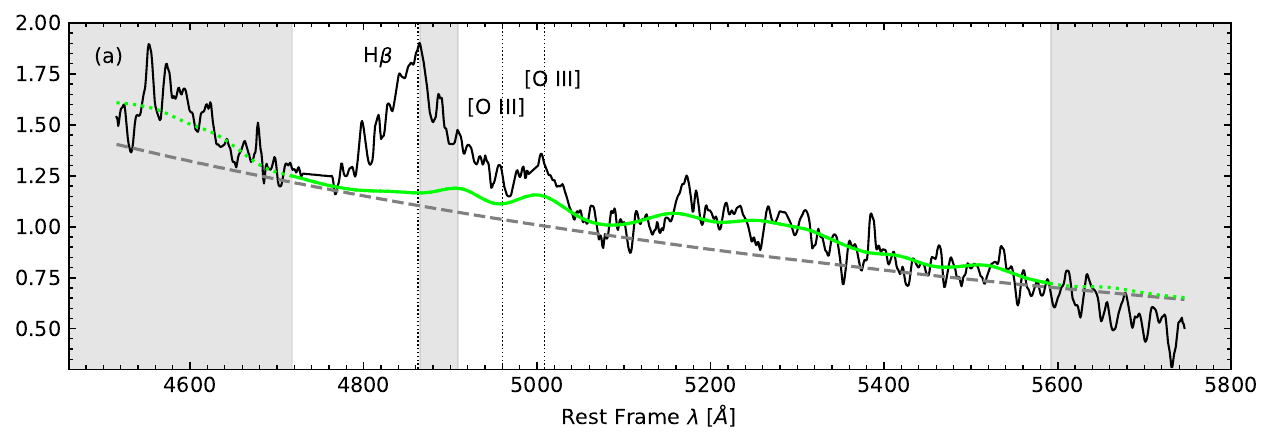}
    \includegraphics[width=0.30\linewidth]{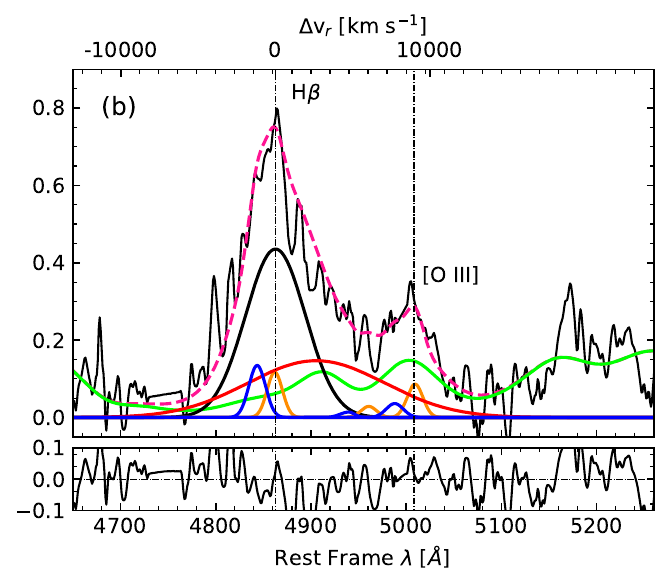}
    \\
    \includegraphics[width=\linewidth]{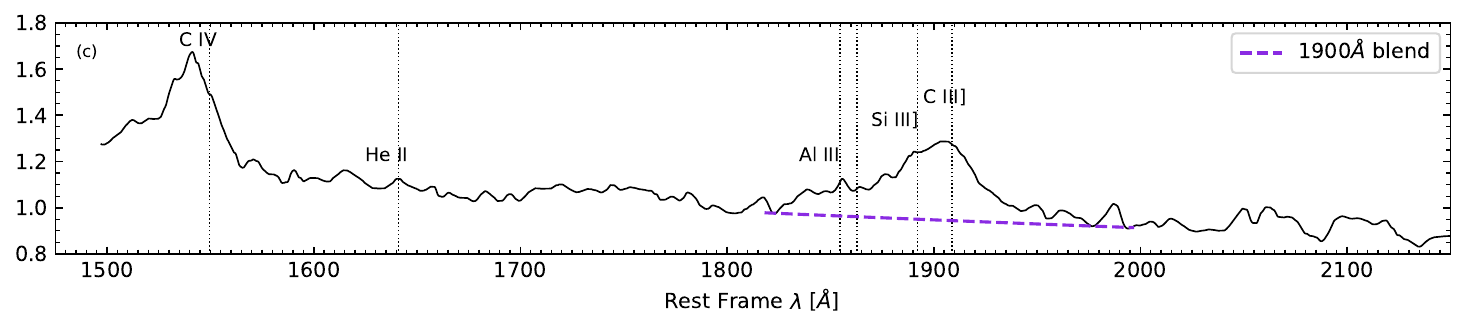}
    \\
    \raggedleft
    \includegraphics[width=0.31\linewidth]{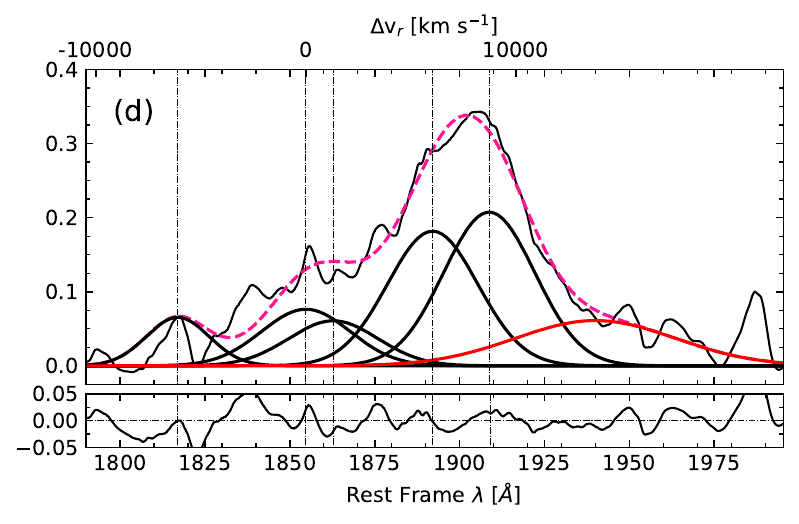}
    \caption{PKS0858-279. Colours and lines as Figure \ref{fig:PKS0226}.}
    \label{fig:PKS0858}
\end{figure}

\clearpage
\subsection{CTSJ01.03}
\begin{figure}[h!]
    \centering
     \includegraphics[width=0.69\linewidth]{figs/plot_cnt_0103_model2.pdf}
    \includegraphics[width=0.305\linewidth]{figs/plot_hb_0103_model3.pdf}
    \\
    \includegraphics[width=\linewidth]{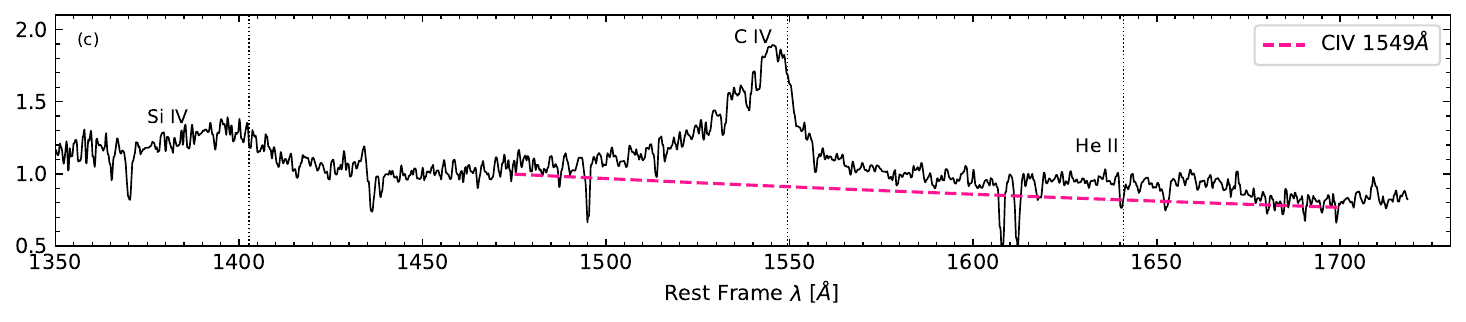}
    \\
    \centering
     \includegraphics[width=0.33\linewidth]{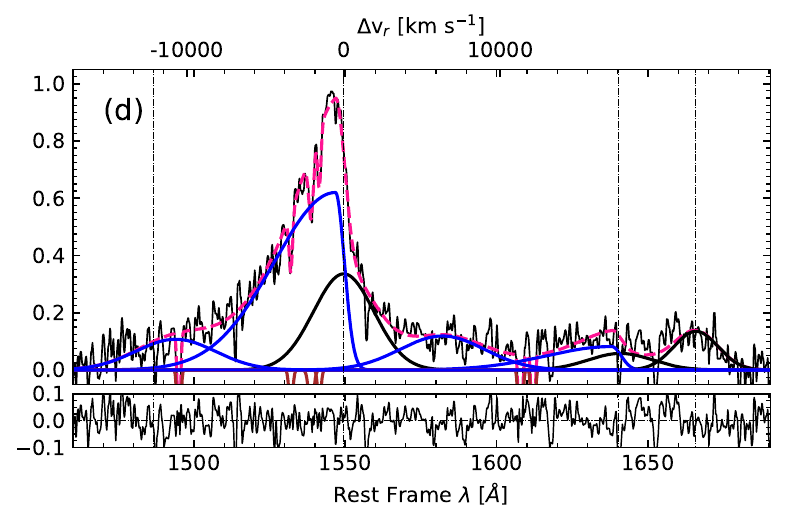}
    \caption{CTSJ01.03. Colours and lines as Figure \ref{fig:PKS0226}.}
    \label{fig:CTSJ0103}
\end{figure}

\clearpage
\subsection{WB J0948+0855}
\label{app:SDSSJ0948}
\begin{figure}[h!]
    \centering
    \includegraphics[width=0.69\linewidth]{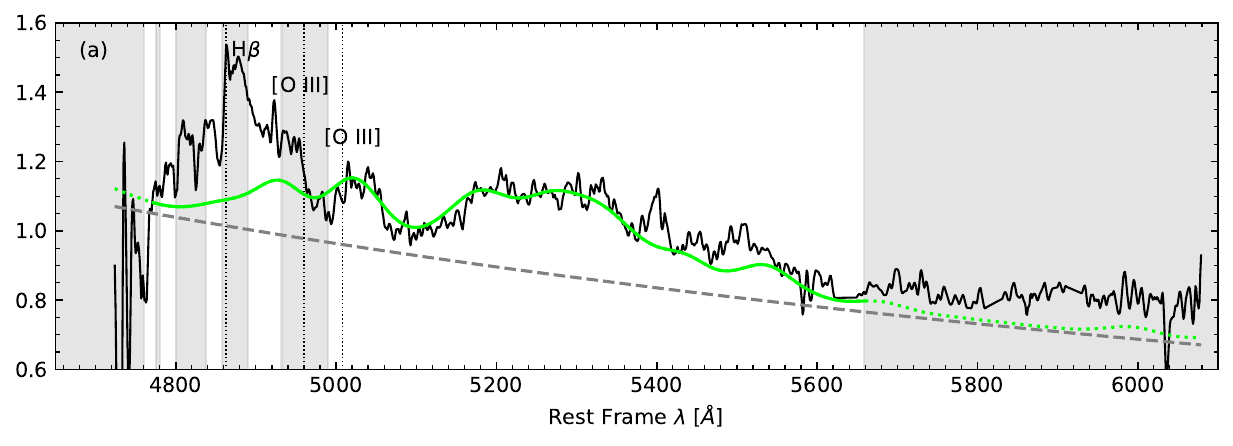}
    \includegraphics[width=0.305\linewidth]{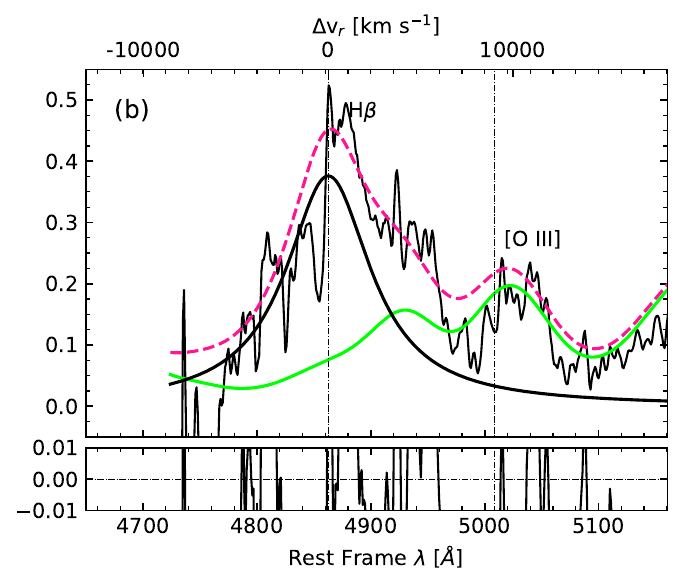}
    \\
    \includegraphics[width=\linewidth]{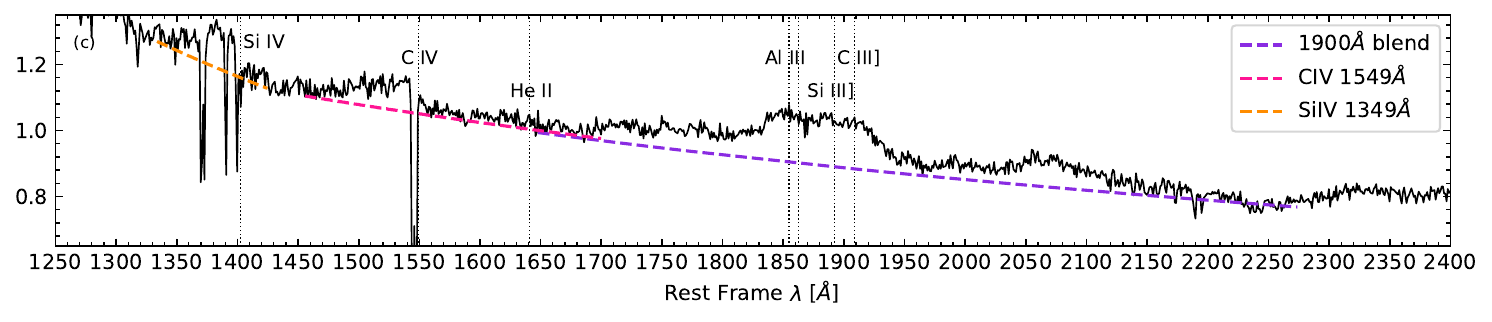}
    \includegraphics[width=0.325\linewidth]{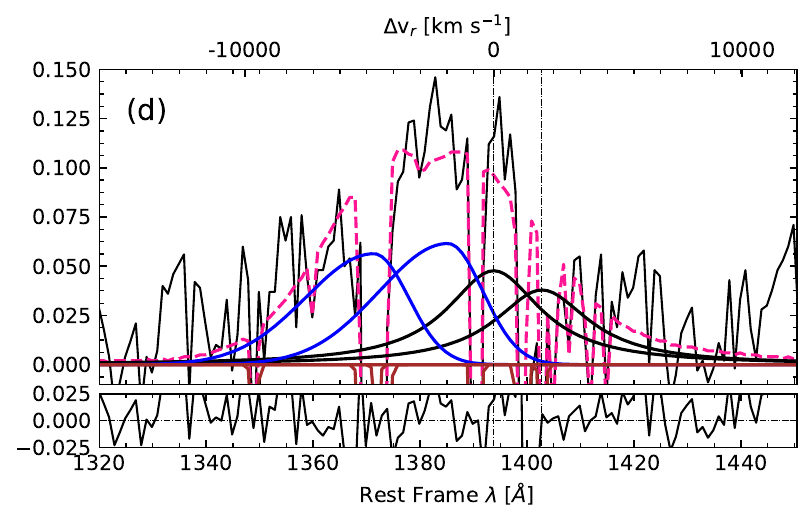}
    \includegraphics[width=0.325\linewidth]{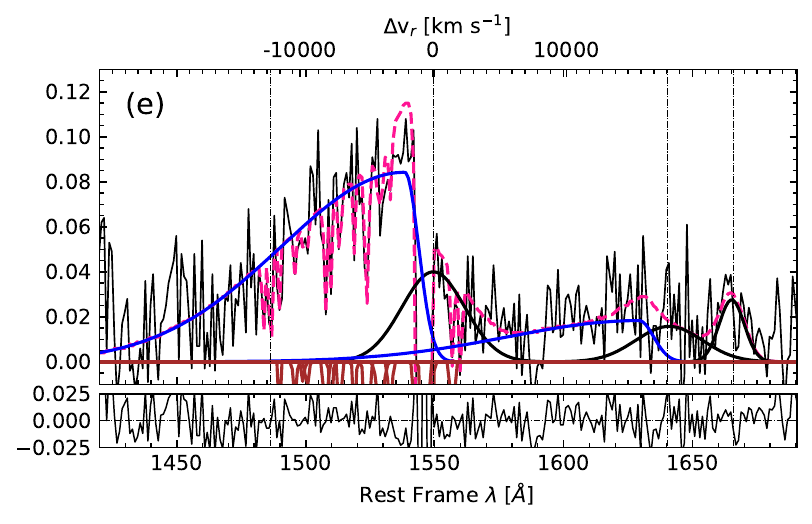}
    \includegraphics[width=0.325\linewidth]{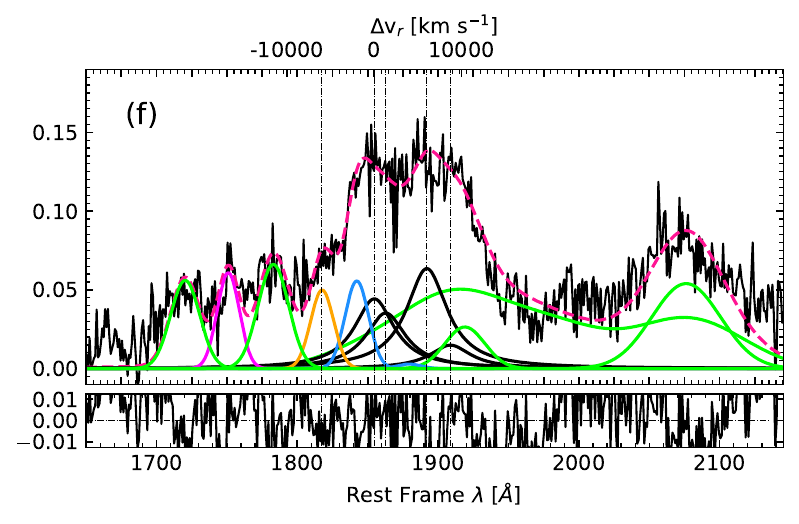}
    \raggedright
    \includegraphics[width=0.325\linewidth]
    {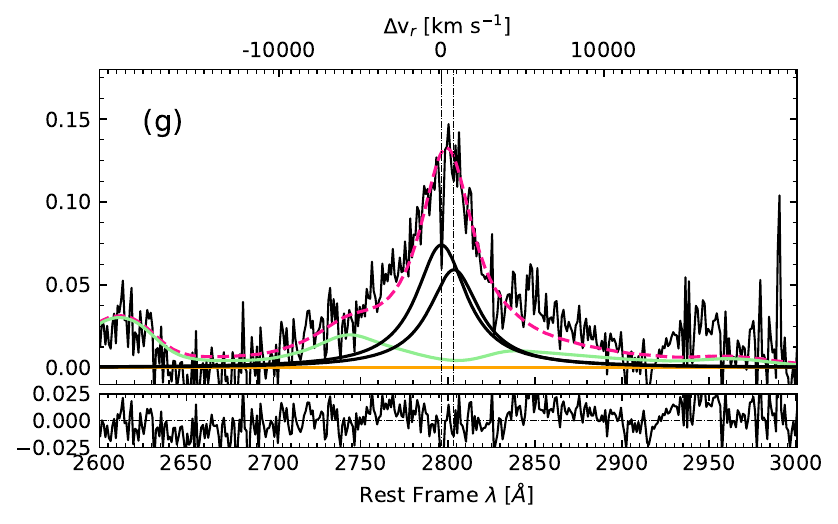}
    \caption{WB J0948+0855. Colours and lines as Figure \ref{fig:PKS0226}.}
    \label{fig:SDSSJ0948}
\end{figure}

An \hb{} BLUE component may be present, however we cannot detect it since the region is located exactly at the border of the spectrum. Due to the difficulty of isolating a narrow component in both \hb{} and \oiiiseven{} emission lines, the redshift of this source has been estimated based on the \ion{Mg}{II}$\lambda \lambda$2796,2803 doublet, since this line could be easily identified in the UV spectra (see panel (g) of Fig. \ref{fig:SDSSJ0948}).

\clearpage
\subsection{CTSJ03.14}

\begin{figure}[h!]
    \centering
    \includegraphics[width=0.69\linewidth]{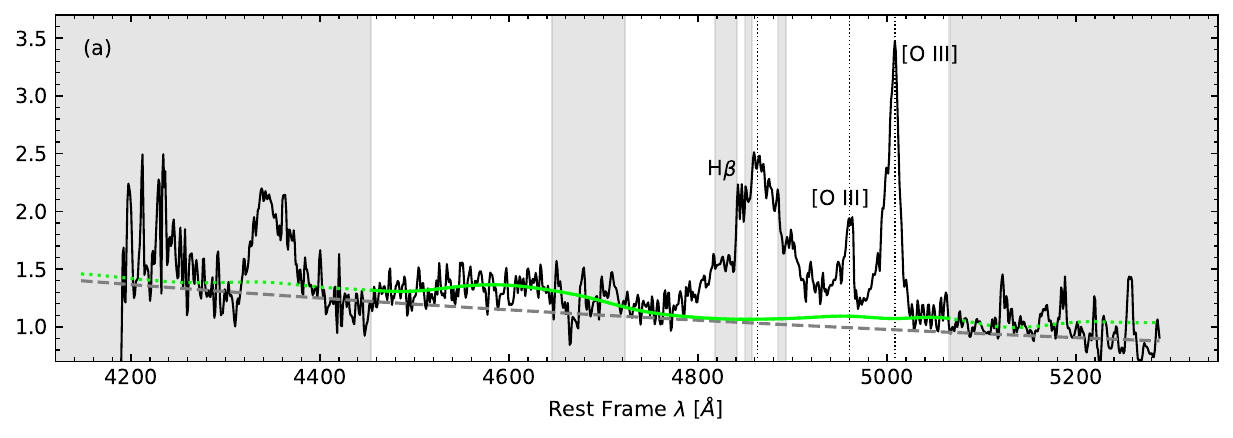}
    \includegraphics[width=0.305\linewidth]{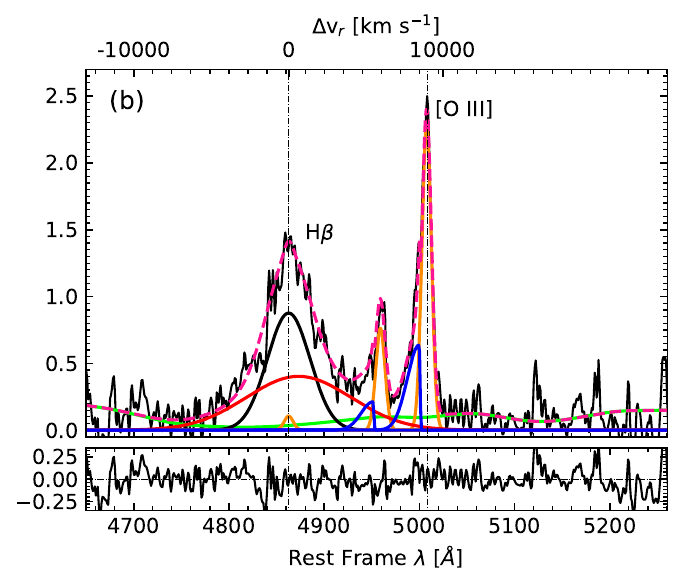}
    \\
    \includegraphics[width=\linewidth]{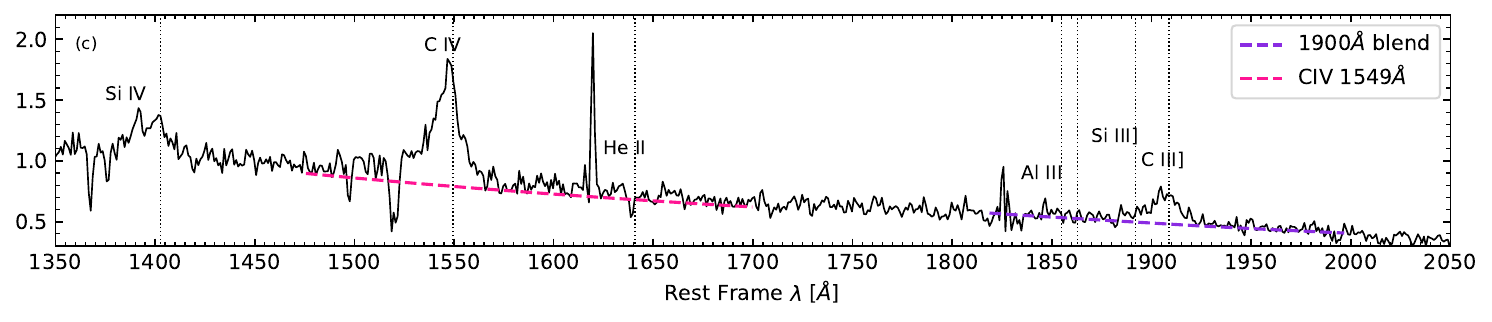}
    \\
    \raggedleft
    \includegraphics[width=0.33\linewidth]{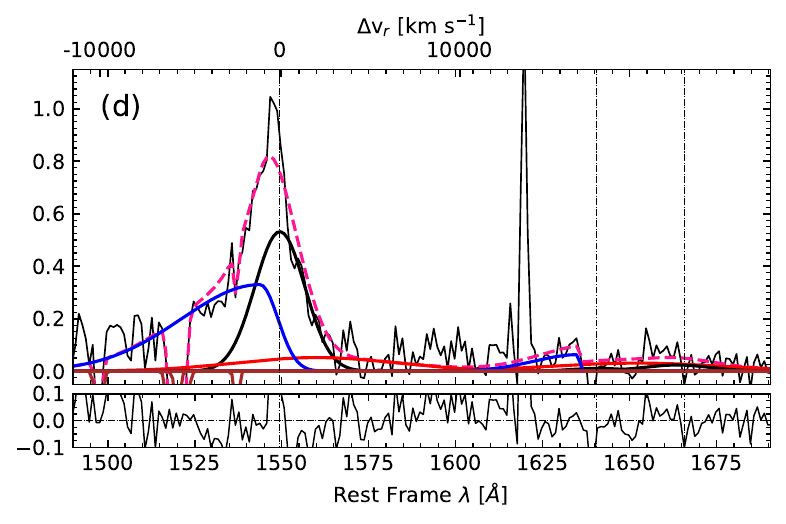}
    \includegraphics[width=0.33\linewidth]{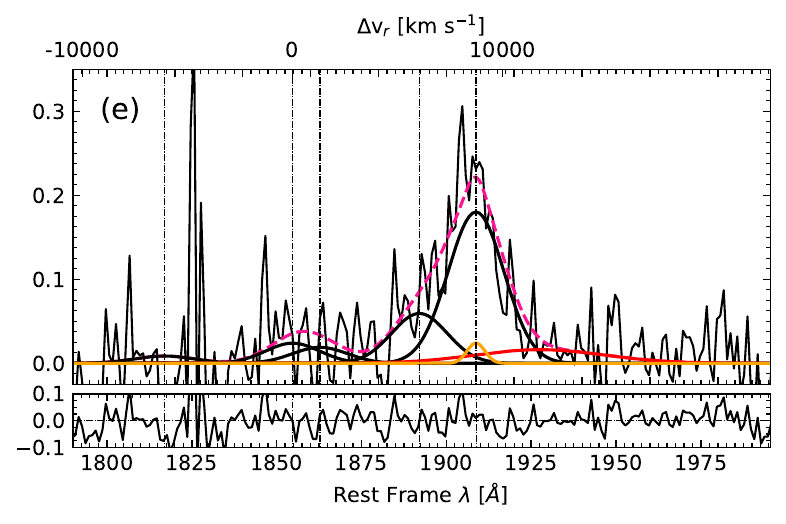}
    \caption{CTSJ03.14. Colours and lines as Figure \ref{fig:PKS0226}.}
    \label{fig:CTSJ0314}
\end{figure}

\subsection{PKS1448-232}

\begin{figure}[h!]
    \centering
    \includegraphics[width=0.69\linewidth]{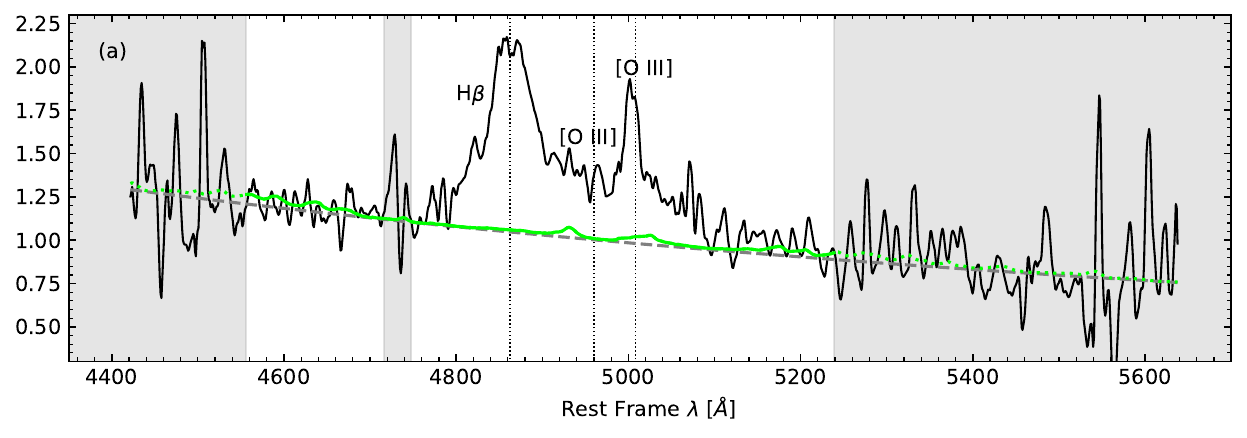}
    \includegraphics[width=0.305\linewidth]{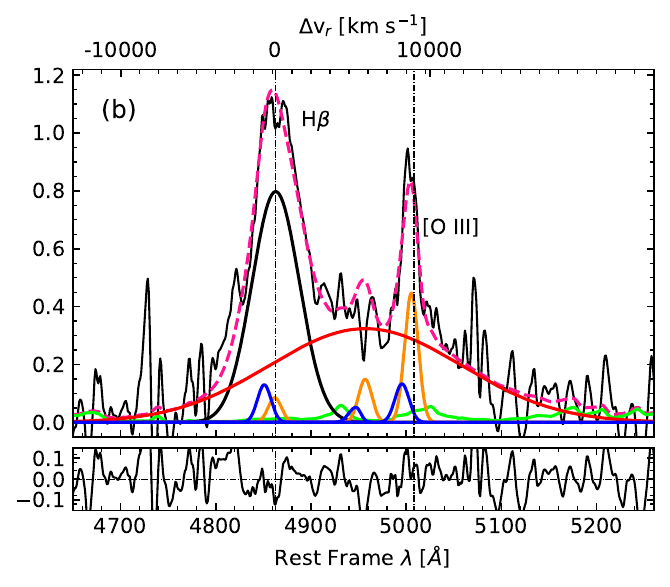}
    \caption{PKS1448-232. Colours and lines as Figure \ref{fig:PKS0226}.}
    \label{fig:PKS1448}
\end{figure}

\clearpage
\subsection{$[\textrm{HB89}]$1559+088}

\begin{figure}[h!]
\centering
    \includegraphics[width=0.69\linewidth]{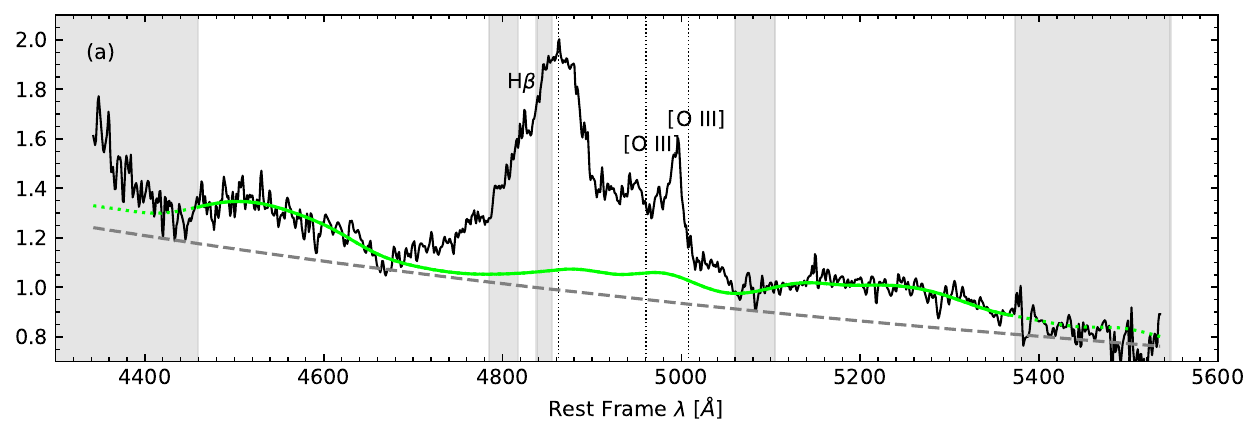}
    \includegraphics[width=0.305\linewidth]{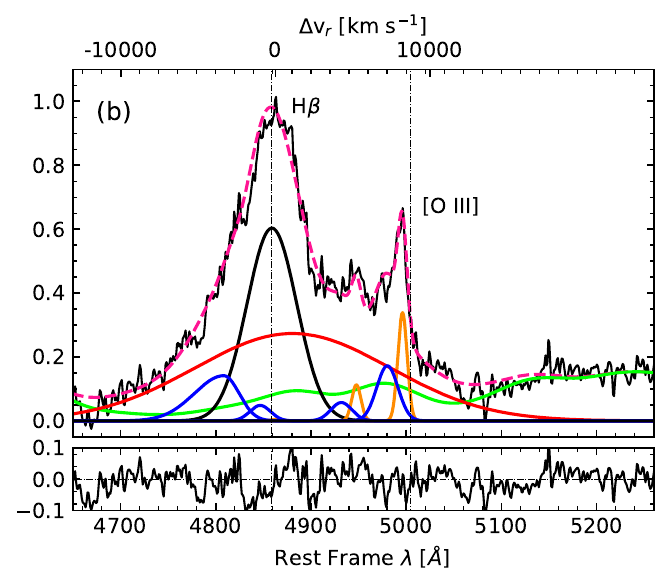}
    \\
    \includegraphics[width=0.995\linewidth]{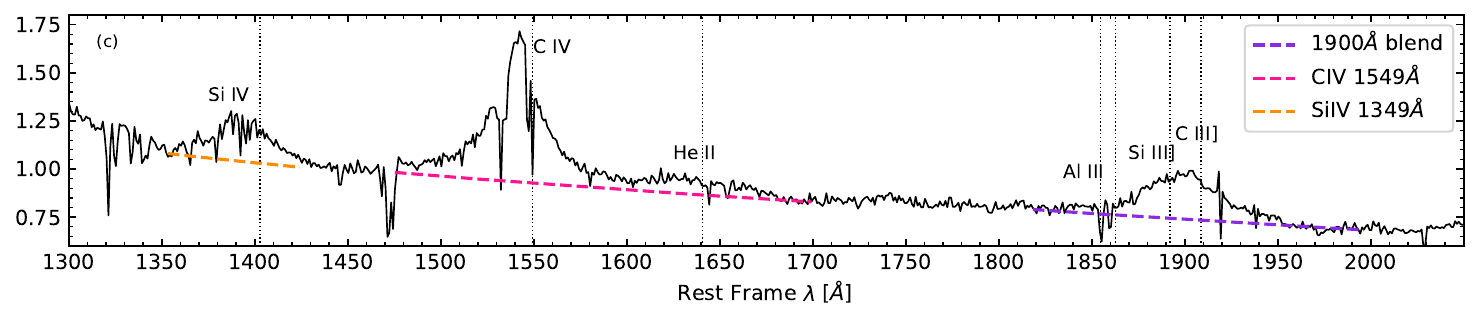}
    \\
    \includegraphics[width=0.325\linewidth]{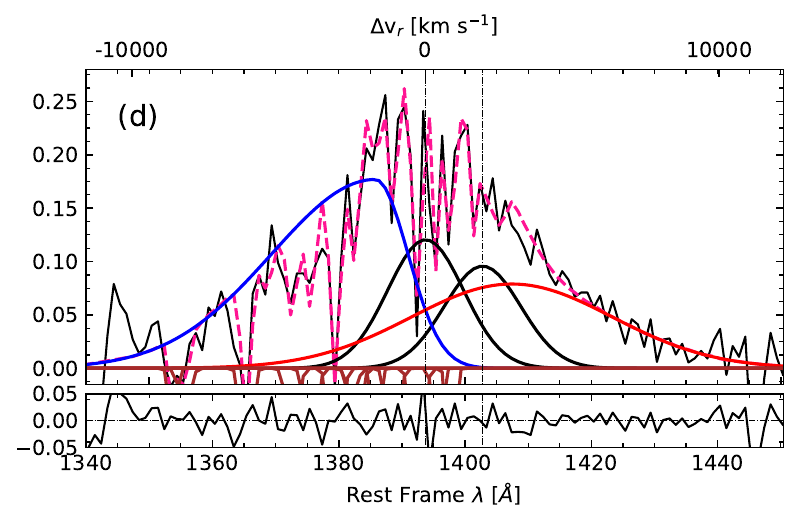}
    \includegraphics[width=0.325\linewidth]{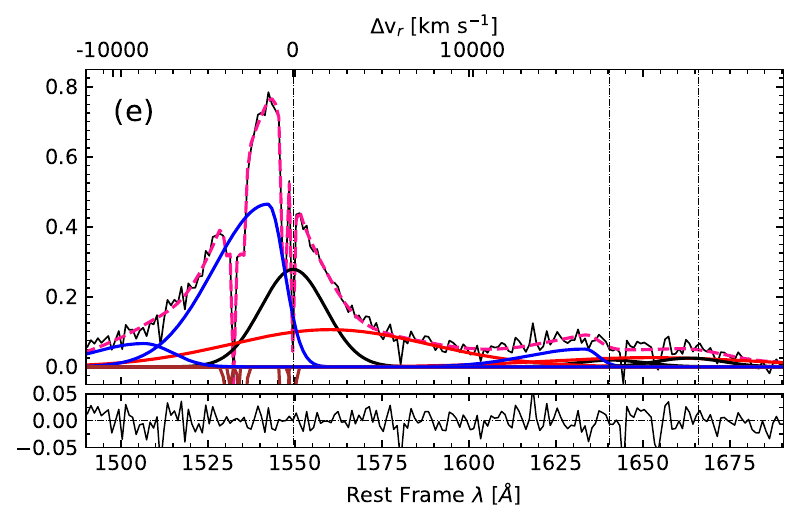}
    \includegraphics[width=0.33\linewidth]{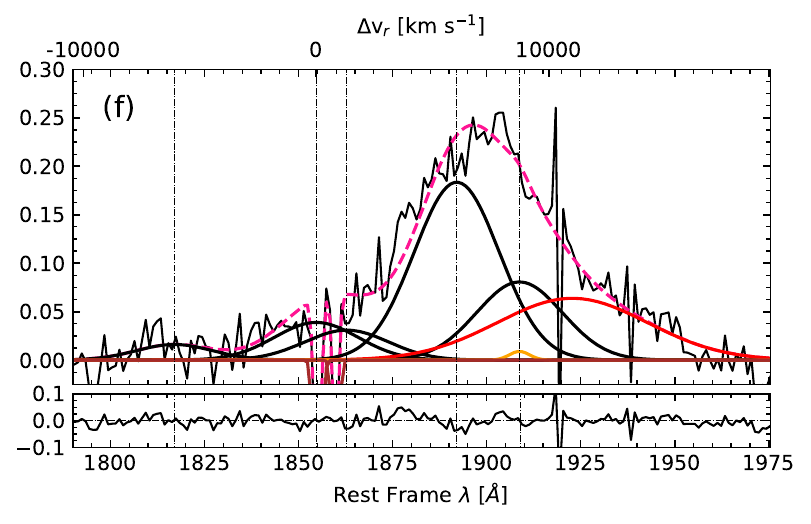}
    \caption{$[\textrm{HB89}]$1559+088. Colours and lines as Figure \ref{fig:PKS0226}.}
    \label{fig:Q1559}
\end{figure}

\subsection{FBQS J2149-0811}

\begin{figure}[h!]
    \centering
    \includegraphics[width=0.69\linewidth]{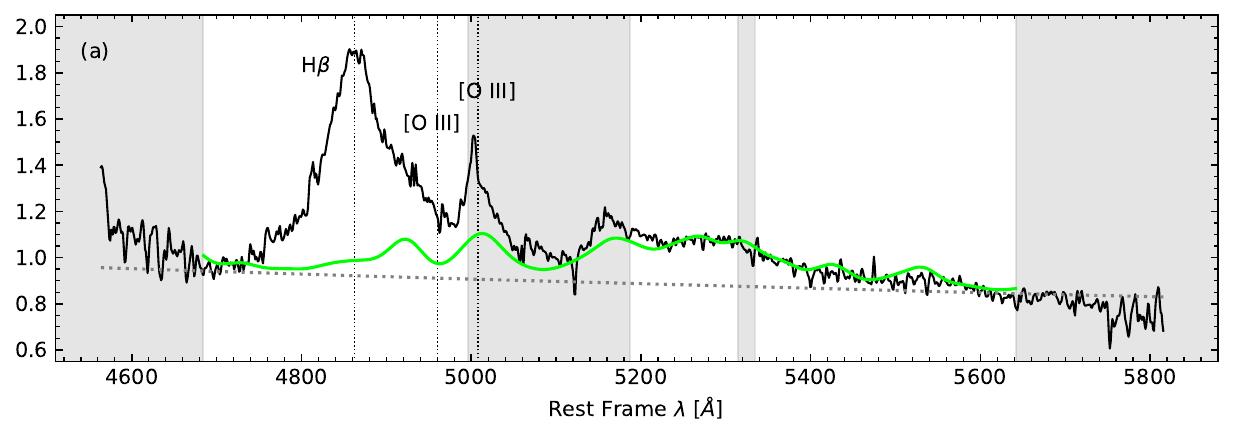}
    \includegraphics[width=0.305\linewidth]{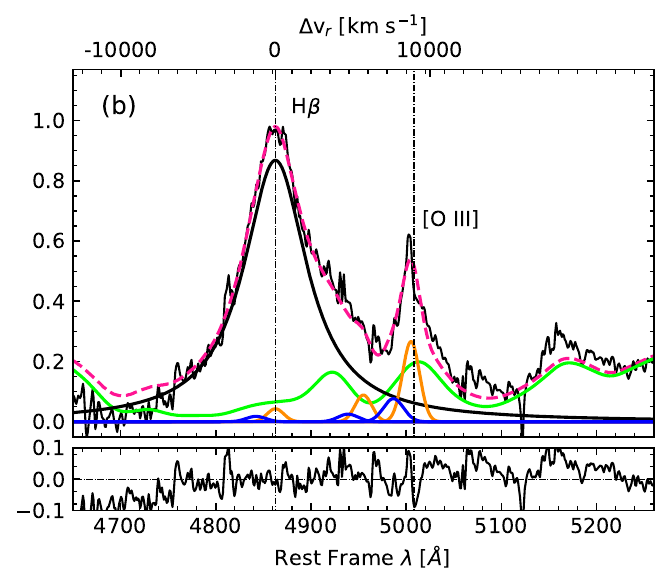}
    \caption{FBQS J2149-0811. Colours and lines as Figure \ref{fig:PKS0226}.}
    \label{fig:2149}
\end{figure}
\clearpage

\section{Radio properties and UV and optical parameters of the samples}
\label{app:radio-hems-fos}
\LTcapwidth=1\textwidth
\begin{longtable}{p{4cm}cccccccccccc}
\caption{ {Identification of the sources, radio properties, and results of the UV and optical parameters obtained through the re-analysis by multi-component fitting and used in this study for both the HEMS and FOS samples.}}\\

    \noalign{\smallskip}
    \hline
    \hline
    &&&&&&& \multicolumn{2}{c}{\oiiionly{}} & &\multicolumn{2}{c}{\civonly{}}\\
    \cline{8-9} \cline{11-12}
    Source & RA & DEC & $z$ & $P_{\textrm{1.4GHz}}$ & log($R_{\textrm{K}}$) & Radio &  $c(1/4)$ & $c(1/2)$ & & $c(1/4)$ & $c(1/2)$\\
    &&&&&&Class. & [km s$^{-1}$] & [km s$^{-1}$] & & [km s$^{-1}$]  & [km s$^{-1}$]\\
      (1) & (2) & (3) & (4) & (5) & (6) & (7) & (8) & (9) & & (10) & (11) \\ 
    \midrule
    \endfirsthead
    \caption{ {Identification of the sources, radio properties, and results of the UV and optical parameters obtained through the re-analysis by multi-component fitting and used in this study for both the HEMS and FOS samples. (cont.)}}\\
    \toprule
    \toprule
    &&&&&&& \multicolumn{2}{c}{\oiiionly{}} & &\multicolumn{2}{c}{\civonly{}}\\
    \cline{8-9} \cline{11-12}
    Source & RA & DEC & $z$ & $P_{\textrm{1.4GHz}}$ & log($R_{\textrm{K}}$) & Radio &  $c(1/4)$ & $c(1/2)$ & & $c(1/4)$ & $c(1/2)$\\
    &&&&&&Class. & [km s$^{-1}$] & [km s$^{-1}$] & & [km s$^{-1}$] & [km s$^{-1}$]\\
      (1) & (2) & (3) & (4) & (5) & (6) & (7) & (8) & (9) & & (10) & (11)\\ 
    \midrule
    \endhead
    \midrule
\endfoot
    \hline
\endlastfoot
\multicolumn{12}{c}{HEMS}\\
\midrule
HE 0035-2853	&	00 38 06.5	&	–28 36 49 	&	1.638	&	< 0.08	&	< 0.31	&	RQ	&	-137 $\pm$ 405	&	-33 $\pm$ 34	&	& -1080 $\pm$ 970	&	-1530 $\pm$ 550\\
HE 0043-2300	&	00 45 39.5	&	–22 43 56	&	1.540	&	12.41	&	2.36	&	RL	&	-312 $\pm$ 144	&	-231 $\pm$ 55 & &	-1110 $\pm$ 720	&	-980 $\pm$ 380\\
HE 0058-3231 & 01 00 39.2 & -32 14 57 & 1.582 & < 0.07 & < 0.35 & RQ &	-109 $\pm$ 47	&	-109 $\pm$ 16	&	& 190 $\pm$ 1190	&	-290 $\pm$ 400\\
HE 0109-3518	&	01 11 43.5	&	–35 03 01	&	2.406	&	< 0.17  &	< -0.09	&	RQ	&	-32 $\pm$ 49	&	-32 $\pm$ 15	&	& -1050 $\pm$ 530	&	-820 $\pm$ 260\\
HE 0122-3759	&	01 24 17.4	&	–37 44 23	&	2.200	&	< 0.14	&	< 0.23	&	RQ	&	-1051 $\pm$ 103	&	-965 $\pm$ 600	& &	-4800 $\pm$ 540	&	-3930 $\pm$ 870\\
HE 0203-4627	&	02 05 52.4	&	–46 13 30	&	1.438	&	15.45	&	2.07	&	RL	&	-318 $\pm$ 259	&	-284 $\pm$ 18	& &	-1020 $\pm$ 880	&	-910 $\pm$ 460\\
HE 0205-3756 & 02 07 27.2 & -37 41 57 & 2.433 & 4.78 & 1.51 & RQ &	-480 $\pm$ 210	&	-480 $\pm$ 102	&	& -3320 $\pm$ 1080	&	-2920 $\pm$ 330\\
HE 0248-3628	&	02 50 55.3	&	–36 16 35 	&	1.536	&	0.77	&	 0.83	&	RQ	&	-331 $\pm$ 286	&	-36 $\pm$ 29	&	& -3950 $\pm$ 810	&	-2790 $\pm$ 400\\
HE 0251-5550	&	02 52 40.1	&	–55 38 32	&	2.351	&	< 0.43	&	< 0.23	&	RQ	&	-480 $\pm$ 220	&	-440 $\pm$ 180	&	& -2170 $\pm$ 1860	&	-1640 $\pm$ 320\\
HE 0349-5249 & 03 50 59.3 & -52 40 35 & 1.541 & < 0.07 & <-0.08 & RQ &	4 $\pm$ 109	&	4 $\pm$ 37	&	& -1960 $\pm$ 900	&	-1750 $\pm$ 460\\
HE 0359-3959	&	04 01 14.0	&	–39 51 33	&	1.521	&	0.24	&	 0.85	&	RQ	&	-	&	-	&	& -5990 $\pm$ 610	&	-5880 $\pm$ 490\\
HE 0436-3709 & 04 38 37.3 & -37 03 41 & 1.445 & < 0.06 & < 0.37 & RQ &	-114 $\pm$ 144	&	-114 $\pm$ 50	&	& -1680 $\pm$ 1070	&	-1880 $\pm$ 580\\
HE 0507-3236	&	05 09 17.8	&	–32 32 45	&	1.577	&	< 0.07	&	< 0.18	&	RQ	&	-540 $\pm$ 175	&	-268 $\pm$ 65	&	& -3100 $\pm$ 970	&	-1670 $\pm$ 430\\
HE 0512-3329	&	05 14 10.8	&	–33 26 23 	&	1.587	&	< 0.07	&	< 0.06	&	RQ	&	-964 $\pm$ 151	&	-436 $\pm$ 73	&	& -2960 $\pm$ 920	&	-2470 $\pm$ 340\\
HE 0926-0201	&	09 29 13.5	&	–02 14 47	&	1.682	&	< 0.01	&	< -0.82	&	RQ	&	-581 $\pm$ 146	&	-403 $\pm$ 99	&	& -3280 $\pm$ 1030	&	-2280 $\pm$ 410\\
HE 0940-1050	&	09 42 53.5	&	–11 04 27	&	3.093	&	< 0.28	&	< 0.45	&	RQ	&	-1432 $\pm$ 121	&	-1378 $\pm$ 50	&	& -3290 $\pm$ 660	&	-2800 $\pm$ 600\\
HE 1039-0724	&	10 42 19.3	&	–07 40 37 	&	1.458	&	< 0.02	&	< 0.15	&	RQ	&	-592 $\pm$ 206	&	-199 $\pm$ 66	&	& -490 $\pm$ 1250	&	-700 $\pm$ 740\\
HE 1104-1805	&	11 06 33.5	&	–18 21 25	&	2.319	&	< 0.16	&	< 0.17	&	RQ	&	-60 $\pm$ 20	&	-40 $\pm$ 18	&	& -1630 $\pm$ 830	&	-1500 $\pm$ 370\\
HE 1120+0154   &	11 23 20.7 	&	01 37 48 	&	1.472	&	< 0.02	&	< -0.53	&	RQ	&	-147 $\pm$ 145	&	-60 $\pm$ 34	&	& -1110 $\pm$ 1730	&	-670 $\pm$ 420 \\
HE 1347-2457 & 13 50 38.8 & -25 12 16 & 2.599 & < 0.20 & < 0.19 & RQ &	-	&	-	&	& -5850 $\pm$ 660 	&	-5490 $\pm$ 440\\
HE 1349+0007	&	13 51 50.5	&	00 07 39 	&	1.444	&	< 0.02	&	< -0.11	&	RQ	&	-589 $\pm$ 303	&	-262 $\pm$ 36	&	& -4090 $\pm$ 840	&	-3040 $\pm$ 460\\
HE 1409+0101	&	14 12 21.7	&	00 47 19	&	1.650	&	0.43	&	 0.65	&	RQ	&	-806 $\pm$ 154	&	-731 $\pm$ 67	&	& -2900 $\pm$ 1040	&	-2790 $\pm$ 420\\
HE 2147-3212	&	21 50 52.3	&	–31 58 26	&	1.543	&	< 0.07	&	< 0.33	&	RQ	&	-844 $\pm$ 189	&	-96 $\pm$ 65	&	& -4440 $\pm$ 1090	&	-3110 $\pm$ 620\\
HE 2156-4020	&	21 59 54.7	&	–40 05 50	&	2.543	&	< 0.19	&	< 0.36	&	RQ	&	-768 $\pm$ 84	&	-724 $\pm$ 36	&	& -1870 $\pm$ 900	&	-1850 $\pm$ 370\\
HE 2202-2557 	&	22 05 29.8	&	–25 42 23 	&	1.535	&	4.24	&	2.13	&	RL	&	-518 $\pm$ 110	&	-339 $\pm$ 126	&	& -1280 $\pm$ 1000	&	-2080 $\pm$ 420\\
HE 2349-3800	&	23 52 10.7	&	–37 43 22	&	1.604	&	4.66	&	2.07	&	RL	&	-653 $\pm$ 116	&	-647 $\pm$ 45	&	& -2810 $\pm$ 1120	&	-2070 $\pm$ 490\\
HE 2352-4010	&	23 55 34.5	&	–39 53 54 	&	1.580	&	< 0.07	&	< 0.28	&	RQ	&	-1544 $\pm$ 162	&	-2113 $\pm$ 219	&	& -4370 $\pm$ 720	&	-2390 $\pm$ 390\\
HE 2355-4621 & 23 58 09.2 & -46 05 00 & 2.382 & < 0.17 & < 0.30 & RQ &	-151 $\pm$ 96	&	-151 $\pm$ 33	&	& 190 $\pm$ 1360 &	10 $\pm$ 350\\

\midrule
\multicolumn{12}{c}{FOS}\\
\midrule
ICRF J000559.2+160949	&	00 05 59.2	&	+16 09 49	&	0.450	&	4.71	&	2.99	&	RL	&	11 $\pm$ 34	&	11 $\pm$ 12	&	& 254 $\pm$ 789	&	138 $\pm$ 157\\
Mrk 335	&	00 06 19.5	&	+20 12 10	&	0.025	&	 $10^{-4}$	&	0.31	&	RQ	&	69 $\pm$ 46	&	67 $\pm$ 16	&	& -313 $\pm$ 468	&	-31 $\pm$ 87\\
LEDA 1790	&	00 29 13.7	&	+13 16 03	&	0.145	&	$10^{-3}$	&	0.67	&	RQ	&	2 $\pm$ 31	&	1 $\pm$ 10	&	& -1253 $\pm$ 1573	&	-69 $\pm$ 181\\
PG 0044+030	&	00 47 05.9	&	+03 19 54	&	0.623	&	0.09	&	1.12	&	RQ	&	-241 $\pm$ 78	&	-141 $\pm$ 53	&	& -1054 $\pm$ 812	&	-729 $\pm$ 227\\
Mrk 1502	&	00 53 34.9	&	+12 41 35	&	0.061	&	$10^{-4}$	&	0.33	&	RQ	&	-1311 $\pm$ 167	&	-857 $\pm$ 158	&	& -2286 $\pm$ 398	&	-1458 $\pm$ 552\\
LBQS 0100+0205	&	01 03 12.9	&	+02 21 09	&	0.394	&	< 0.01 &	< 0.71	&	RQ	&	-62 $\pm$ 58	&	-61 $\pm$ 20	&	& 12 $\pm$ 744	&	-102 $\pm$ 165\\
3C 057 	&	02 01 57.1	&	-11 32 33	&	0.671	&	37.92	&	4.31	&	RL	&	-496 $\pm$ 98	&	-341 $\pm$ 84	& & 	-943 $\pm$ 599	&	-1007 $\pm$ 185\\
3C 84	&	03 19 48.1	&	+41 30 42	&	0.018	&	0.17	&	3.65	&	RL	&	-119 $\pm$ 35	&	-120 $\pm$ 12	& &	-122 $\pm$ 25	&	-190 $\pm$ 78\\
$[\textrm{HB89}]$ 0403-132 	&	04 05 34.0	&	-13 08 13	&	0.571	&	41.00	&	4.28	&	RL	&	-3 $\pm$ 38	&	-3 $\pm$ 13	& &	455 $\pm$ 1171	&	239 $\pm$ 111\\
LEDA 75249	&	04 52 30.0	&	-29 53 35	&	0.286	&	0.02	&	1.27	&	RQ	&	-76 $\pm$ 53	&	-62 $\pm$ 18	& &	-1501 $\pm$ 463	&	-1021 $\pm$ 233\\
Mrk 1095	&	05 16 11.4	&	-00 08 59	&	0.032	&	$10^{-4}$ &	0.46	&	RQ	&	65 $\pm$ 40	&	65 $\pm$ 13	& &	-566 $\pm$ 213	&	-236 $\pm$ 180\\
ICRF J074541.6+314256	&	07 45 41.6	&	+31 42 56	&	0.461	&	3.90	&	2.87	&	RL	&	-16 $\pm$ 51	&	5 $\pm$ 12	& &	349 $\pm$ 1079	&	-240 $\pm$ 182\\
ICRF J084047.5+131223	&	08 40 47.5	&	+13 12 23	&	0.680	&	37.05	&	5.27	&	RL	&	-61 $\pm$ 86	&	4 $\pm$ 17	&& 	459	$\pm$ 668 &	225 $\pm$ 131\\
$[\textrm{HB89}]$ 0850+440	&	08 53 34.2	&	+43 49 02	&	0.515	&	< $10^{-3}$ 	&	< 0.07	&	RQ	&	-1073 $\pm$ 203	&	-329 $\pm$ 212	& &	-1669 $\pm$ 669	&	-1423 $\pm$ 298\\
LB 9308	&	09 06 31.8	&	+16 46 11	&	0.411	&	7.66	&	4.91	&	RL	&	204 $\pm$ 99	&	246 $\pm$ 18	&	& 420 $\pm$ 1845	&	142 $\pm$ 187\\
ICRF J092703.0+390220	&	09 27 03.0	&	+39 02 20	&	0.695	&	42.84	&	4.19	&	RL	&	-54 $\pm$ 46	&	-55 $\pm$ 15	& &	1189 $\pm$ 1462	&	424 $\pm$ 138\\
PG 0947+396	&	09 50 48.3	&	+39 26 50	&	0.207	&	< $10^{-4}$	&	< 0.13	&	RQ	&	-97 $\pm$ 89	&	30 $\pm$ 19	& &	2 $\pm$ 682	&	-231 $\pm$ 134\\
PG 0953+414	&	09 56 52.3	&	+41 15 22	&	0.235	&	< $10^{-4}$	&	< -0.73	&	RQ	&	-123 $\pm$ 85	&	-89 $\pm$ 25	& &	-207 $\pm$ 682	&	-58 $\pm$ 128\\
Ton 469	&	09 58 20.9	&	+32 24 02	&	0.530	&	10.34	&	3.21	&	RL	&	65 $\pm$ 93	&	102 $\pm$ 28	& &	1566 $\pm$ 1669	&	969 $\pm$ 465\\
SDSS J100402.61+285535.3	&	10 04 02.6	&	+28 55 35	&	0.330	&	< $10^{-3}$	&	< -0.01	&	RQ	&	-583 $\pm$ 97	&	-585 $\pm$ 25	& &	-1725 $\pm$ 1034	&	-1149 $\pm$ 152\\
LEDA 29208	&	10 04 20.1	&	+05 13 00	&	0.161	&	< $10^{-4}$		&	< -0.09	&	RQ	&	-187 $\pm$ 75	& 	-176 $\pm$ 23	& &	{-592 $\pm$ 726}	&	{-226 $\pm$ 144}\\
FBQS J101027.5+413238	&	10 10 27.5	&	+41 32 39	&	0.613	&	5.64	&	3.24	&	RL	&	53 $\pm$ 98	&	47 $\pm$ 18	& &	-213 $\pm$ 1008	&	-33 $\pm$ 198\\
NGC 3227	&	10 23 30.5	&	+19 51 54	&	0.004	&	$10^{-5}$	&	1.27	&	RQ	&	{40 $\pm$ 56	}&	{30 $\pm$ 17}	& &	-155 $\pm$ 89	&	-195 $\pm$ 105\\
7C 1028+3118	&	10 30 59.0	&	+31 02 55	&	0.178	&	0.19	&	2.73	&	RL	&	90 $\pm$ 33	&	85 $\pm$ 8	& &	-321 $\pm$ 1078	&	35 $\pm$ 201\\
PG 1049-005	&	10 51 51.4	&	-00 51 17	&	0.359	&	< $10^{-3}$ &	< -0.13	&	RQ	&	-79 $\pm$ 84	&	-42 $\pm$ 18	& &	-102 $\pm$ 793	&	-152 $\pm$ 131\\
2MASX J10523275+6125211	&	10 52 32.7	&	+61 25 20	&	0.421	&	5.01	&	3.67	&	RL	&	101 $\pm$ 37	&	101 $\pm$ 12	& &	1143 $\pm$ 1574	&	149 $\pm$ 182\\
PG 1100+772	&	11 04 13.8	&	+76 58 58	&	0.312	&	6.25	&	3.26	&	RL	&	-54 $\pm$ 53	&	-46 $\pm$ 17	& &	37 $\pm$ 1144	&	38 $\pm$ 286\\
NGC 3516	&	11 06 47.4	&	+72 34 07	&	0.009	&	$10^{-5}$	&	1.31	&	RQ	&	-58 $\pm$ 36	&	-42 $\pm$ 11	& &	-861 $\pm$ 617	&	68 $\pm$ 43\\
ICRF J110715.0+162802	&	11 07 15.0	&	+16 28 02	&	0.632	&	10.51	&	3.60	&	RL	&	110 $\pm$ 50	&	101 $\pm$ 10	& &  	614 $\pm$ 755	&	323 $\pm$ 163\\
3C 254	&	11 14 38.7	&	+40 37 20	&	0.737	&	52.56	&	4.49	&	RL	&	211 $\pm$ 207	&	266 $\pm$ 17	& &	1282 $\pm$ 1349	&	509 $\pm$ 116\\
PG 1114+445	&	11 17 06.3	&	+44 13 33	&	0.143	&	< $10^{-4}$	&	< -0.34	&	RQ	&	-61 $\pm$ 57	&	-35 $\pm$ 20	& &	343 $\pm$ 734	&	145 $\pm$ 100\\
LEDA 34570	&	11 18 30.2	&	+40 25 54	&	0.154	&	$10^{-3}$	&	0.60	&	RQ	&	-171 $\pm$ 47	&	-170 $\pm$ 16	& &	-784 $\pm$ 710	&	-241 $\pm$ 150\\
PG 1116+215 &	11 19 08.6	&	+21 19 17	&	0.177	&	$10^{-3}$	&	0.42	&	RQ	&	-280 $\pm$ 155	&	-106 $\pm$ 49	& &	-1238 $\pm$ 722	&	-848 $\pm$ 235\\
2MASS J11243917+4201450	&	11 24 39.1	&	+42 01 45	&	0.225	&	< $10^{-4}$	&	< -0.09	&	RQ	&	67 $\pm$ 50	&	69 $\pm$ 16	& &	-44 $\pm$ 389	&	-30 $\pm$ 76\\
LBQS 1138+0204	&	11 41 21.7	&	+01 48 03	&	0.382	&	< $10^{-3}$	&	< 0.09	&	RQ	&	15 $\pm$ 27	&	23 $\pm$ 9	& &	-1076 $\pm$ 936	&	-871 $\pm$ 248\\
LBQS 1144-0115	&	11 47 18.0	&	-01 32 07	&	0.383	&	< $10^{-3}$	&	< -0.03	&	RQ	&	 {$-455 \pm 55$**}	&	 {$-377 \pm 30$**}	& &	189 $\pm$ 983	&	267 $\pm$ 158\\
LB 2136	&	11 53 24.4	&	+49 31 08	&	0.334	&	4.86	&	4.04	&	RL	&	-29 $\pm$ 31	&	-21 $\pm$ 12	& &	884 $\pm$ 777	&	416 $\pm$ 130\\
LEDA 38224	&	12 04 42.1	&	+27 54 11	&	0.165	&	< $10^{-4}$	&	< -0.16	&	RQ	&	-42 $\pm$ 74	&	-11 $\pm$ 16	& &	{-637 $\pm$ 937}	&	{-174 $\pm$ 111}\\
PB 3894	&	12 14 17.6	&	+14 03 13	&	0.081	&	$10^{-4}$	&	0.46	&	RQ	&	-45 $\pm$ 37	&	-38 $\pm$ 13	& &	-198 $\pm$ 441	&	-138 $\pm$ 86\\
NGC 4253	&	12 18 26.5	&	+29 48 46	&	0.012	&	$10^{-4}$	&	1.28	&	RQ	&	32 $\pm$ 31	&	32 $\pm$ 11	& &	-91 $\pm$ 385	&	-89 $\pm$ 76\\
PG 1216+069	&	12 19 20.9	&	+06 38 38	&	0.332	&	$10^{-3}$	&	0.25	&	RQ	&	10 $\pm$ 35	&	10 $\pm$ 12	& &	-761 $\pm$ 1458	&	137 $\pm$ 158\\
3C 273	&	12 29 06.6	&	+02 03 08	&	0.158	&	23.86	&	3.45	&	RL	&	{-312 $\pm$ 280}	&	{-268 $\pm$ 62}	& &	-618 $\pm$ 951	&	-549 $\pm$ 107\\
SBS 1250+568	&	12 52 26.3	&	+56 34 19	&	0.320	&	6.47	&	4.42	&	RL	&	110 $\pm$ 39	&	110 $\pm$ 13	& &	133 $\pm$ 799	&	-60 $\pm$ 110\\
SBS 1259+593	&	13 01 12.9	&	+59 02 06	&	0.478	&	< $10^{-3}$	&	< -0.35	&	RQ	&	-102 $\pm$ 482	&	-81 $\pm$ 36	& &	-3034 $\pm$ 597	&	-2681 $\pm$ 239\\
ICRF J130533.0-103319	&	13 05 33.0	&	-10 33 19	&	0.278	&	1.50	&	2.71	&	RL	&	-261 $\pm$ 91	&	-143 $\pm$ 69	& &	619 $\pm$ 1585	&	9 $\pm$ 127\\
LEDA 45656	&	13 09 47.0	&	+08 19 48	&	0.155	&	< $10^{-4}$	&	< -0.33	&	RQ	&	-73 $\pm$ 67	&	-23 $\pm$ 18	& &	103 $\pm$ 1049	&	117 $\pm$ 147\\
FBQS J131217.7+351521	&	13 12 17.7	&	+35 15 21	&	0.182	&	0.04	&	1.56	&	RQ	&	{-128 $\pm$ 135}	&	{-33 $\pm$ 31}	& &	-1099 $\pm$ 1392	&	190 $\pm$ 464\\
2XMM J135315.8+634546	&	13 53 15.8	&	+63 45 45	&	0.088	&	$10^{-3}$	&	1.22	&	RQ	&	-159 $\pm$ 98	&	-79 $\pm$ 24	& &	195 $\pm$ 761	&	152 $\pm$ 107\\
PB 4142 	&	13 54 35.6	&	+18 05 17	&	0.151	&	< $10^{-4}$	&	< -0.15	&	RQ	&	-227 $\pm$ 66	&	-122 $\pm$ 54	& &	{-376 $\pm$ 972}	&	{-101 $\pm$ 174}\\
$[\rm HB89]$ 1354+195	&	13 57 04.4	&	+19 19 07	&	0.720	&	41.34	&	3.87	&	RL	&	-53 $\pm$ 41	&	-53 $\pm$ 14	& &	731 $\pm$1199	&	13 $\pm$ 153\\
FBQS J1405+2555	&	14 05 16.2	&	+25 55 34	&	0.163	&	$10^{-4}$ &	0.12	&	RQ	&	-179 $\pm$ 118	&	-195 $\pm$ 39	& &	-853 $\pm$ 730	&	-454 $\pm$ 250\\
LEDA 50313	&	14 06 21.8	&	+22 23 46	&	0.097	&	$10^{-4}$	&	0.42	&	RQ	&	-339 $\pm$ 84	&	-248 $\pm$ 35	& &	-1571 $\pm$ 876$^{\dag}$	&	-1537 $\pm$ 875$^{\dag}$\\
$[\textrm{HB89}]$ 1415+451	&	14 17 00.8	&	+44 56 06	&	0.115	&	$10^{-4}$	&	0.36	&	RQ	&	-593 $\pm$ 40	&	-593 $\pm$ 13	& &	-1103 $\pm$ 436	&	-974 $\pm$ 144\\
NGC 5548	&	14 17 59.5	&	+25 08 12	&	0.017	&	$10^{-4}$&	0.88	&	RQ	&	-15 $\pm$ 44	&	-2 $\pm$ 14	& &	199 $\pm$ 993	&	274 $\pm$ 170\\
Mrk 813	&	14 27 25.0	&	+19 49 52	&	0.111	&	$10^{-4}$ &	0.39	&	RQ	&	187 $\pm$ 70	&	176 $\pm$ 20	& &	-161 $\pm$ 1561	&	-180 $\pm$ 411\\
$[\textrm{HB89}]$ 1425+267	&	14 27 35.6	&	+26 32 14	&	0.363	&	0.40	&	2.40	&	RL	&	-244 $\pm$ 76	&	32 $\pm$ 30	& &	-964 $\pm$ 1275	&	-908 $\pm$ 298\\
2MASS J14294306+4747262	&	14 29 43.0	&	+47 47 26	&	0.220	&	< $10^{-4}$	&	0.29	&	RQ	& 	-160 $\pm$ 200	&	-143 $\pm$ 150 & &	175 $\pm$ 511	&	175 $\pm$ 103\\
Mrk 478	&	14 42 07.4	&	+35 26 22	&	0.077	&	$10^{-4}$	&	0.33	&	RQ	&	-15 $\pm$ 53	&	-10 $\pm$ 20	& &	-955 $\pm$ 494	&	-755 $\pm$ 142\\
$[\textrm{HB89}]$ 1444+407	&	14 46 45.9	&	+40 35 05	&	0.267	&	< $10^{-4}$	&	< -0.32	&	RQ	&	-169 $\pm$ 50	&	-174 $\pm$ 15	& &	-1411 $\pm$ 670	&	-1284 $\pm$ 207\\
ICRF J145427.4-374733	&	14 54 27.4	&	-37 47 33	&	0.314	&	2.56	&	3.59	&	RL	&	21 $\pm$ 41	&	24 $\pm$ 13	& &	1172 $\pm$ 814 	&	698 $\pm$ 110\\
ICRF J151443.0+365050	&	15 14 43.0	&	+36 50 50	&	0.371	&	3.83	&	3.29	&	RL	&	99 $\pm$ 74	&	120 $\pm$ 11	& &	{-198 $\pm$ 891}	&	{-48 $\pm$ 271}\\
$[\textrm{HB89}]$ 1538+477	&	15 39 34.8	&	+47 35 31	&	0.772	&	0.17	&	1.26	&	RQ	&	-124 $\pm$ 52	&	-111 $\pm$ 17	& &	-1338 $\pm$ 1231	&	-546 $\pm$ 306\\
$[\rm HB89]$ 1543+489	&	15 45 30.2	&	+48 46 08	&	0.398	&	0.01	&	0.70	&	RQ	&	-1985 $\pm$ 170	&	-1777 $\pm$ 88	& &	-2287 $\pm$ 569	&	-1911 $\pm$ 228\\
ICRF J154743.5+205216	&	15 47 43.5	&	+20 52 16	&	0.266	&	0.91	&	2.88	&	RL	&	41 $\pm$ 56	&	37 $\pm$ 16	& &	352 $\pm$ 1400	&	-114 $\pm$ 183\\
Mrk 493	&	15 59 09.6	&	+35 01 47	&	0.031	&	$10^{-4}$	&	0.84	&	RQ	&	-21 $\pm$ 28	&	-22 $\pm$ 9	& &	53 $\pm$ 313	&	51 $\pm$ 62\\
Ton 256	&	16 14 13.2	&	+26 04 16	&	0.131	&	$10^{-3}$	&	1.29	&	RQ	&	-30 $\pm$ 40	&	-19 $\pm$ 13	& &	1395 $\pm$ 1244	&	304 $\pm$ 111\\
3C 334	&	16 20 21.8	&	+17 36 23	&	0.556	&	2.18	&	2.88	&	RL	&	-31 $\pm$ 73	&	3 $\pm$ 21	& &	109 $\pm$ 1402	&	-77 $\pm$ 374\\
SBS 1626+554	&	16 27 56.1	&	+55 22 31	&	0.133	&	< $10^{-4}$	&	< -0.39	&	RQ	&	-29 $\pm$ 62	&	-29 $\pm$ 24	& &	10 $\pm$ 957	&	-208 $\pm$ 147\\
FBQS J163020.7+375656	&	16 30 20.7	&	+37 56 56	&	0.395	&	0.10	&	1.62	&	RQ	&	6 $\pm$ 76	&	70 $\pm$ 19	& &	1573 $\pm$713	&	1174 $\pm$ 190\\
3C 345	&	16 42 58.8	&	+39 48 36	&	0.593	&	69.71	&	5.01	&	RL	&	16 $\pm$ 97	&	-2 $\pm$ 21	& &	389 $\pm$ 245	&	250 $\pm$ 213\\
3C 351.0	&	17 04 41.3	&	+60 44 30	&	0.372	&	< $10^{-3}$	&	< 0.34	&	RQ	&	-45 $\pm$ 53	&	-27 $\pm$ 14	& &	251 $\pm$ 1753	&	120 $\pm$ 238\\
4C 73.18	&	19 27 48.4	&	+73 58 01	&	0.302	&	9.89	&	3.79	&	RL	&	-4 $\pm$ 48	&	-2 $\pm$ 13	& &	320 $\pm$ 859	&	76 $\pm$ 144\\
Mrk 509	&	20 44 09.7	&	-10 43 24	&	0.034	&	$10^{-4}$	&	0.44	&	RQ	&	-19 $\pm$ 40	&	-15 $\pm$ 13	& &	-820 $\pm$ 601	&	-481 $\pm$ 154\\
PG 2112+059	&	21 14 52.5	&	+06 07 42	&	0.461	&	0.01	&	0.27	&	RQ	&	307 $\pm$ 108	&	307 $\pm$ 37	& &	-1774 $\pm$ 584	&	-1554 $\pm$ 209\\
$[\rm HB89]$ 2128-123	&	21 31 35.2	&	-12 07 04	&	0.500	&	12.94	&	3.22	&	RL	&	-19 $\pm$ 42	&	-9 $\pm$ 10	& &	1019 $\pm$ 1392	&	400 $\pm$ 168\\
4C 31.63	&	22 03 14.9	&	+31 45 38	&	0.295	&	6.86	&	3.40	&	RL	&	 {$-741 \pm 76$**}	&	 {$-728 \pm 26$**}	& &	-664 $\pm$ 1243	&	-135 $\pm$ 238\\
UGC 12163	&	22 42 39.3	&	+29 43 31	&	0.025	&	$10^{-4}$ &	1.05	&	RQ	&	32 $\pm$ 31	&	32 $\pm$ 11	& &	5 $\pm$ 323	&	-61 $\pm$ 318\\
$[\rm HB89]$ 2243-123	&	22 46 18.2	&	-12 06 51	&	0.626	&	22.28	&	3.96	&	RL	&	{-244 $\pm$ 64}	&	{-221 $\pm$ 18}	& &	879 $\pm$ 1001	&	212 $\pm$ 95\\
MR 2251-178	&	22 54 05.8	&	-17 34 55	&	0.064	&	$10^{-3}$ &	0.53	&	RQ	&	-23 $\pm$ 56	&	-23 $\pm$ 19	& &	-613 $\pm$ 1153	&	-539 $\pm$ 189\\
4C 11.72	&	22 54 10.4	&	+11 36 38	&	0.326	&	4.21	&	3.18	&	RL	&	106 $\pm$ 65	&	113 $\pm$ 19	& &	-312 $\pm$ 1513	&	-206 $\pm$ 127\\
NGC 7469	&	23 03 15.6 	&	+08 52 26	&	0.016	&	$10^{-3}$	&	0.87	&	RQ	&	-83 $\pm$ 48	&	-59 $\pm$ 12	& &	-93 $\pm$ 258	&	386 $\pm$ 59\\
4C 09.72	&	23 11 17.7	&	+10 08 15	&	0.434	&	4.21	&	3.16	&	RL	&	-55 $\pm$ 114	&	-17 $\pm$ 18	& &	368 $\pm$ 1839	&	114 $\pm$ 158 \\
ICRF J234636.8+093045	&	23 46 36.8	&	+09 30 45	&	0.672	&	24.90	&	3.61	&	RL	&	42 $\pm$ 83	&	120 $\pm$ 28	& &	-252 $\pm$ 1036	&	-129 $\pm$ 143\\
$[\textrm{HB89}]$ 2349-014 &	23 51 56.1	&	-01 09 13	&	0.174	&	1.26	&	3.50	&	RL	&	187 $\pm$ 70	&	176 $\pm$ 20	& &	-1585 $\pm$ 771 &	-985 $\pm$ 302\\

    \label{tab:radio_properties}
    \label{tab:radio_fos_hems}
\end{longtable}
\tablefoot{ {(1) Source identification. (2) Right ascension. (3) Declination. (4) Redshift. (5) Radio power at 1.4 GHz, in units of $10^{26}$ W Hz$^{-1}$. (6) Radioloudness parameter. (7) Radio classification. (8), (9) Velocity centroids at $\frac{1}{4}$ and $\frac{1}{2}$ intensities of the \oiiionly{} full profile. (10), (11) Same for the \civonly{} full profile. ** \oiiionly{} emission strongly affected by atmospheric lines.$^{\dag}$ \civonly{} emission strongly affected by absorptions.}}
\clearpage

\twocolumn
\section{Details on the estimations of the outflow parameters}

\label{app:outflows}
\subsection{\oiii{}}
\label{app:oiii_outflow}

\par  The expressions for the \oiiiseven{} outflow parameters  were derived by \citet{Marziani_2017} assuming a bipolar outflow structure  {\citep[c.f. ][and references therein]{Kim_2023}}. All the \oiiiseven{} outflow equations reported here are in accordance with the work of \citet{canodiazetal12} and \citet{Fiore_2017}.

The \oiiiseven{} luminosity can be related to the outflow mass by
\begin{equation}
    M_{\textrm{ion}}=1\times 10^7\ L^\mathrm{out}_{\textrm{[OIII],44}}\left(\frac{Z}{5Z_{\odot}}\right)^{-1}n_\mathrm{H,3}^{-1} \ \  \mathrm{[M_\odot]},
\end{equation}
where $L^\mathrm{out}_{\textrm{[OIII],44}}$ is the outflow-emitted luminosity of \oiiiseven{} emission line in units of $10^{44}$\,erg\,s$^{-1}$, the density ($n_\mathrm{H}$) and the metallicity ($Z$) have been scaled to $10^3$\,cm$^{-3}$ and 5 times solar, respectively.\\

If we assume that the outflow is confined to a solid angle $\Omega$, then the mass outflow rate $\dot{M}^{\textrm{out}}_\mathrm{ion}$ at a distance $r$ can be written as:
\begin{equation}
    \label{eq:mout_oiii}    \dot{M}_{\textrm{ion}}=30\,L_{\textrm{[OIII],44}}^{\textrm{out}}\ \ v_{\textrm{o,1000}}\ \ r_{\textrm{1kpc}}^{-1}\left(\frac{Z}{5Z_{\odot}}\right)^{-1} n_\textrm{H,3}^{-1}\ [\textrm{M}_{\odot} \textrm{yr}^{-1}].
\end{equation}

Even if part of a nuclear outflow, the \oiii\ emitting gas is probably not being anymore  accelerated by radiation pressure, and so we consider as maximum outflow velocity $v_{\textrm{o}}$ the centroid at 1/2 intensity of the BLUE component. If we assume the \oiiiseven{} outflow to be nuclear, we can then suppose a super-solar chemical composition $Z=5Z_{\odot}$ { at low redshift, that is also consistent with the metallicity derived for the most metal rich bulge stars. At high redshift, there is no estimation we are aware of the chemical composition of the outflowing gas. A proxy is provided by the metallicity of the stellar component of the host galaxy \citep{xuetal18}, expected to be twice the solar value}.  \\

The thrust might be written as:
\begin{equation}
    \dot{M}_{\mathrm{ion}}v_{\mathrm{o}}=1.9\times10^{35}L_{\mathrm{[OIII],44}}^{\mathrm{out}}\ v^2_{\mathrm{o,1000}}\ r^{-1}_{\mathrm{1kpc}}\left(\frac{Z}{5Z_{\odot}}\right)^{-1}n_\mathrm{H,3}^{-1}\ [\textrm{g cm s}^{-2}].
\end{equation}
  
Similarly, the kinetic power $\dot{E}_{\mathrm{out}}$ is then given by $\dot{E}_{\mathrm{out}}\sim \frac{1}{2}\dot{M}_{\mathrm{out}}v_{\mathrm{o}}^2$, which leads to
\begin{equation}
    \dot{\epsilon}_{\mathrm{kin}}=9.6\times 10^{42}L_{\mathrm{[OIII],44}}^{\mathrm{out}}\  v_{\mathrm{o,1000}}^3\ r_{\mathrm{1kpc}}^{-1}\left(\frac{Z}{5Z_{\odot}}\right)^{-1}n_\mathrm{H,3}^{-1}\ [\textrm{erg s}^{-2}].
\end{equation}

\subsection{\civ}
\label{appendix:civ_outflows}

Both \civ\ and \oiii\ are predominantly produced by collisional excitation, for which the collisional excitation rate of Eq. 3.20 from \citet{osterbrockferland06} applies:
\begin{equation}
    \label{eq:q12}
    q_{12} = \frac{8.629\cdot 10^{-6}}{T^\frac{1}{2}} \frac{\Upsilon}{g_1} \exp{(-h\nu/{kT})}\, \, \mathrm{[cm^{+3} s^{-1}]}
\end{equation}

\noindent where $h\nu$\ is the energy of the photon emitted in the transition (1.28$\times10^{-11}$ erg in the case of \civ), and $k$ is the Boltzmann's constant in erg\,K$^{-1}$. $\Upsilon$\ is the collision strength and is almost independent from temperature (\citet{osterbrockferland06} yield $\Upsilon \approx 8.91$). The $g_1 = (2J+1) = 2$, as the lower level of the transition is  the ground state ($^2 S_\frac{1}{2}$). The exponential argument is $h\nu/kT_{10000} \approx 9.28 T^{-1}_{10000}$, and dominates over the factor $T^{-\frac{1}{2}}$.\\

The mass of ionized gas is given by 

\begin{equation}
 M_\mathrm{c^{+3}} \sim \frac{L_\mathrm{CIV}}{(h\nu)_\mathrm{CIV} q_{12} n_\mathrm{H} \mathrm{[C/H]} X_{C^{+3}} } \cdot \mu\ m_\mathrm{P}. 
 \label{eq:mion}
\end{equation}

\noindent Here $L_\mathrm{CIV}$\ is the \civ\ luminosity, $(h\nu)_\mathrm{CIV}$ the \civ\ photon energy,  [C/H] the relative abundance of Carbon with respect to Hydrogen, in number: $8.43 \approx 12 + \log \mathrm{[C/H]}$\ for solar abundance, implying [C/H] $\approx 2.69 \cdot 10^{-4}$. The factor  $X_{C^{+3}} = {[C^{+3}]}/{\sum_\mathrm{i} [C^{+\mathrm{i}}]} \le 1$\ is the {\bf ionic} fraction of triply ionized Carbon, and  $\mu m_\mathrm{P}$\ is the molecular mass of the gas, with $\mu\approx 1.4$ (neglecting the mass of the metals), and $m_\mathrm{P}$\ the proton mass.

 

The mass of the emitting gas strongly depends on temperature.  A single value of the dynamical temperature could be assumed, in keeping with the simplest assumptions. However, the $T$\ value should be chosen carefully. 

\subsubsection{Observational constraints}

To estimate a typical value of the dynamical temperature, we can apply two main observational constraints on the blueshifted emission due to the outflow:

\begin{itemize}
\item{ \civ\ equivalent width: in super-Eddington candidates, the equivalent width is typically very low $W \sim 10$\, \AA\ and in Population A  $W \lesssim 30$\, \AA\ \citep{kinneyetal90,sulenticetal00a,marzianietal16,martinezaldama_2018}. The sources that are radiating at high Eddington ratio are the ones showing the lowest $W$\ along the E1 main sequence, with a large fraction of them being weak lined quasars (WLQs) \citep{shemmeretal10,wuetal12,marzianietal16,martinezaldama_2018,jinetal23}. Even for the Pop. B sources of the HEMS and ISAAC samples $W$ \civ\ is usually below $\approx$ 50 \AA.}

\item{ \civ/\hb\ ratio: its value, due to the absence of a strong \hb\ blueshifted component, can be assumed to be \civ/\hb\ $\gtrsim 5 \sim 10$ for blueshifted components. }
\end{itemize}

The decomposition of the \civ\ profile in two components reveals very different properties and  physical conditions for the outflow and virial component. In the outflow case, the \civ\ equivalent width is in the range from the detectability limit of a few \AA\  to a few tens  \AA. 
This observational constraint  suggests that the outflow  component of the \civ\ profile is not  produced by gas in physical conditions optimized for its  emission. 

\subsubsection{Photoionization modeling}
\label{appendix:civ_modeling}

{The diagrams of Fig. \ref{fig:civew} show the $W$ of \civ\, (left panels) and the \civonly/\hb{} ratio (right panels) as a function of the ionization parameter $U$, according to CLOUDY 17.02 \citep{ferlandetal17} computations, assuming two cases: (1) \mbh\,=\,10$^9$\,M$_\odot$, five times solar metallicity, and a standard AGN SED representative for low-redshift sources that have Eddington ratios $\gtrsim 0.1 - 0.2$\,\citep{mathewsferland87} (upper panels); and (2) in bottom panels for \mbh\,=\,10$^{9.5}$\,M$_\odot$, five times solar metallicity and the SED defined by \citet{krawczyketal13}, more appropriate for representing our high-z high luminosity sources (see more details below}).
Very low $W$\ is possible in conditions of Carbon under-ionization and over-ionization  with respect to C$^{+3}$. For the blueshifted component, a high-ionization solution has been found with    typical values $\log U \sim -0.5 -  0$ \citep[][]{sniegowska_2021,garnicaetal22}. 

The {upper left panel} of Fig. \ref{fig:civew}  shows that the expected $W$(\civ) at $U \sim 1$\ is   several tens of \AA, consistent with the observations if the covering factor is moderate ($\sim 0.2 - 0.5)$, with a marginal dependence on density. The  \civ/\hb\ ratio is constrained to be $\gtrsim 5$, with  \civ/\hb\ $\sim 10$\ for the density range $10^{9.5} - 10^{10}$ cm$^{-3}$. 

The electron temperature associated with  these conditions is high ($T \sim 20\,000$ K), implying a very high efficiency for the collisional excitation. The point here is that, even if the collisional efficiency is high, the ionic fraction of C$^{+3}$ is very low ($\sim 0.01$), yielding a $\tilde{q}X$\ factor $\approx (3 - 4) \cdot 10^{-11}$\ cm$^{3}$ s$^{-1}$. 

{At high luminosity, however, the softening of the accretion disk emission due to large masses makes the SED significantly different from that of \citet{mathewsferland87}. The peak of the disk emission can be displaced toward lower frequency by a factor $\sim 10$ \citep{durasetal17,durasetal20}, reducing significantly the amount of ionizing photons. We therefore considered a different SED, defined for intermediate redshift quasars with $W$(\civonly) $\lesssim 30$\,\AA\ \citep{krawczyketal13}. The simulations predict lower $W$(\civonly) by a factor $\approx 5$ (bottom panels on Fig. \ref{fig:civew}) with respect to the \citet{mathewsferland87} SED. The $W$(\civonly) and the \civonly/\hb\ ratio provide some constraints over the ionization parameter and density. The \civonly/\hb\  {ratio falls within a range of $\sim$ 8 - 33} for the ISAAC sources where the BLUE of \hb\ has been detected, and the $W \sim 10$\AA. For $\log n_\mathrm{H}\approx9.5$, these conditions imply a $\log U$  between -0.7 and 0.3. In this case, for $\log U \approx -0.2$, and computing a weighted average over $r^2$, we obtain a much higher efficiency, $\overline{Xq(T)}$ to $ \approx 1.22\cdot10^{-10}$\,cm$^3$\,s$^{-1}$. Much lower efficiencies are possible if $\log U \rightarrow 0 $ \ and $\log n_\mathrm{H} \rightarrow 9$. In this case the predicted $W$(\civonly)\ might be below 10 \AA, as found for weak lined quasars \citep{diamond-stanicetal09,shemmeretal10,marziani_2016a}.}


\begin{figure}[t!]
    \centering
    \includegraphics[width=0.49\linewidth]{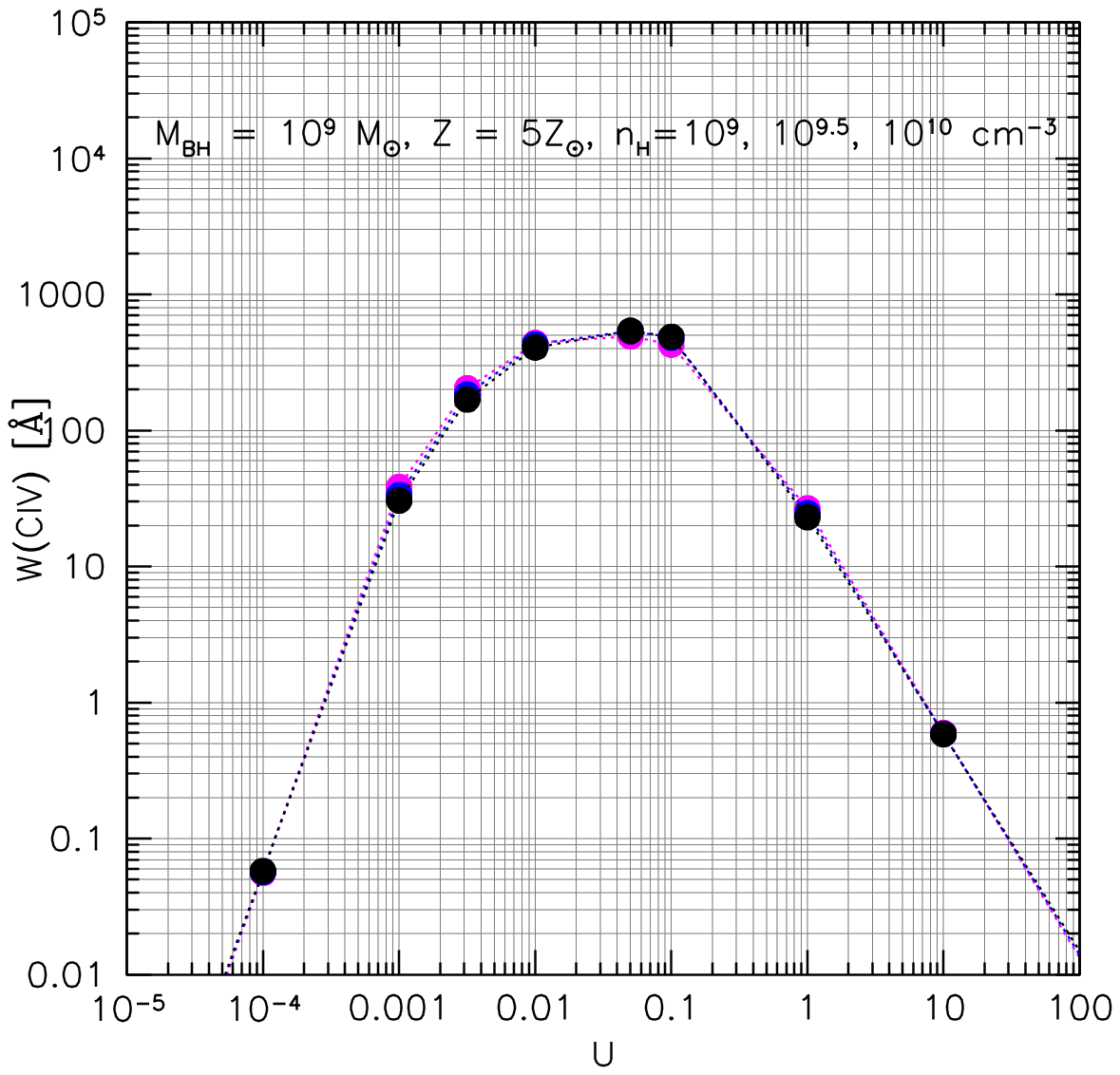}
    \includegraphics[width=0.49\linewidth]{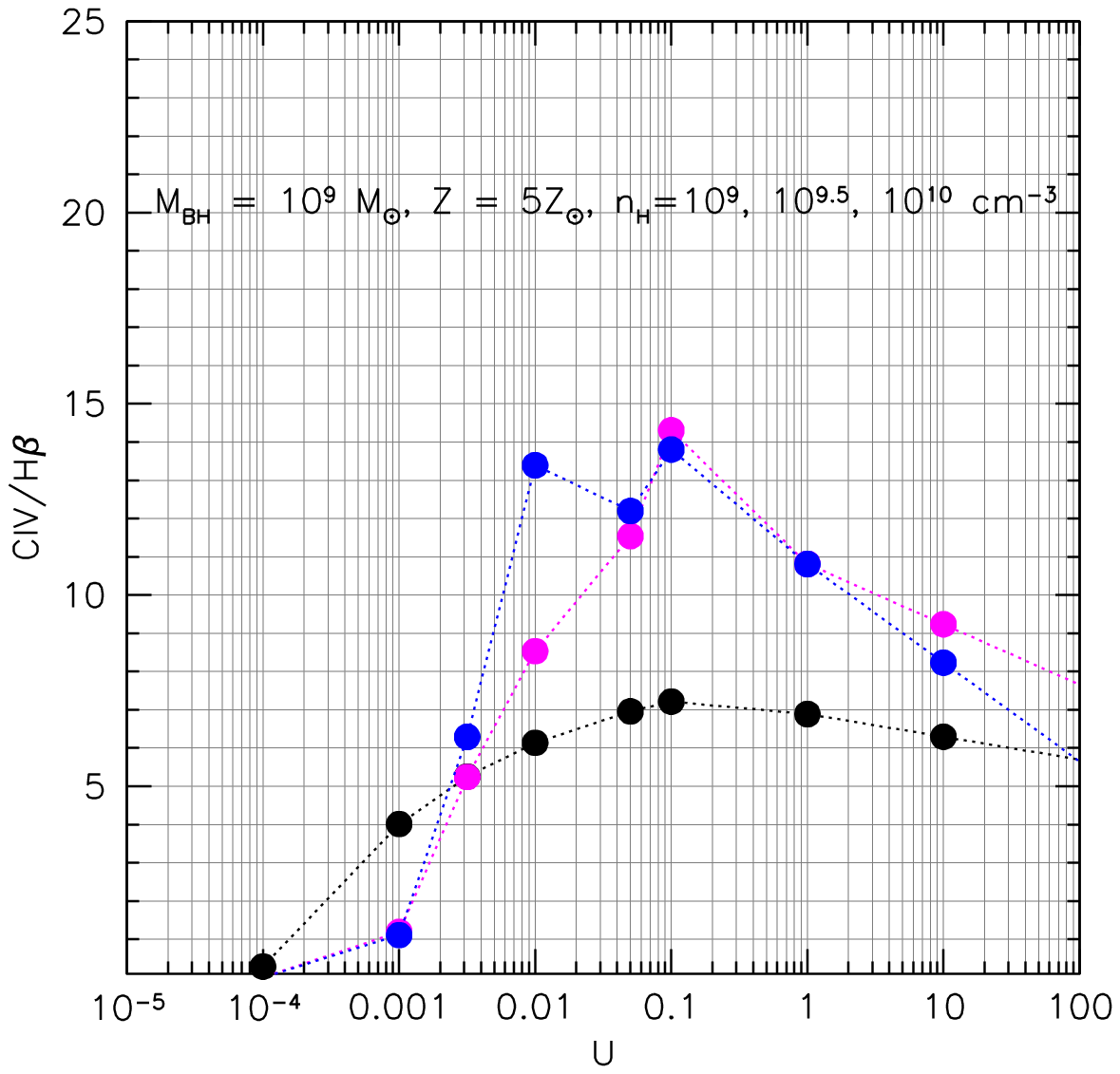}\\
    \includegraphics[width=0.49\linewidth]{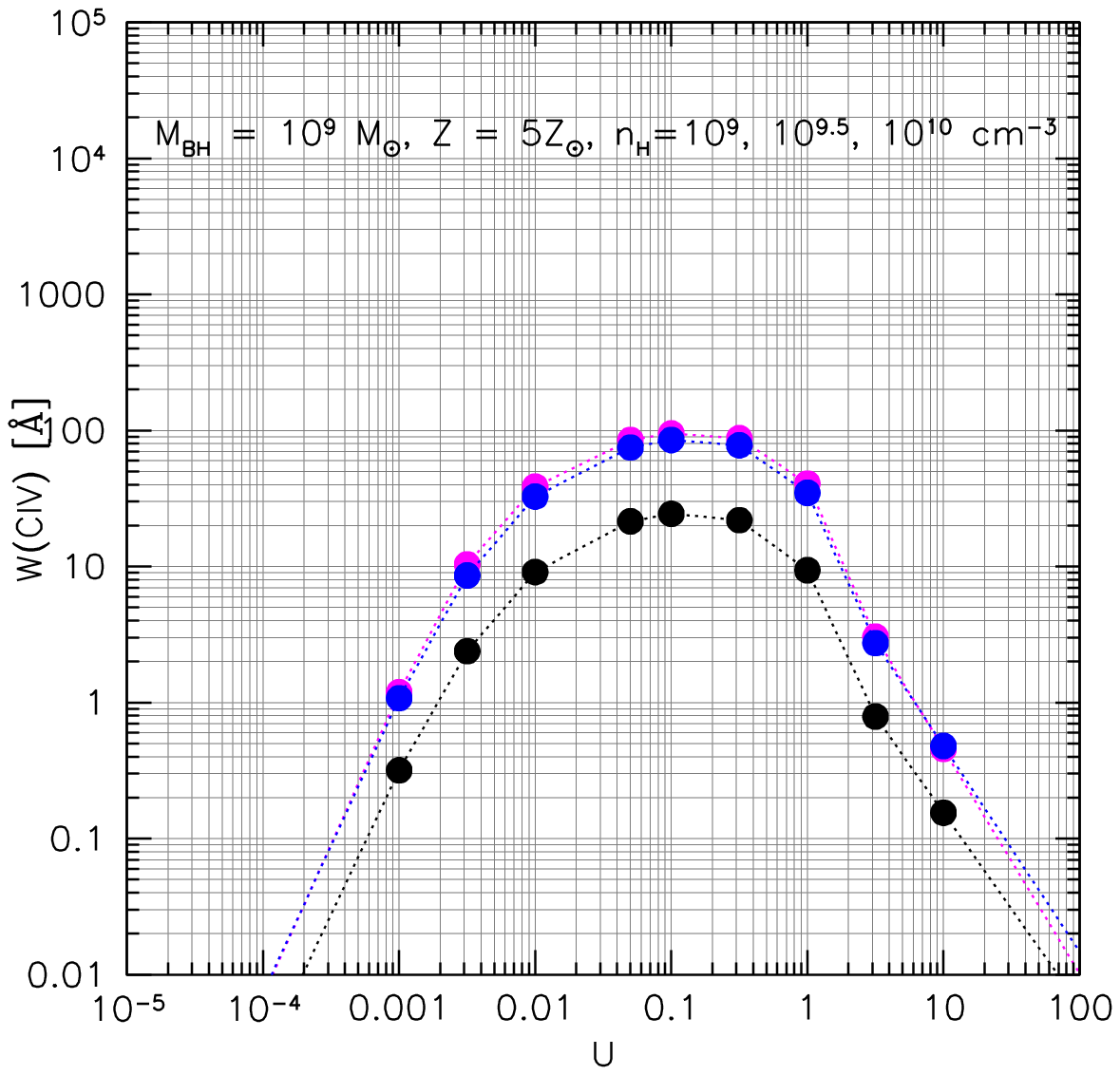}
    \includegraphics[width=0.49\linewidth]{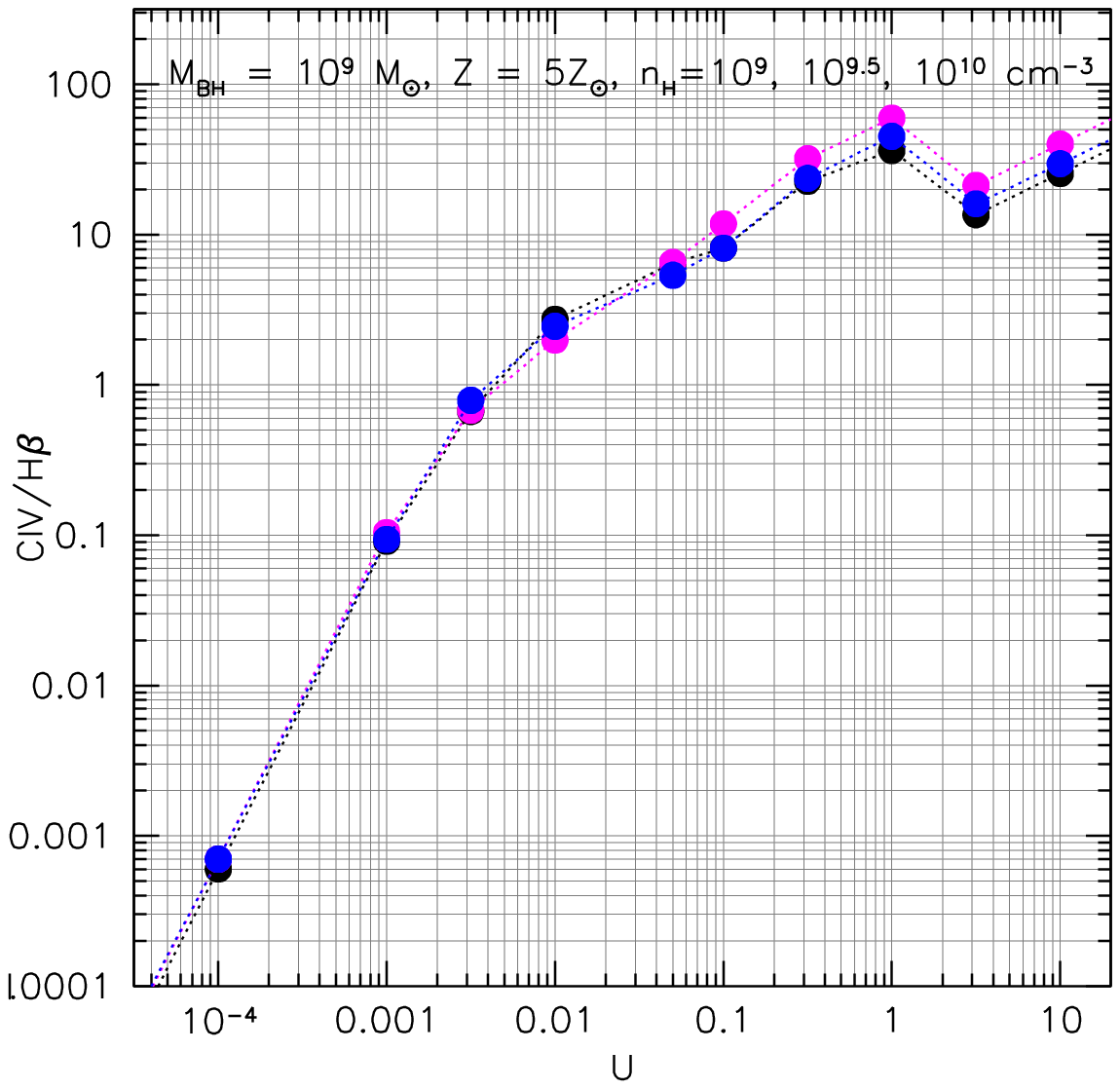}\\
    \caption{Relation between $W$ \civ{}  (left panels) and intensity ratio \civ/\hb\ (right panels) with the ionization parameter $U$\ for three density values $\log$   $n_\mathrm{H}=9,9.5,10$ [cm$^{-3}$] (black, blue, and magenta symbols, respectively), column density $N_\mathrm{c} = 10^{22}$\ cm$^{-2}$, 5 times solar metallicity. Top: For $\log$\,\mbh =\,$10^9$\,[$M_\odot$], and a \citet{mathewsferland87} SED. 
    Bottom: $\log$\,\mbh = $10^{9.5}$\,[$M_\odot$], and a \citet{krawczyketal13} SED.
    }
    \label{fig:civew}
\end{figure}
\begin{figure}[t!]
    \centering
    \includegraphics[width=0.8\linewidth]{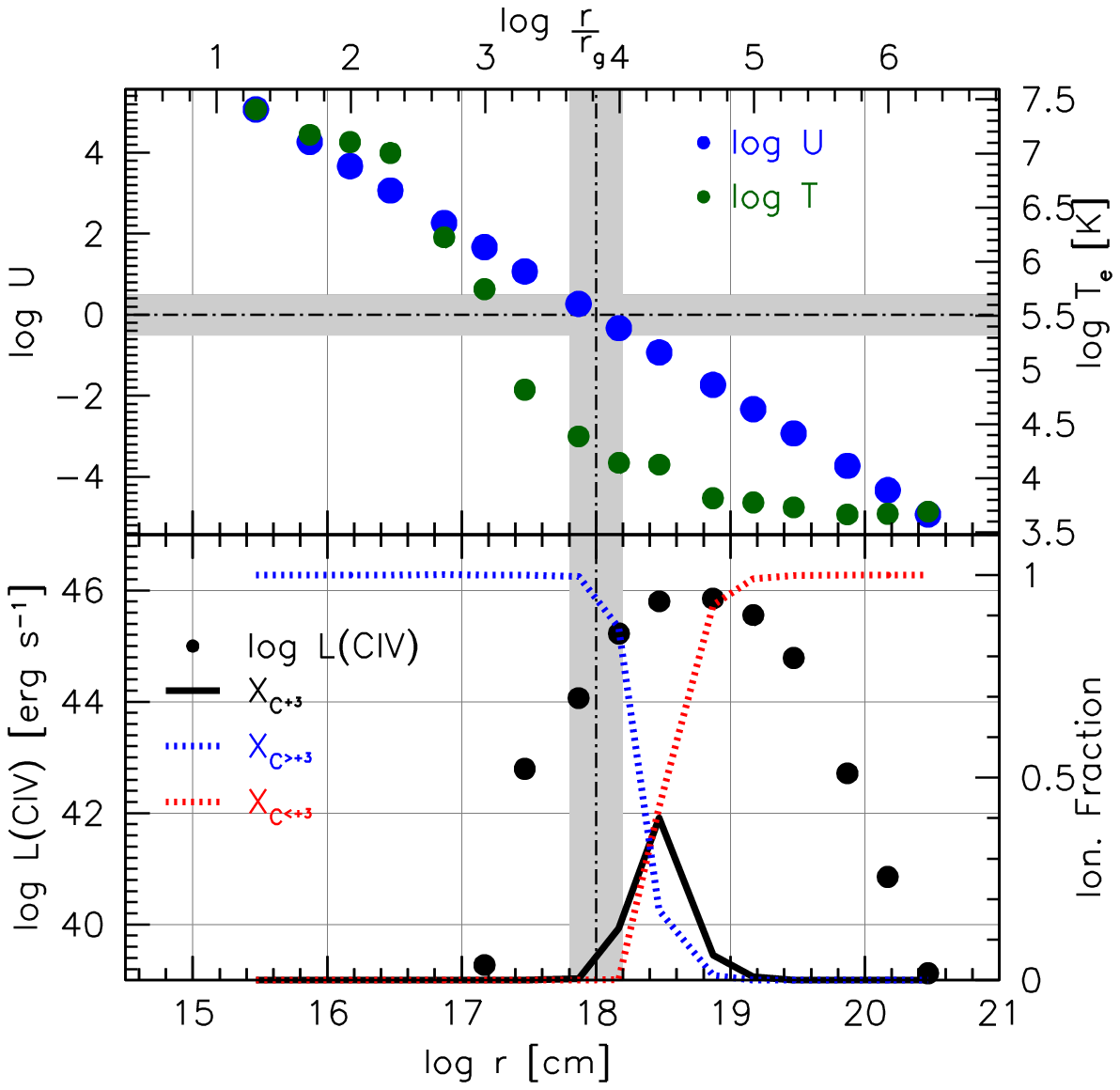}\\
    \includegraphics[width=0.8\linewidth]{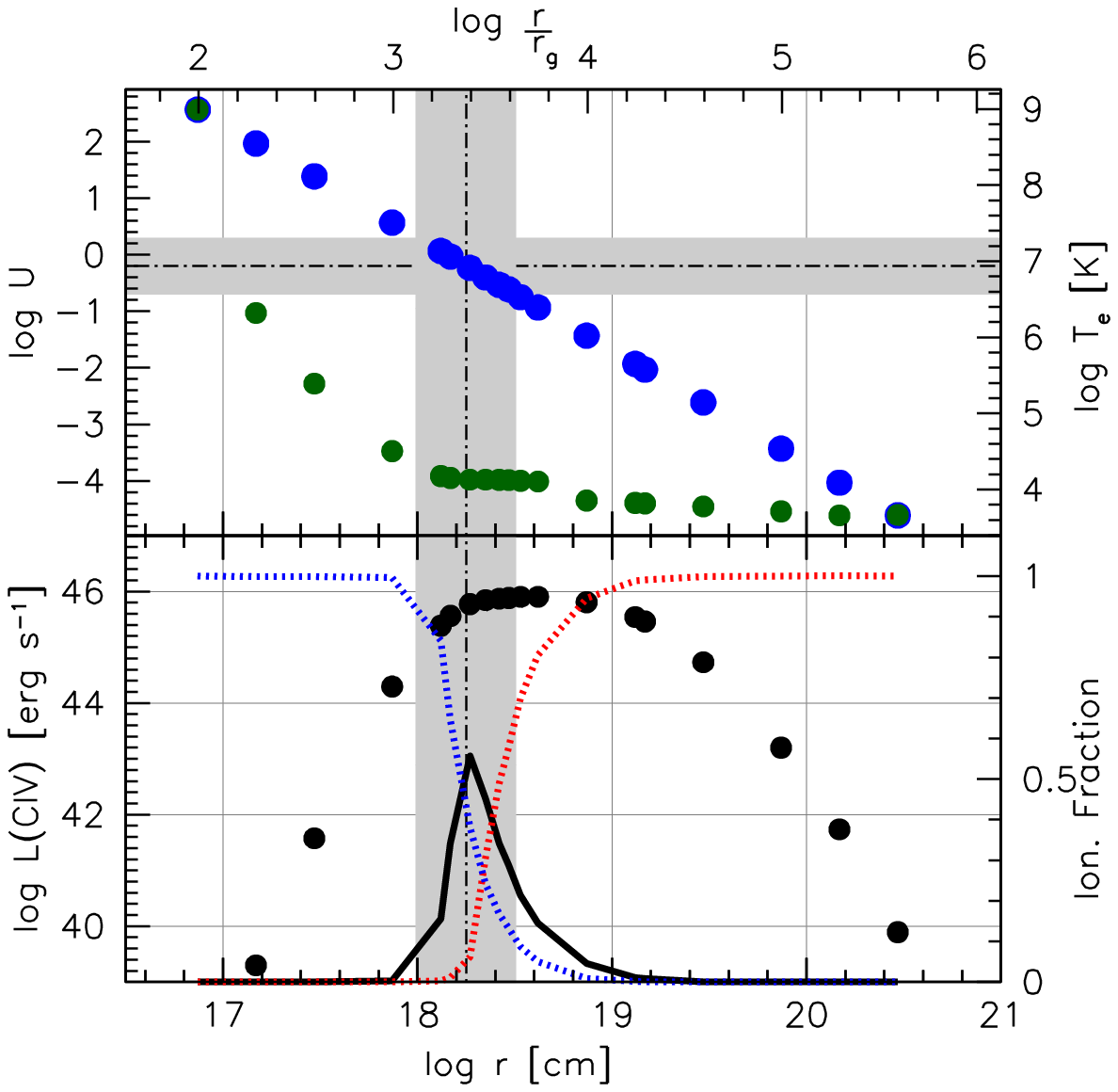}

    \caption{Behavior of ionization parameter, electron temperature, \civ\ luminosity, and ionic fraction as a function of radius for density value $\log n_\mathrm{H} = 9.5$ [cm$^{-3}$], {and for the two adopted SEDs for the low (top panels) and high-z (bottom panels) samples}.  \textit{Top, upper panel:} behavior of the ionization parameter (left axis scale, blue dots) and electron temperature (right axis scale, green dots) as a function of radius for a \mbh\,=10$^9$ M$_\odot$ radiating at Eddington limit and illuminating a slab of gas.  
    We assume column density $N_\mathrm{c} = 10^{22}$\ cm$^{-2}$, { \citet{mathewsferland87} SED,} and the metallicity five times solar in all cases. \textit{Top, lower panel:} logarithm of \civ\ luminosity (black dots) and ionic fraction as a function of radius. The blue and red line represent the ionic stages higher and lower than 3, respectively, and the { black circles}  ionic stage 3.
    \textit{Bottom, upper and lower panels}: same as top panels, for $\log M_\mathrm{BH} =$ 9.5 [M$_\odot$], and SED from \citet{krawczyketal13}.  
    { In both the top and bottom panels, the horizontal grey bands indicate the range of ionization parameters consistent with the observed \civonly\ equivalent width and the \civonly/\hb\ ratio. The vertical bands represent the corresponding range of permitted radii.}
     \label{fig:tr09}}
 \end{figure}


Fig. \ref{fig:tr09} shows the trends expected for the ionization parameter and electron temperature  as a function of radius, for the value of the density $\log n_\mathrm{H} = 9.5$  [cm$^{-3}$].  The intercepts between the expected range of ionization parameter and radius does not fall in correspondence of the peak ionization fraction   { in the case of the \citet{mathewsferland87} SED, while it does in the high-luminosity case due to the ``softer'' \citet{krawczyketal13} SED  } (bottom panels of Fig. \ref{fig:tr09}). The lower limit assumed for $\log U \approx -0.5$\ implies a condition of over-ionization for the line emitting gas. Electron temperature is around  $T_\mathrm{e} \approx 20\,000$ K.   



The photoionization computations predict an equivalent width more than an order of magnitude higher than the observed value if  \civ\ emission is due to gas at $\log U \sim -1.3$ { in the case of the \citet{mathewsferland87}} SED appropriate for low-$z$\ sources (upper right panel of Fig. \ref{fig:civew}).   In addition, the distance expected for $\log U = -1.3$\ and $\log n_\mathrm {H} \approx 9.0$ is $\log r \approx 18.8$ [cm],   more than one order of magnitude higher than the scaling radius of \citet{Kaspi_2021},  $\log r \approx 17.43$ [cm] { (see the ionic fraction distribution in the lower half of the upper panel of Fig. \ref{fig:tr09}). } 


{Taking into account previous considerations, we use for the ionized mass}  the following expression: 
\vspace{-0.2cm}
\begin{equation}
      M_\mathrm{ion} \approx 6.5 \cdot 10^2\ L_\mathrm{CIV,45}^{\mathrm{out}} \left[\frac{Z}{5Z_\odot}\right]^{-1} n_\mathrm{H,9.5}^{-1}\ [\mathrm{M_\odot}] \label{eq:mion1}
\end{equation}

\noindent which incorporates the average over density in the range $10^{9.5} - 10^{10.5}$\ cm$^{-3}$\ of $Xq(T)$\ for $\log U = 0$,
$\overline{Xq(T)} \approx 3.3 \cdot 10^{-11}$\ cm$^3$\ s$^{-1}$. Averaging over $\log U\,=\,-0.5,\,0,\,+0.5$, would imply an increase of $\overline{Xq(T)}$ to $ \approx 5.65\cdot10^{-11}$\, cm$^3$\,s$^{-1}$, by a factor $\approx 1.7$. Such increase in the efficiency in \civonly\ production would only reinforce the conclusion on the weakness of the outflows at low-$z$.


\subsubsection{Wind dynamical parameters}
\label{appendix:civ_wind_parameters}

To estimate the wind dynamical parameters we assume the  same simple model utilized for \oiii. 
The mass outflow rate is therefore given by:

\vspace{-0.2truecm}
\begin{equation}
    \dot{M}_\mathrm{ion} \approx 10\, L^\mathrm{out}_{\rm CIV,45}\ v_{5000}\ r_{\mathrm{1pc}}^{-1} \left(\frac{Z}{5Z_\odot}\right)^{-1}  n_\mathrm{H,9.5}^{-1}\ [\textrm{M}_\odot\ \mathrm{yr}^{-1}]
    \label{eq:mout}
\end{equation}
 {where $L^\mathrm{out}_{45}$ is the luminosity of the blueshifted component in units of 10$^{45}$\,erg\,s$^{-1}$, the outflow velocity $v_{5000}$ is scaled to 5000\,\kms, the gas density is in units of $10^{\,9.5}$, and the radius is now in parsec units. For this paper, we assume that the \civonly{} outflow is accelerated to a final, terminal velocity $v\,\approx\,c(1/2)\,+\,2\sigma\,\approx\,c(1/2)\,+\,{\mathrm{FWHM}_{\mathrm{BLUE}}}/1.18$, where $c(1/2)$ corresponds to the centroid velocity at half intensity of the \civonly{} BLUE component.}
\\

The  thrust $\dot{M}_\mathrm{ion}^\mathrm{out} v $\ and kinetic power can then  be computed  as:

\vspace{-0.2cm}
\begin{equation}
\dot{M}_\mathrm{ion}v\approx 3.15 \cdot 10^{35} L^\mathrm{out}_\mathrm{CIV,45}\,v_{5000}^2\,r_{\rm 1pc}^{-1}\left(\frac{Z}{5Z_\odot}\right)^{-1}n_\mathrm{H,9.5}^{-1}\  [\mathrm{g\,cm\,s^{-2}}]\label{eq:thrust}
\end{equation} 
\begin{equation}
 {\dot{E}_{kin}=\frac{1}{2}\dot{M}_\mathrm{ion} v^{2} \approx 7.9\cdot10^{43} L^\mathrm{out}_\mathrm{CIV,45}v_{5000}^3 r_{\rm 1pc}^{-1}\left(\frac{Z}{5Z_\odot}\right)^{-1}n_\mathrm{H,9.5}^{-1}\ [\textrm{erg s}^{-2}}]\label{eq:kp}
\end{equation}

These expressions, {used  to calculate the dynamical parameters of the outflows for our low-z FOS sample}, are quantitatively similar to the ones reported in earlier works \citep{marzianietal16,Marziani_2017}. Those authors assumed an ad hoc temperature to account for the low \civ\ radiative efficiency of the emitting gas suggested by the observed low equivalent width of \civ.  


{At high luminosity, for the ISAAC and HEMS samples, and by considering the results from the CLOUDY computations by using the more appropriate SED from \citet{krawczyketal13} and for an estimated $\log U \approx -0.2$, and an efficiency, $\overline{Xq(T)} \approx 1.22\cdot10^{-10}$\,cm$^3$\,s$^{-1}$ (see section \ref{appendix:civ_modeling}), Eqs.} 
\ref{eq:mion1}, \ref{eq:mout}, \ref{eq:thrust}, and \ref{eq:kp} should be divided by a factor $\approx  3.63$.  
Eqs. \ref{eq:mion1}, once the constants are divided by the factor reported above yields estimates that are in basic agreement with those of \citet{vietrietal20}: they differ by  a factor $\sim$ 2-3 in $M_\mathrm{ion}$, and the difference comes from the assumed  efficiency of the \civ\ emission. Additional  factors  accounting for differences in the estimated values concern the geometry: a factor $3$\ from the assumption of a thin flat layer in place of a spherical surface element,  and {an additional factor of} 2 for the unseen contribution of the receding cone enters in Eqs. \ref{eq:mion1}, \ref{eq:mout},  \ref{eq:thrust} and \ref{eq:kp}. 

\end{appendix}

\end{document}